\pdfoutput=1
\documentclass{article}

\PassOptionsToPackage{usenames,dvipsnames,svgnames,table}{xcolor}
\usepackage{tikz}
\usepackage{tkz-graph}  
\usepackage{graphicx}
\usetikzlibrary{positioning,arrows.meta,math,shapes.geometric,decorations.pathmorphing,decorations.pathreplacing,calc,trees}

\usepackage[utf8]{inputenc}
\usepackage[margin=1in]{geometry}
\usepackage[normalem]{ulem}
\usepackage{mathtools,bm}
\usepackage[inline]{enumitem}
\usepackage{amsmath,amsfonts,amssymb,amsthm,hyperref,algorithm,eqparbox,array,ragged2e,stmaryrd,bbm}
\usepackage{xspace}
\usetikzlibrary{arrows.meta,decorations.markings}

\usepackage{tocloft}
\usepackage[nottoc]{tocbibind} %

\usepackage[usenames,dvipsnames,svgnames,table]{xcolor}

\usepackage{ifthen}
\usepackage{subfig}

\usepackage{thm-restate}
\usepackage{enumitem}
\usepackage[capitalise,nameinlink,noabbrev]{cleveref}

\hypersetup{
	colorlinks=true,
	urlcolor=MidnightBlue,
	linkcolor=MidnightBlue,
	citecolor=MidnightBlue,
	unicode
}
\usepackage{algorithm}
\usepackage{algpseudocode}
  \DeclareSymbolFont{extraup}{U}{zavm}{m}{n}
  \DeclareMathSymbol{\vardiamond}{\mathalpha}{extraup}{87}

\newtheorem{theorem}{Theorem}
\newtheorem*{theorem*}{Theorem}
\newtheorem{lemma}[theorem]{Lemma}

\newtheorem{claim}[theorem]{Claim}
\newtheorem{corollary}[theorem]{Corollary}
\newtheorem*{corollary*}{Corollary}

\newtheorem{fact}{Fact}
\theoremstyle{remark}
\newtheorem{remark}{Remark}
\crefname{claim}{Claim}{Claims}
\Crefname{claim}{Claim}{Claims}
\crefname{lemma}{Lemma}{Lemmas}
\Crefname{lemma}{Lemma}{Lemmas}

\theoremstyle{definition}
\newtheorem{definition}{Definition}

\makeatletter
\def\th@example{%
  \thm@notefont{}%
  \normalfont %
}
\def\th@definition{%
  \thm@notefont{}%
  \normalfont %
}
\makeatother

\theoremstyle{example}

\renewcommand{\setminus}{\smallsetminus}
\renewcommand{\hat}{\widehat}

\DeclareMathOperator*{\Exp}{\mathbb{E}}

\newcommand{\y}{\bm{y}}

\renewcommand{\epsilon}{\varepsilon}

\newcommand{\fix}{\textsf{\upshape fix}}

\newcommand{\lb}{\llbracket}
\newcommand{\rb}{\rrbracket}

\newcommand{\poly}{\mathrm{poly}}
\newcommand{\polylog}{\mathrm{polylog}}

\newcommand{\err}{\mathrm{err}}
\newcommand{\calH}{\mathcal{H}}

\newcommand{\spn}{\mathrm{span}}

\newcommand{\newmeasure}[2]{\newcommand{#1}{{\textup{\sffamily #2}}\xspace}}
\newmeasure{\C}{C}

\newcommand{\TFNP}{\textsf{TFNP}}
\newcommand{\UEoPL}{\textsf{UEoPL}}
\newcommand{\EoPL}{\textsf{EoPL}}
\newcommand{\PLS}{\textsf{PLS}}

\newcommand{\PPP}{\textsf{PPP}}
\newcommand{\PPA}{\textsf{PPA}}
\newcommand{\PPAD}{\textsf{PPAD}}

\newcommand{\act}{\textsf{act}}

\newcommand{\ZeroEdge}{\textsc{ZeroEdge}}
\newcommand{\FP}{\textsf{FP}}
\newcommand{\NP}{\textsf{NP}}
\newcommand{\SAT}{\textsf{SAT}}
\newcommand{\SSAT}{\textsf{\#}\SAT}

\renewcommand{\fix}{\textsf{fix}}
\newcommand{\free}{\textsf{free}}
\renewcommand{\dim}{\mathrm{dim}}
\newcommand{\cube}{\textsf{cube}}
\newcommand{\range}{\textsf{range}}
\newcommand{\mate}{\textsc{Mate}}

\newcommand{\PWPP}{\textsf{PWPP}}

\newcommand{\extensionProp}{\hyperref[def:extension]{extension property}\xspace}

\newcommand{\F}{\mathbb F}

\newcommand{\uNS}{\mathsf{uNS}}
\newcommand{\NS}{\mathsf{NS}}

\let\oldprod=\prod %

\RenewDocumentCommand{\prod}{e{_^}}{ %
  \vphantom{\oldprod_{.}} %
  \mathop{\smash{\oldprod %
    \IfValueT{#1}{_{#1}} %
    \IfValueT{#2}{^{#2}}
  }}
}

\newcommand{\artur}[1]{}%

\newcommand{\chatgpt}[1]{}%

\newcommand{\noah}[1]{}%

\newcommand{\toni}[1]{}%

\newcommand{\ziyi}[1]{}%

\newcommand{\mitro}[1]{}%

\newcommand{\weiqiang}[1]{}%

\newcommand{\oliver}[1]{}%

\newcommand{\tianqi}[1]{}%

\newcommand{\dima}[1]{}%

\newcommand{\Pigeon}{\textsc{Pigeon}}
\newcommand{\Succ}{\textsc{Succ}}
\newcommand{\Mate}{\textsc{Mate}}
\newcommand{\Pred}{\textsc{Pred}}
\begin{document}
\title{TFNP in the Random Oracle Model}
\author{\normalsize Noah~Fleming\thanks{Lund University. \texttt{noah.fleming@cs.lth.se}} \and
  Mika~G\"o\"os\thanks{EPFL. \texttt{mika.goos@epfl.ch}} \and
  Ziyi~Guan\thanks{MIT. \texttt{ziyiguan@mit.edu}} \and
  Oliver~Korten\thanks{IAS. \texttt{korten@ias.edu}} \and
  Daniel~Mitropolsky\thanks{Tufts University, MIT. \texttt{mitropol@mit.edu}} \and
  Toniann~Pitassi\thanks{Columbia University. \texttt{tp2684@columbia.edu}} \and
  Artur~Riazanov\thanks{Columbia University. \texttt{tunyash@gmail.com}} \and
  Dmitry~Sokolov\thanks{Universit\'e de Montr\'eal. \texttt{dmitry.sokolov@umontreal.ca}} \and
  Tianqi~Yang\thanks{Columbia University. \texttt{tianqi@cs.columbia.edu}} \and
  Weiqiang~Yuan\thanks{University of Copenhagen. \texttt{weiqiang.yuan@di.ku.dk}}}
\date{}
{\makeatletter\renewcommand*{\@fnsymbol}[1]{\@arabic{#1}}\renewcommand\and{\end{tabular}\hskip 0.8em \@plus.17fil\begin{tabular}[t]{c}\normalsize}\makeatother\maketitle}

\begin{abstract}
We determine which major $\TFNP$ classes are hard in the random oracle model (ROM). Prior work establishes average-case hardness of highly structured subclasses of $\TFNP$ from strong cryptographic assumptions such as LWE and indistinguishability obfuscation. It is not known whether weaker assumptions such as one-way functions imply hardness of any of the standard $\TFNP$ subclasses, even in the worst case. The ROM, where all parties have query access to a random function, lies, morally, between these two regimes: a random function is one-way, but public-key cryptography cannot be built from it in a black-box way. We refer to this setting as \emph{Romantica}.

Our first result settles a long-standing open problem: $\PPAD$, the class of total search problems reducible to finding a Nash equilibrium, is hard on average in Romantica. Combined with the known hardness of PLS (local search), this gives average-case hardness of $\EoPL = \PPAD \cap \PLS$. In contrast, we show that $\UEoPL$ can be solved in Romantica with polynomially many oracle queries by a computationally unbounded solver, a sharp distinction between $\EoPL$ and $\UEoPL$. This resolves the black-box hardness landscape of all major $\TFNP$ classes in Romantica.

Finally, we initiate the study of separations between $\TFNP$ subclasses in Romantica: $\PWPP$ and $\PPA$ do not reduce to $\PLS$ under \emph{public-coin} reductions. 
\end{abstract}

{\setcounter{tocdepth}{1}\setlength{\cftbeforesecskip}{4pt}\renewcommand{\contentsname}{Table of Contents}\tableofcontents}
\clearpage

\section{Introduction}

$\TFNP$ is the class of \emph{total search problems} in $\NP$, those for which solutions always exist and can be efficiently verified. $\TFNP$ is studied through its subclasses, each capturing a \emph{principle of totality}. The smallest, $\FP$, consists of the total problems solvable in polynomial time; every other subclass is defined by reduction to a canonical complete problem. Four well-studied classes are $\PPP$, $\PWPP$, $\PLS$ and $\PPAD$. In $\PPP$, for the \emph{pigeonhole principle}~\cite{Pap94}, the complete problem is to find a collision $C(x) = C(x')$ of a circuit $C\colon [2^n] \rightarrow [2^{n}-1]$. $\PWPP$, for the \emph{weak} pigeonhole principle~\cite{Jer16}, has the same complete problem but now, $C\colon [2^n] \rightarrow [2^{n-1}]$ has twice as many inputs as outputs. The third, $\PLS$, captures \emph{local search}~\cite{JPY88}: %
given a monotonically \emph{increasing} function $f\colon [2^n]\rightarrow [2^n]$, find a point at which it does not increase. Finally, $\PPAD$ captures the hardness of finding a Nash equilibrium in bimatrix games~\cite{DGP09,CDT09}, although the original complete problem rests on a \emph{parity argument on directed graphs}~\cite{Pap94}. Formal definitions are in \cref{glossary}. %

\paragraph{TFNP hardness and cryptography.} The theory of total search problems studies under what conditions $\TFNP$ subclasses are \emph{hard}, or \emph{hard on average}, as evidence for the computational intractability of total problems such as finding a Nash equilibrium. It is unlikely that $\NP$-hardness alone will suffice: if $\TFNP$ contains an $\NP$-hard problem, then $\NP = \text{co-}\NP$ \cite{MP91}. Over the past decade, researchers have studied \emph{cryptographic} assumptions as a source of hardness for $\TFNP$. Conversely, this maps to studying which total problems are hard in each of Impagliazzo's ``worlds''~\cite{Imp95}. %
All cryptographic assumptions trivially imply the hardness of $\NP$, and are believed to in fact imply more, e.g. hardness of some $\NP$-\emph{intermediate} classes. $\TFNP$ is a candidate for such a class.%

The main line of work in this field has centered on proving $\UEoPL$ average-case hardness from \emph{strong} cryptographic assumptions, i.e. in the higher worlds. $\UEoPL$~\cite{FGMS20} is the lowest of the standard $\TFNP$ subclasses: it is contained in every class mentioned so far. %
Average-case hardness of $\UEoPL$ was first shown from indistinguishability obfuscation \cite{BPR15} (see also \cite{cryptoeprint:2026/2333}), that is, in Obfustopia, and later from public-key assumptions \cite{GPS16, LV20, KPY20, JKKZ21, BCHKLPR22, KLV23, CFJW25}, that is, in Cryptomania. %

In Minicrypt, the world where \emph{one-way functions} (OWFs) exist but no public-key assumptions hold, $\TFNP$ itself is hard on average, either non-uniformly or under an additional derandomization assumption~\cite{HNY17}. However, no problem in any of its standard subclasses is known to be hard in Minicrypt, not even in the worst case, whereas the stronger assumptions above give average-case hardness of $\UEoPL$, the most structured class of all. %
Barriers against restricted black-box constructions suggest that OWFs may not suffice \cite{RSS21,HKKS20,FGHMY24}. 

\paragraph{The random oracle model.} Is public-key cryptography necessary? A natural starting point for this question is the random oracle model (ROM)~\cite{BR93}, a setting often studied in cryptography when OWFs appear to be insufficient. Briefly, in the ROM, all parties (instance samplers, solvers, verifiers, etc.) have query-access to a randomly sampled function (see \cref{sec:tfnp-rom-def-short} for a complete definition). %
A random function is itself one-way, so morally, the ROM ``contains'' Minicrypt. On the other hand, public-key cryptography cannot be built in the ROM in a black-box way~\cite{IR89} (and hence not from OWFs either). The world with a random oracle, which we call {\color{DeepPink}\textbf{Romantica} $\heartsuit$}, therefore sits roughly between Minicrypt and Cryptomania.\footnote{Neither world contains the other. A random oracle is a one-way function, so everything built from one-way functions in a black-box way exists in Romantica, but not everything Minicrypt may contain: for instance, no one-way permutation can be built from a random oracle in a black-box way~\cite{Rudich,KSS10}. Conversely, Romantica contains unconditional primitives that are not known to follow from public-key cryptography, such as succinct arguments~\cite{Micali00} and proofs of sequential work~\cite{DBLP:conf/eurocrypt/CohenP18}.} 
A natural question is whether $\TFNP$ or its subclasses are hard in Romantica. Hardness in Romantica means an efficient sampler with oracle access whose output instances, which may contain oracle gates, no poly-time solver with the same oracle can solve. Three $\TFNP$-subclasses were previously known to be hard on average in Romantica: $\PPP$ and $\PWPP$ (whose hard distributions are simple), as well as $\PLS$, which requires a non-trivial construction based on proofs of sequential work \cite{BG20}. If moreover $\#\text{SAT}$ is hard, then $\UEoPL$, and hence every class above it, is hard on average in Romantica \cite{CHKPRR19,BCHKLPR22}. Whether any other subclass, in particular $\PPAD$, $\UEoPL$, $\PPA$ or $\EoPL$, is hard in Romantica without additional assumptions has been open until now.

\subsection{Our results}
We resolve, for every standard $\TFNP$ subclass, whether it is hard on average in Romantica; \cref{fig:summary-of-results} summarizes the results. Overall we show: $\PPAD$ is hard on average, and hence so is $\EoPL = \PPAD \cap \PLS$, whereas $\UEoPL$ collapses to $\FP$ in the \emph{black-box} sense: a solver making polynomially many oracle-queries exists. Finally, we initiate the study of separations between $\TFNP$ subclasses in Romantica: neither $\PWPP$ nor $\PPA$ reduces to $\PLS$ under public-coin reductions. We now state these results more fully in turn.

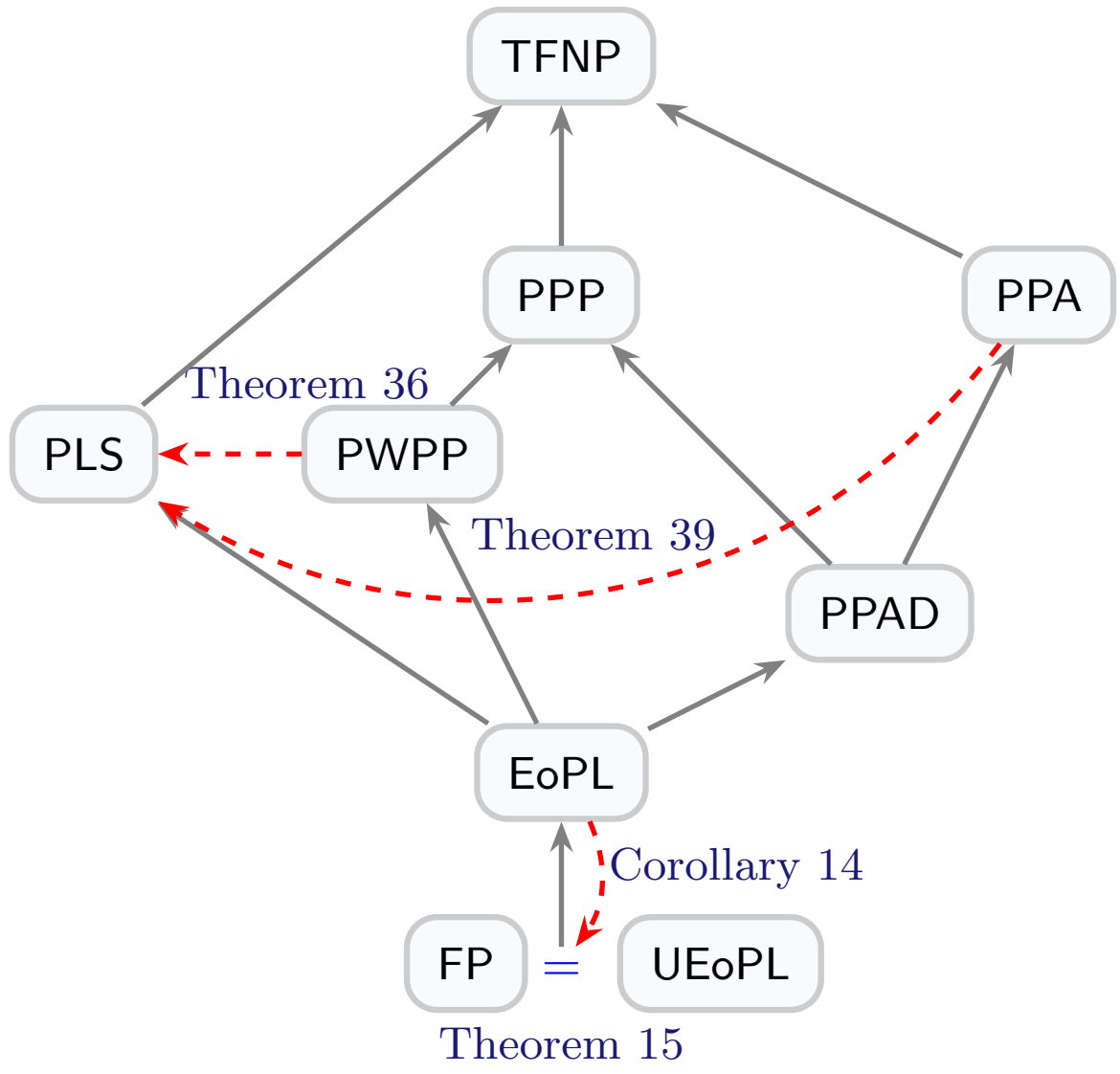
\begin{figure}[t]
    \centering

\begin{tikzpicture}[scale=1.1]
\tikzset{inner sep=0,outer sep=3}

\tikzstyle{a}=[inner sep=6pt, inner ysep=6pt,outer sep=0.5pt,
draw=black!20!white, fill=Cerulean!3!white, very thick, rounded corners=6pt, align=center]
\tikzstyle{b}=[color=red]
\tikzstyle{c}=[color=blue]

\node[a] (TFNP) at (0,6.5) {\small $\TFNP$};

\node[a] (PLS) at (-3,4) {\small $\PLS$};

\node[a] (PPP) at (0,5) {\small $\PPP$};

\node[a] (PWPP) at (-1,4) {\small $\PWPP$};

\node[a] (PPA) at (3,5) {\small $\PPA$};

\node[a] (PPAD) at (2,3) {\small $\PPAD$};

\node[a] (EOPL) at (0,2) {\small $\EoPL$};

\node[a] (UEOPL) at (1,0.8) {\small $\UEoPL$};

\node[a] (FP) at (-0.6,0.8) {\small $\FP$};

\node[c] (equals) at (0,0.75) {$=$};

\node[b] (thmPWPPvsPLS) at (-1.6,4.45) {\small \Cref{thm:PWPP_not_in_PLS}};

\node[b] (thmPPAvsPLS) at (0.2,3.5) {\small \cref{thm:ppa-vs-pls}};

\node[c] (thmFP=UEOPL) at (0,0.3) {\small \cref{thm:ueopl}};

\node[b] (thmEOPLsepFP) at (1.1,1.4) {\small \cref{cor:eopl-hardness}};

\path[-{Stealth[length=6pt]},line width=1pt,gray]
(equals) edge (EOPL)
(EOPL) edge (PPAD)
(EOPL) edge (PWPP)
(EOPL) edge (PLS)
(PPAD) edge  (PPP)
(PPAD) edge (PPA)
(PPP) edge (TFNP)
(PLS) edge (TFNP)
(PWPP) edge (PPP)
(PPA) edge (TFNP);
\path[-{Stealth[length=6pt]},line width=1pt,red, dashed]
(PWPP) edge (PLS);

\path[-{Stealth[length=6pt]},line width=1pt,red, dashed, bend left = 30]
(EOPL) edge (equals);

\path[-{Stealth[length=6pt]},line width=1pt,red, dashed, bend left = 42]
(PPA) edge (PLS);

\end{tikzpicture}

\caption{Separations, inclusions, and equivalences in the ROM (\emph{Romantica}), red arrows are average-case \emph{separations} and are the results of this paper. Average-case hardness is equivalent to average-case separation from $\FP$.}
    \label{fig:summary-of-results}
\end{figure}

\begin{theorem*}[\cref{thm:main-ppad-hardness}, Informal]
\label{thm:ppad-hard-informal}
$\PPAD$ is hard on average in the random oracle model. The hard distribution is efficiently samplable, and any solver making polynomially many oracle queries finds a solution with negligible probability.
\end{theorem*}

The proof modifies the Yamakawa--Zhandry hypergraph~\cite{YZ24}, originally used to separate quantum from classical query algorithms, so that the search problem becomes total while remaining hard; see \cref{subsection:PPAD}.

\cref{thm:main-ppad-hardness} and the known hardness of $\PLS$~\cite{BG20} imply, via a simple or-composition, average-case hardness of $\PLS \cap \PPAD$. This intersection equals $\EoPL$~\cite{FGHS,GHJ22}, the class of problems reducible to finding the end of a potential line~\cite{FGMS20}.

\begin{corollary*}[\cref{cor:eopl-hardness}, Informal]
$\PPAD \cap \PLS = \EoPL$ is hard on average in the random oracle model.
\end{corollary*}

Since $\PPA \supseteq \PPAD$, every class in \cref{fig:summary-of-results} except $\UEoPL$ is now known to be hard on average in Romantica. The remaining question is whether $\UEoPL$, the lowest widely studied non-trivial $\TFNP$ subclass (and in particular a subclass of $\EoPL =\PPAD \cap \PLS$), is also hard on average in Romantica. We answer this essentially in the \emph{negative}, highlighting a key structural difference between $\UEoPL$ and $\EoPL$ with respect to random oracles:

\begin{theorem*}[\cref{thm:ueopl}, Informal] \label{thm:ueopl-easy-informal}
In the random oracle model, given any poly-time samplable distribution over instances of $\UEoPL$, there exists an algorithm that finds solutions with probability $\geq 2/3$ and uses at most polynomially many queries to the random oracle.
\end{theorem*}

\cref{thm:ueopl} concerns query efficiency, not time efficiency: the solver makes polynomially many oracle queries, but its computation need not be efficient. Indeed, as mentioned previously, $\UEoPL$ is known to be hard on average in Romantica with the additional assumption that $\#\text{SAT}$ is hard on average~\cite{CHKPRR19}. This suggests that in Romantica, a computational assumption is necessary (and sufficient) to establish $\UEoPL$-hardness. The proof rests on a recent dichotomy lemma~\cite{BCG26}; see \cref{subsec:UEoPL-overview}.

To our knowledge, all previous average-case hardness results for $\PPAD$ and $\EoPL$ establish hardness of $\UEoPL$ and inherit it by containment. In Romantica, \cref{thm:ueopl} rules out this route, so \cref{thm:main-ppad-hardness} is the first $\PPAD$ hardness result that does not go through $\UEoPL$.

Finally, we also initiate the study of separations between different $\TFNP$ subclasses in Romantica: $\PWPP$ and $\PPA$, the class for \emph{parity arguments} on undirected graphs~\cite{Pap94}, \emph{cannot} be reduced to $\PLS$ under \emph{public reductions}, whose only randomness is the random oracle itself:
\begin{theorem*}[\cref{thm:PWPP_not_in_PLS}, Informal]
In the random oracle model, there is no efficient public reduction from a $\PWPP$-complete problem to a $\PLS$-complete problem.
\end{theorem*}

\begin{theorem*}[\cref{thm:ppa-vs-pls}, Informal]
In the random oracle model, there is no efficient public reduction from a $\PPA$-complete problem to a $\PLS$-complete problem.
\end{theorem*}

Our proofs reduce these separations to proof-complexity lower bounds; see \cref{subsec:pls-separations-overview}. This step relies on the reductions being public. The oracle is visible to the solver of the target problem, so a public reduction hides nothing from the solver, who may choose which solution to return with full knowledge of the reduction. A \emph{private-coin} reduction may in addition toss coins hidden from the solver, which may help the reduction. Separating $\TFNP$ subclasses in Romantica under private-coin reductions remains open, as do separations between other pairs of subclasses. 

\paragraph{A proof-complexity perspective.}\sloppy
Black-box $\TFNP$ subclasses correspond exactly to proof
systems~\cite{BIKPP94,BussFI23,GHJ22}, and passing to the random oracle model, a reduction may
 now fail on
a small fraction of oracles. From the proof side, this corresponds to proofs that
are  \emph{approximately sound}, meaning that they are correct on all but a small fraction of
\emph{uniformly random}  inputs. 
In particular, a ROM reduction to $\PLS$ is a narrow
random resolution refutation with respect to a locally samplable distribution
(see \cref{sec:PLS_separations})
so \cref{thm:PWPP_not_in_PLS,thm:ppa-vs-pls} are lower bounds for
approximately sound proofs.
We note that for stronger proof systems such as bounded-depth Frege systems, a recurring issue is how to define approximate soundness in a meaningful way. Note that for any Frege proof (even of low depth) of a tautology, every line in the proof is semantically equivalent to ``1'' so any useful notion of an approximate proof must incorporate the local nature of the rules in order to avoid this pitfall.

More generally lower bounds for approximate proofs have many applications outside of cryptography: (i) they underlie lower bounds on the extension complexity of
LP and SDP relaxations for approximating NP-hard problems; (ii) they are the object
of study in random resolution, with applications to theories of approximate
counting; (iii) they are exactly the missing ingredient in the Razborov--Smolensky
approach to $\mathrm{AC}^0[p]$-Frege lower bounds that motivated the
introduction of Nullstellensatz and Polynomial Calculus; and (iv) they are the mirror
image of the algorithmic problem of refuting random 3SAT, which reduces to
certifying that a random 3XOR instance is highly unsatisfiable. We discuss the first three connections in \cref{sec:approx-sound}.

\subsection{Related work}

\subsubsection{Average-case TFNP and cryptography}
\paragraph{Unconditional hardness of $\TFNP$ subclasses in Romantica.}

It is not hard to see that the existence of collision-resistant hash functions (CRHFs) implies the average-case hardness of the weak pigeonhole principle. Hence, the average-case hardness of $\PPP$ and $\PWPP$ in the ROM follows from the fact that CRHFs exist in the ROM. Apart from this, the only $\TFNP$-subclass that is known to be hard (and in fact, average-case hard) prior to our work is $\PLS$ \cite{BG20}. Their result is obtained by combining the construction of proofs of work in the ROM \cite{DBLP:conf/eurocrypt/CohenP18} and ideas from the Fiat-Shamir heuristics. 

\paragraph{On the pursuit of $\PPAD$ hardness.}

Showing hardness of $\PPAD$ in the ROM has been a long-standing open problem. Prior to our work, the best known results were conditional hardness for $\UEoPL$ (and hence $\PPAD$) which, in addition to the random oracle, involved a complexity-theoretic assumption like $\SSAT$ being hard on average \cite[Theorem~2]{CHKPRR19}, or hardness of $\EoPL$ from the iterated squaring assumption \cite{CHKPRR19b}. As mentioned previously, this does not contradict our new $\UEoPL$ upper-bound since \cref{thm:ueopl} only proves a \emph{query}-complexity upper bound (i.e., a black-box upper-bound); the algorithm we give is not computationally efficient in the white-box sense.

\cref{thm:main-ppad-hardness} is proved in a very different way compared to previous conditional hardness results. \cite[Theorem~2]{CHKPRR19} is proved by applying Fiat-Shamir to the sum-check protocol for $\SSAT$. It is worth noting that several works have studied the instantiation of the Fiat-Shamir heuristics using concrete cryptographic hash functions, obtaining conditional hardness of $\PPAD$ in the plain model (see, e.g., \cite{LV20,JKKZ21,BCHKLPR22,KLV23,CFJW25}). It remains an open problem whether one can replace the random oracle model in \cref{thm:main-ppad-hardness} with well-studied cryptographic assumptions; black-box hardness does not translate into white-box hardness in general~\cite{KNY19}.

\subsubsection{Average-case TFNP and proof complexity}
\label{sec:approx-sound}
 
Proof complexity and black-box $\TFNP$ classes are intimately connected: each
$\TFNP^{\rm dt}$ subclass is characterized by a proof system~\cite{BussFI23}, with $\PLS$
corresponding to resolution width~\cite{BKT14,kamath2019some}, $\PPA$ to Nullstellensatz
over $\mathbb{F}_2$~\cite{GKRS19}, and $\PPAD$ to unary
Nullstellensatz~\cite{GHJ24}.

When extending these equivalences to the ROM, the relevant new feature is error, which we model by a
low-width ``advice'' DNF $B$ which is satisfied with probability at most $1/3$ over the
instance distribution $\omega(u)$ induced by the uniformly random oracle $u$. In particular,
a ROM reduction from some problem $F$ to $\PLS$ becomes a narrow resolution refutation of $F \vee B$ for an appropriate choice of $B$ satisfying the aforementioned properties.
This is closely related to ``random resolution width''
of~\cite{BKT14}, and more generally to the notion of randomized reductions in $\TFNP$ \cite{FlemingG00RS026}. The key distinction is that in our case, the distribution $\omega(u)$ must be samplable by a low depth decision tree from a random oracle (i.e. a string of random independent bits); this is a significant restriction and rules out many candidate ``hard distributions'' one may attempt to use to prove a lower bound.

Our separations are thus lower bounds for approximately sound proofs. We briefly survey three settings in which such proofs arise.

\paragraph{Extension complexity.}
Lower bounds on the size of LP and SDP relaxations for approximating
CSPs~\cite{CLRS13,LRS15,KMR17} convert, via query-to-communication lifting, a
small relaxation into a low-degree Sherali--Adams or Sum-of-Squares certificate.
The extracted certificate is correct only up to a small error, measured on
average over assignments, which is absorbed into the approximation guarantee.
At the core of such a result is therefore a degree lower bound for approximate
low-degree refutations of a particular gap statement. %

\paragraph{Random resolution.}
Random resolution, defined in~\cite{BKT14} following a suggestion of Dantchev,
formalizes resolution refutations that may err on a small fraction of
assignments; it arose as the propositional counterpart of Je\v{r}\'abek's
theories of approximate counting. The first lower bounds for random resolution were obtained for the colored polynomial local search principle CPLS, and for the pigeonhole principle~\cite{PT19,KT22}.

The relation to the random oracle
model is as follows. A ROM reduction from a totality principle $F$ to $\PLS$
produces %
a narrow resolution
refutation of $F$ together with advice that is correct with high probability over the instance distribution $\omega(u)$, where $u$ is the uniformly random oracle and $\omega$ is a forest of shallow decision trees. That is, a ROM
reduction to $\PLS$ is a narrow random resolution refutation with respect to a \emph{locally samplable} distribution, and a ROM lower bound against $\PLS$ is a random resolution lower bound over that particular distribution. The converse
does not always hold. For example, the previously established  lower bounds for random resolution~\cite{PT19,KT22} yield 
distributions that are hard for random resolution,  
but these distributions are not known to be locally samplable
from 
uniform bits, so they do not imply ROM separations. 
Separations between $\TFNP$ subclasses under deterministic black-box reductions are well understood~\cite{BCEIP98,Mor01,BOM04,HY20,GKRS19,BT22,GHJ24}.
Similarly the randomized separations between pairs of $\TFNP$ subclasses in \cite{Li24,JLRX24,Kis26}, and the oracle separations of the multi-collision classes in \cite{KNY18}, do not apply to the ROM.
In contrast, Bitansky and Gerichter~\cite{BG20} obtained a ROM separation
of $\PLS$ from $\FP$; that is, they constructed a
distribution over $\PLS$ instances that is
efficiently samplable from a uniformly random oracle, but that is hard for $\FP$. 
Similarly we obtain the following ROM separations:  (i) $\PWPP$ vs. $\PLS$ (\cref{thm:PWPP_not_in_PLS}), (ii) $\PPA$ vs. $\PLS$ (\cref{thm:ppa-vs-pls}) and (iii) $\EoPL$ vs. $\FP$ (\cref{cor:eopl-hardness}).

\paragraph{$\mathrm{AC}^0[p]$-Frege and the Razborov--Smolensky method.}
Nullstellensatz~\cite{BIKPP94} and Polynomial Calculus~\cite{CEI96} were
introduced in large part to obtain an analogue, for $\mathrm{AC}^0[p]$-Frege,
of the Razborov--Smolensky method~\cite{Raz87,Smo87}, which replaces each gate
of a circuit by a low-degree polynomial over $\mathbb{F}_p$ agreeing with it on
most inputs. Applied to a proof, the method replaces each line by a low-degree
polynomial correct on most assignments, yielding an approximately sound
low-degree algebraic refutation. \cite{BIK+96} made this precise:
$\mathrm{AC}^0[p]$-Frege lower bounds follow from degree lower bounds for
Nullstellensatz over $\mathbb{F}_p$ with extension axioms formalizing the
approximation step, axioms that are sound only with high probability over a
random choice of auxiliary variables. Strong degree lower bounds for the exact
systems followed~\cite{BIKPP94,Raz98,IPS99,BGIP01}, and these underlie the
Nullstellensatz characterizations of $\PPA$ and $\PPAD$ used in this paper; the robust
half of the program is what remains open. %

\section{Techniques}

We overview the techniques behind our results. \cref{sec:tfnp-rom-def-short} recalls the decision-tree formulation of the random oracle model used throughout. \cref{subsection:PPAD} sketches the proof of $\PPAD$ hardness (\cref{thm:main-ppad-hardness}), \cref{subsec:UEoPL-overview} the $\UEoPL$ algorithm (\cref{thm:ueopl}), and \cref{subsec:pls-separations-overview} the separations from $\PLS$ (\cref{thm:PWPP_not_in_PLS,thm:ppa-vs-pls}). 

\subsection{$\TFNP$ in the ROM briefing}
\label{sec:tfnp-rom-def-short}

In the cryptographic formulation, a $\TFNP$ problem in the ROM is specified by a \emph{verifier}, a polynomial-time algorithm with oracle access that, given an instance and a candidate solution, decides whether the candidate is a solution; totality means that every instance has at least one solution. An instance is a circuit $C$ with oracle gates mapping $\{0,1\}^{\poly(n)}$ to $\Sigma$, with $|\Sigma| \le 2^{\poly(n)}$. The problem is hard on average if some efficient \emph{sampler} with oracle access outputs instances that no polynomial-time \emph{solver} with oracle access solves with probability $2/3$ over the oracle. We view the oracle as a random string $u \in \{0,1\}^{\mathbb{N}}$. Let the truth table of $C$ be $x \in \Sigma^N$, $N = 2^{\poly(n)}$. Then for a fixed oracle $u$, each entry of $x$ is one evaluation of $C$, which reads $\poly(n) = \polylog(N)$ bits of $u$, and checking a candidate solution means evaluating $C$ at a few points. Please see Appendix B for a formal definition of the cryptographic formulation of the ROM model. 

Note that only these query counts matter for our lower bound proofs, which leads to the following \emph{decision-tree formulation}, used throughout the paper.
A \emph{$\TFNP^{\rm dt}$ problem} is a relation $S \subseteq \Sigma^N \times \mathcal{O}$, with $|\mathcal{O}| \le 2^{\polylog(N)}$, that is total and such that for every $o \in \mathcal{O}$ the predicate $(x,o) \in S$ is decided by a decision tree of depth $\polylog(N)$ over $x$. An \emph{instance distribution} is a forest $\omega = (\omega_1,\dots,\omega_N)$ of decision trees $\omega_i\colon \{0,1\}^{\mathbb{N}} \to \Sigma$ of depth $\polylog(N)$, mapping the oracle $u$ to the instance $\omega(u) \in \Sigma^N$.
A \emph{solver} is a decision tree $A\colon \{0,1\}^{\mathbb{N}} \to \mathcal{O}$.

We say that $S \in \FP$ \emph{in the black-box ROM} if for every decision forest $\omega$ there is a solver $A$ of depth $\polylog(N)$ with
\[ \Pr_{\bm u \sim \{0,1\}^{\mathbb{N}}}[(\omega(\bm u), A(\bm u)) \in S] \ge 2/3, \]
and that $S$ is \emph{hard on average in Romantica} otherwise. A polynomial-time solver with oracle access makes polynomially many queries, so it is a solver of depth $\polylog(N)$, and black-box hardness implies hardness in the cryptographic formulation. 

Since $\omega$ is itself a shallow forest over $u$, it can be absorbed into the problem: $S' = \{(u,o) \mid (\omega(u),o) \in S\}$ is again a $\TFNP^{\rm dt}$ problem, now over the oracle string. Throughout the paper the instance is therefore the oracle itself, and for convenience we take $\bm u \sim \Sigma^{\mathbb{N}}$.\footnote{It is not hard to see that this is without loss of generality: for every $\Sigma$ the uniform distribution $\Sigma^N$ can be sampled with $O(\log (N|\Sigma|))$ queries to $\bm u$ per coordinate within total variation distance $(N|\Sigma|)^{-\Omega(1)}$, see e.g. \cite{HLSVZ26}.}

\subsection{$\PPAD$ lower bound}\label{subsection:PPAD}

Consider the following search problem template: given a large hypergraph with a labeling of the vertices, find an edge whose vertex labels satisfy a constraint. This yields a search problem for every choice of hypergraph and constraint. For several simple choices of the two, it is easy to show that the problem is \emph{total}, but then it is also easy to solve. Our proof takes the \emph{Yamakawa--Zhandry hypergraph}, originally used to separate quantum and classical algorithms~\cite{YZ24}. The \emph{constraint} part of their search problem does not guarantee totality, but the search problem is hard for classical algorithms. The key idea is to tweak their construction, which sits at the hard-but-not-total end of this spectrum, toward the total-but-easy end, stopping at a problem that is total (in fact, in $\PPAD$) while still hard.

Which class the resulting problem lands in is determined by how its totality is proved. We first show hardness-on-average in Romantica of the class $\PPA$, for \emph{parity arguments}, which contains the subclass $\PPAD$. We then refine the construction to obtain a hard-on-average problem in $\PPAD$. Informally, problems in $\PPA$ are exactly the problems whose totality can be proven by an efficient parity argument, and the problems in $\PPAD$ are the ones whose totality can be proven by an efficient signed counting argument. In our construction, these arguments simply sum local constraints modulo $2$ or over $\mathbb{Z}$ with signs $\pm 1$, respectively. These connections have been made formal in \cite{GKRS19} and \cite{GHJ24} respectively. The corresponding polynomial refutations must have suitably bounded degree and size, with coefficients measured in unary over $\mathbb{Z}$; the formal proof verifies these bounds and provides efficient pairings witnessing the cancellations. In this exposition, we suppress these efficiency requirements, making the task:
\begin{center}
    Find a problem that is hard on average given random access to a uniformly random string, and whose totality can be proven by counting in $\mathbb{F}_2$ (for $\PPA$) or $\mathbb{Z}$ (for $\PPAD$).
\end{center}

\paragraph{$\PPA$ worst-case hardness.} A canonical $\PPA$ problem is finding a \emph{falsified equation} in an unsatisfiable system of linear equations over $\mathbb{F}_2$. Let $N = 2^n$, and let $x_1,\dots, x_N$ be the variables of the system. We identify each equation $\sum_{v \in e} x_v = 1$ (over $\mathbb{F}_2$) with a \emph{hyperedge} $e \subseteq [N]$, and an \emph{evaluation} $x \in \{0,1\}^N$ of all $N$ variables with a binary labeling of the nodes of a hypergraph $G = ([N], E)$.  Given any hypergraph $G$ with edges of size $|e| \leq \poly(n)$ and an efficiently-accessible labeling of its nodes, we obtain the following $\NP$-relation:
\begin{equation}
\label{eq:ppa-easy-prob}
    \textsc{Lin}_G \coloneqq \Big\{(x, e) \,\big|\, \sum\nolimits_{v \in e} x_v = 0\Big\}\tag{$\star$}
\end{equation}

Of course, this $\NP$-relation is only total if we can guarantee the unsatisfiability of the linear system. One way is to enforce that each node $v \in [N]$ is contained in an \emph{even} number (at most $\poly(n)$) of hyperedges, and that the number of hyperedges $|E|$ is \emph{odd}. Then, one can show that the problem lies in $\PPA$.

However, these conditions do not make the problem $\textsc{Lin}_G$ hard, even in the worst-case (e.g. $G$ can be an odd cycle). Previous work has shown that \emph{worst}-case hardness holds if $G$ has strong \emph{expansion} properties \cite{Tse1970,Urquhart87}. However, it is not clear how to extend this to average-case hardness; for instance, if the labeling $\bm x \sim \{0,1\}^N$ is sampled uniformly (a distribution that can be easily realized in the ROM), then finding a solution of \eqref{eq:ppa-easy-prob} is easy for \emph{every} $G$: any edge is going to be a solution with probability $1/2$.

\paragraph{Yamakawa--Zhandry problem.} Our second key ingredient is an $\NP$-search problem defined by Yamakawa and Zhandry~\cite{YZ24} in order to separate quantum and classical algorithms for search problems in the ROM. Let $G = ([N], E)$ be a fixed hypergraph with $E \subseteq \binom{[N]}{\ell}$, where $\ell = \poly(n)$. The generic version of the problem $\textsc{YamakawaZhandry}_G$ is: given a labeling $x \in \{0,1\}^N$ find a hyperedge $e \in E$ such that $x_v = 0$ for all $v \in e$. Note that this is unlike the condition that makes an edge a solution in the previous section, and makes this problem very far from total.  However, Yamakawa and Zhandry construct a specific hypergraph for which it is hard to find a solution given \emph{uniformly random} $\bm x \sim \{0,1\}^N$. In other words, in the previous section we saw constructions of a hypergraph and edge-condition which yielded a total problem, but solutions were easy to find for random labelings. On the other end of the spectrum, Yamakawa-Zhandry's hypergraph and edge-condition yields a hard search problem over uniform labelings, but not all instances will have solutions.

The key property of their construction is \emph{well-spreadness}: $G = ([N], E)$ is $(r, a, b)$-well-spread if for every set $S \in \binom{[N]}{a}$ the number of hyperedges $e \in E$ such that $|e \cap S| \ge r$ is at most $b$. Why does well-spreadness make $\textsc{YamakawaZhandry}_G$ hard? Suppose a decision tree queried at most $a$ labels in $\bm x$. Then there are at most $b$ hyperedges for which the tree queries more than $r$ labels. For each of those hyperedges, consider the moment when exactly $r$ labels are queried. Then the conditional probability that the remaining queries make the edge all-$0$ is $2^{r - \ell}$. So if $b \cdot 2^{r - \ell} \ll 1$, we are unlikely to discover any solution.

\paragraph{$\PPA$-hard problem.} One obvious issue with $\textsc{YamakawaZhandry}_G$ for us is that it is not total, e.g. the labeling $x = 1^N$ has no solutions. Hence we will need to modify the \emph{condition} for an edge to be a solution. For the problem to additionally be in $\PPA$, a \emph{parity argument} must guarantee that at least one edge satisfies our chosen condition. For a given graph $G$, let $p_e\colon \{0,1\}^N \to \{0,1\}$ be the condition \emph{verifier} of $e$, i.e. $p_e(x) = 1$ iff $e$ is a solution for $x$. One way to guarantee totality is to choose $p_e$ such that
\[ \sum_{e \in E} p_e(x) \equiv 1 \bmod 2. \]
As witnessed by $\textsc{Lin}_G$, a choice of condition $p_e = 1 + \sum_{v \in e} x_v$ over $\mathbb{F}_2$ satisfies the equation above (i.e. yields a total problem) whenever every node is contained in an even number of hyperedges and the total number of edges is odd. However, as we also discussed, $\Pr[p_e(\bm x) = 1] = 1/2$, so even if $G$ is well-spread, finding the solution will be easy.

There is a natural fix for that: suppose that instead of binary labels, we used labels $x_1,\dots,x_N \in \{0,1\}^r$, and we set $p_e(x) = \lb \oplus_{v \in e} x_v = 0^r \rb$. Then over a uniformly-random labeling, $\Pr[p_e(\bm x) = 1] = 2^{-r}$, so for every well-spread $G$ finding $e$ with $p_e(\bm x) = 1$ is hard. 

In order to see what is needed from $G$ to make the problem total, it is convenient to plant a forbidden solution. Take $e^* \in E$ and fix all its labels to $0^r$, then forbid $e^*$ as a solution. That makes the condition on the hypergraph $G$ we need more convenient: $\sum_{e \in E} p_e(x) \equiv 0 \bmod 2$. Since $p_{e^*}(x)=1$, this gives $\sum_{e \neq e^*}p_e(x) \equiv 1 \bmod 2$, guaranteeing a non-forbidden solution. To see what we need from $G$, write $p_e$ as a polynomial in the \emph{bits} of the labels; the target expression becomes
\[ \sum_{e \in E} \prod_{i \in [r]} (1 + \sum_{v \in e} x_{v,i}) \equiv 0 \bmod 2.\]

Expanding the brackets we get that the coefficient of monomial $x_{v_1, i_1} \cdot \dots \cdot x_{v_k,i_k}$, with distinct bit coordinates $i_1,\dots,i_k$, is the number of edges containing $\{v_1,\dots,v_k\}$ modulo $2$. That gives us a sufficient condition on the hypergraph: we say that $G=([N],E)$ is $r$-even if for every $S \in \binom{[N]}{\le r}$ the number of edges $e \supseteq S$ is even. $S=\emptyset$ ensures cancellation of the constant term. In \cref{sec:ppad-construction}, we show that a variant of the Yamakawa--Zhandry hypergraph is $r$-even for $r=\Theta(\ell)$.

\paragraph{$\PPAD$-hard problem.} To obtain a problem in $\PPAD$, we replace the parity argument above with an integer cancellation argument. The input to the new hyperedge-search problem is a labeling $x \in ([r]\cup\{\bot\})^N$, where $\bot$ is a void label that we do not count. We fix all labels on the distinguished edge $e^*$ to $\bot$ and again forbid $e^*$ as a solution. The polynomial formulation of this problem has essentially the same structure:
\[ p_e \coloneqq \prod_{i \in [r]} (1-\sum_{v \in e} \lb x_v = i \rb). \]
Thus, $p_e(x) \neq 0$ iff every non-$\bot$ label occurring on $e$ occurs at least twice. The values of $p_e$ are now integers rather than solution indicators. We choose signs $\epsilon_e \in \{\pm 1\}$ so that $\sum_{e \in E} \epsilon_e p_e \equiv 0$. In the expansion, the coefficient of a degree-$k$ indicator monomial on distinct vertices $S=\{v_1,\dots,v_k\}$ and distinct labels is $(-1)^k\sum_{e\supseteq S}\epsilon_e$, for $0\le k\le r$; terms assigning different labels to the same vertex vanish. Hence, it suffices to require that for every $S \in \binom{[N]}{\le r}$, including the empty set, the signs of the edges containing $S$ sum to zero. We show in \cref{sec:ppad-construction} that the Yamakawa-Zhandry hypergraph admits such signing.

Since all labels on $e^*$ are $\bot$, we have $p_{e^*}(x)=1$. Normalizing the signing so that $\epsilon_{e^*}=1$ and removing this planted contribution gives
\[ -\sum_{e \in E\setminus\{e^*\}}\epsilon_e p_e(x)=1. \]
At least one non-forbidden edge therefore has $p_e(x)\ne0$, proving totality via \emph{counting over integers}. We make this formal in \cref{sec:ppad-containment}.

\subsection{$\UEoPL$ upper bound}\label{subsec:UEoPL-overview}
 
\paragraph{The problem setup.}
It is convenient to use the grid form of $\UEoPL$ from \cite{GHJ24}.  The nodes
are the cells $(i,j)$ of an $N\times N$ grid, $i$ the row.  Every node has a
successor pointer $\Succ_{(i,j)}\in[N]\cup\{\bot\}$, read as naming the
node $(i+1,\Succ_{(i,j)})$ of the next row, and every node below row~1 has a
predecessor pointer.  A node is \emph{active} if its successor's predecessor points back to
it: $\Succ_{(i,j)}=k$ and $\Pred_{(i+1,k)}=j$ (in the last row, active just means
$\Succ\neq\bot$).  If $(1,1)$ is active, following successors from it traces
a \emph{line} that descends one row per step, so it must end at an inactive node
that an active node points to (a \emph{proper sink}) or at an active node of row
$N$.  A solution is any of: $(1,1)$ inactive; a proper sink; an active node of
row $N$; an active node other than $(1,1)$ with no active predecessor; or
\emph{two active nodes in the same row}.  The last kind is what makes this
$\UEoPL$ rather than $\EoPL$: proof of a second line is also a solution.  Activity
is the only notion our algorithm uses, and testing it costs at most two pointer reads.
 
In the random oracle model an instance is generated from the oracle
$u\in\{0,1\}^{\mathbb{N}}$ by decision trees: each pointer is a tree of depth $q$ over the
bits of $u$.  We show (\cref{thm:ueopl}) that
there exists a decision tree of depth $O(q^5\log N)$, querying the bits of $u$, that outputs a solution with
probability $2/3$ over $\bm u$, and never outputs a wrong answer.  This is a
\emph{black-box} statement: it counts queries, not computation. 

\paragraph{Jumping and a win-win argument.}
The obvious strategy is to begin at the active cell $(1,1)$ in row 1, and move down one row at a time until we either reach a  proper sink or an active node in the last row. 
The query cost is $2q$ queries per row, hence $\Theta(qN)$ queries.
To get to $\mathrm{poly}(q)\log N$ queries, we jump: a jump of $2^r$ rows is two jumps of
$2^{r-1}$ rows.  Composed naively this doubles the cost at each of the $\log N$
levels, which is no better. 
The proof therefore rests on a \emph{dichotomy lemma} (\cref{lem:dichotomy}, originally proved in \cite{BCG26}): an algorithm that finds solutions, whatever its query cost, can be collapsed either to a query-efficient one or to one that outputs \emph{two} solutions. In our setting it says the following about any tree $D$, however deep, that almost
always outputs an active node of a target row $b$: either there is a tree of
depth $O(q^2)$, independent of the depth of $D$, that outputs an active node of
row $b$ with probability $1/2$; or there is a tree, about twice as deep as $D$,
that outputs \emph{two} active nodes of row $b$, hence a solution, with
probability $\Omega(1/q)$.  The first case lets us replace the composition of
two half-jumps by a tree whose depth does not grow with the level; the second
ends the search.

\paragraph{Why the dichotomy holds.}
Think of each accepting leaf of the activity tree of a node $y$ in row $b$ as a
\emph{certificate}: at most $2q$ oracle bits that, once seen, guarantee that
$y$ is active.  Since $D$ almost always outputs an active node, almost every
oracle lies in at least one certificate.  Either the certificates fit together
neatly, each oracle $u$   lying in essentially one of them, or they do not.  If they
do, $D$'s depth was never needed: a greedy procedure locates the certificate
containing the oracle with $O(q^2)$ queries, by repeatedly querying the bits of
the candidate certificate least likely to overlap the others.  If they do not,
then many oracles lie in two certificates for different nodes, so they have two
active nodes in row $b$, which is a solution; and the second node is not
hidden, because on nearby oracles inside its certificate $D$ outputs it, so
running $D$ once more on a slightly perturbed copy of the oracle tends to reveal
it.  Either way we win.  %

\paragraph{The algorithm.}
A subroutine $\textsc{Jump}(v,r)$ takes an active node $v$ and tries to reach
the row $2^r$ below.  It \emph{defines}, but does not run, the tree $C$ that
composes two half-jumps, and applies the dichotomy lemma to $C$ and the
activity trees of the target row.  In the first case it runs the shallow tree
and jumps; in the second it returns the pair-finding tree, which is as deep as
two runs of $C$, all the way up the recursion, and the main loop runs it once
and restarts on failure.  $C$ itself is run only when it is likely to produce a
solution or such a tree, which ends the computation, so this happens $O(q)$
times per level.  The cost of a jump is thus linear in $r$ rather than
exponential, and $r=\lceil\log_2 N\rceil$ levels give the bound.  %

\subsection{Separations from $\PLS$}\label{subsec:pls-separations-overview}

The known worst-case separations from $\PLS$ are obtained by proving \emph{resolution-width} lower bounds. A low-width resolution refutation can be viewed as a generalization of a low-height decision tree. In particular, let $F \subseteq \Sigma^n \times \mathcal{O}$ be a total relation defining the search problem ``given $x \in \Sigma^n$, find $o \in \mathcal{O}$ such that $(x,o) \in F$.'' A height-$w$ decision tree solving $F$ queries indices of an input $x \in \Sigma^n$ adaptively for $w$ rounds and determines some $o \in \mathcal{O}$ such that $(x,o) \in F$. We can view this ``success condition'' in the following way: for each $x$, the decision tree queries $x$ on $w$ indices and hence at the end knows some partial assignment $\rho \in (\Sigma \cup \{ \star\})^n$ of width $w$ (meaning it has $w$ non-star values) which $x$ extends. It must then be the case that the partial assignment $\rho$ already ``fixes a solution,'' meaning that there exists $o_{\rho} \in \mathcal{O}$ such that $(x', o_{\rho}) \in F$ for all $x'$ extending $\rho$.

We now describe a width-$w$ resolution refutation of $F$. This is another kind of query procedure for solving $F$, which also has the property that its current ``knowledge'' about an input $x$ can be described purely in terms of a width $\leq w$ partial assignment $\rho$ which $x$ extends. Unlike a decision tree, a resolution refutation is permitted to run for an unbounded number of rounds; however, we require that the next query at each step, or the choice to halt and output some solution $o$, is \emph{purely a function of the current known partial assignment} (and cannot depend otherwise on the history of the query procedure). To allow this process to run for more than $w$ rounds, we permit the query procedure to ``forget'' the answers to queries it has made in the past in order to free space for the value of new queries while staying within the total budget $w$ on the size of the partial assignment representing its current state. Up to increasing the width by at most $1$, we may assume that whenever the current state is a partial assignment of width exactly $w$, a solution must be determined, and whenever its width is $<w$, a further query is made (possibly forgetting some past queries in the process).
With this convention we formally define the model as follows.
\begin{definition} \label{def:res}
A resolution refutation of width $w$ consists of a set $\mathcal{F} \subseteq (\Sigma \cup \{ \star\})^n$ of partial assignments
together with a function $S\colon \mathcal{F} \times \Sigma \to \mathcal{F}$ such that $\star^n \in \mathcal{F}$ and for all $\rho \in \mathcal{F}$:
\begin{itemize}[noitemsep]
    \item $\rho$ has width at most $w$.
    \item If $|\rho| = w$ it \emph{fixes a solution}: there is some $o \in \mathcal{O}$ such that $(x,o)\in F$ for all $x \in \Sigma^n$ extending $\rho$.
    \item If $|\rho|<w$, there is some $i=i(\rho) \in [n]$, $\rho(i) = \star$, such that for every $a \in \Sigma$, we have that $S(\rho,a)\subseteq \rho \cup \{ i \mapsto a\}$.
\end{itemize}
Finally, there is a strict total ordering $\prec$ on $\mathcal{F}$ such that $\rho \prec S(\rho,a)$ for all $\rho \in \mathcal{F}$ with $|\rho| < w$ and $a \in \Sigma$.
\end{definition}
In the terminology of our previously described query procedure, $\mathcal{F}$ is the set of allowable ``states'' (partial assignments), and $S$ maps each non-terminal state to its successor based on the value of a query. The acyclicity condition is in fact a strict strengthening of the requirement that the query process reaches a terminal state and outputs a solution on every input; we require this termination condition to be enforced ``syntactically'' over the transition operations.
The connection to $\PLS$ is given as follows:

\begin{fact}[\cite{BKT14,kamath2019some}]
For any $F \in \TFNP^{\rm dt}$, $F \in \PLS^{\rm dt}$ iff there is a width-$\poly(\log n)$ pair $(\mathcal{F},S)$ that solves $F$.
\end{fact}

With this perspective we can easily describe the classical separation $\PPP^{\rm dt} \not\subseteq \PLS^{\rm dt}$ \cite{BCEIP98}. Let $\Sigma = \{0,1\}^n \setminus \{0^n\}$ and $N = 2^n$. A $\PPP$ instance is naturally represented as an element $x\in \Sigma^N$, listing the values $x(v)$ for each $v \in \{0,1\}^n$ in lexicographical order. We want to show that no width-$w = \polylog (N)$ refutation can find a collision for $x$. We can actually prove this for width as large as $N-2$: we simply observe that for any collision-free partial assignment $\rho$ with at least two $\star$ values, and any $v$ with $\rho(v) = \star$, there exists $a \in \Sigma$ such that $\rho \cup \{ v \mapsto a\}$ does not fix any solution; we simply take $a$ to be any value in the set $\Sigma \setminus \{\rho(v') \mid v' \leq N\}$, which must be nonempty.

To prove our actual separation, we need to somehow extend this argument to show that it holds ``on average over a random $\bm x \sim \Sigma^N$.'' Consider some candidate width $w=\polylog(N)$ resolution refutation $\mathcal{F}, S$, satisfying all conditions above except the terminal state condition. Our task will be to show that \emph{with high probability} over $\bm x \sim \Sigma^N$, there exists some terminal state $\rho \in \mathcal{F}$, $\bm x \supseteq \rho$, such that $\rho$ does not determine a $\PPP$ solution (meaning there are no two values $v\neq v' \in [N]$ such that $\rho(v) = \rho(v') \in \Sigma$). To approach this, we observe that our previous worst-case lower bound did not exploit a useful property of $\PPP$: if $\rho$ is a partial assignment of width $\polylog (N)$, then for any additional query $v \in [N]$, $\rho \cup \{ v \mapsto a\}$ will not fix a $\PPP$ solution \emph{with overwhelming probability} over a random $\bm a \sim \Sigma$; in particular the failure probability is at most $\frac{\polylog(N)}{N-1}$. We call this property \emph{robustness}; it implies that for any nonterminal $\rho$ which does not fix a solution, 
\[
\Pr_{\bm x \sim \Sigma^N}[ S(\rho, \bm x_{i(\rho)}) \text{ fixes a solution } \mid \bm x \supseteq \rho] \leq \frac{\polylog(N)}{N-1}
\]
A naive strategy would be to start at $\rho = \star^N$ and iteratively apply this bound as we traverse the refutation using $S$, arguing at each step that the probability we fix a solution is small. The obvious issue here is that the number of steps may far exceed $N$ and so a union bound will fail to establish that a solution-fixing state is never reached. Indeed, requiring the refutation to fail when we specifically follow the path starting at the state $\star^N$ far overshoots our intended goal, which is to find \emph{any} terminal $ x\supseteq \rho \in \mathcal{F}$ which does not fix a solution; this will directly correspond to a $\PLS$ solution which fails to determine any $\PPP$ solution, hence ruling out the correctness of the reduction on this $x$.

Our approach will instead be more global. Let $\mathcal{E} \subseteq \mathcal{F}$ be the terminal states in $\mathcal{F}$ which do not fix a $\PPP$ solution and let $B\subseteq \Sigma^N$ be the ``bad'' inputs for which there exists $\rho \in \mathcal{E}$ such that $x \supseteq \rho$. We aim to show that $\Pr[\bm x \in B] \geq 1 - (\log N)^{-\omega(1)}$ for a uniformly random $\bm x$, i.e. the reduction to $\PLS$ fails on all but a negligible fraction of inputs. To prove this, we will define the auxiliary relation/search problem $F'$: given $x \in \Sigma^N$, find either a $\PPP$ solution for $x$, or else find $\rho \in \mathcal{E}$ such that $x \supseteq \rho$. We then observe that $\mathcal{F}$ now serves as a \emph{worst-case resolution refutation} for the relation $F'$. It hence suffices to prove the following: if $F \subseteq \Sigma^N \times \mathcal{O}$ is any $\TFNP^{\rm dt}$ search problem with the aforementioned ``robustness'' property, and we augment it to another $\TFNP^{\rm dt}$ problem $F' \subseteq \Sigma^N \times (\mathcal{O} \sqcup \mathcal{O'})$ where (1) for $x \in \Sigma^N, o \in \mathcal{O}$, $(x,o) \in F' \longleftrightarrow (x,o) \in F$ and (2) $\Pr[\exists o' \in \mathcal{O'}, (x,o') \in F'] \leq 1 - (\polylog(N))^{-1}$, then $F'$ does not have a $\polylog(N)$-width refutation. The key is that we have offloaded the ``average case'' condition into a modification of the search problem itself, and the remaining task is to prove a ``worst case'' resolution width lower bound for this modified search problem. 

To prove this resolution bound we utilize a linear-algebraic embedding originally developed by Razborov \cite{RazborovWFPHP} as a key step in his \emph{pseudo-width} method. To prove a resolution lower bound for $F'$, it is natural to try to focus on the set of ``safe states'' $\rho$ which fix no $\PPP$ solution and which admit at least one extension with no $\mathcal{O'}$ solution, and argue that this set is closed under ``query/forget'' operations. This would immediately yield the width lower bound for $F'$, since any width-$w$ refutation would fail to ``escape'' this set of safe partial assignments (and none of them fix a solution for $F'$).  While this naive safeness property holds for the empty partial assignment and is downward closed (i.e. closed under ``forgetting'' previous queries), it need not be closed under performing additional queries. The linear-algebraic embedding will provide us with a special downward-closed (and nonempty) subset of these safe states which is in fact perfectly closed under queries: for any $|\rho|<w$ in this set and any $i$ with $\rho(i) = \star$, there exists $a \in \Sigma$ such that $ \rho \cup \{ i \mapsto a\}$ also lies in the set. 

As alluded to above, the only property of $\PPP$ we will use is its ``robustness'' (we have not defined robustness formally at this point). This property is easily shown to hold for the subclass $\PWPP$, and so in \cref{subsec:pwpp-notin-pls} we prove the stronger ROM separation $\PWPP \not\subseteq \PLS$ directly. In \cref{subsec:ppa-notin-pls} we exhibit a problem in the class $\PPA^{\rm dt}$ which satisfies the same robustness property and hence obtain the ROM separation $\PPA\not\subseteq \PLS$ as well. Unlike in the case of $\PWPP$ it is not obvious how to efficiently construct a robust distribution of $\PPA$; here we require a modification of the previously-discussed construction showing $\PPAD \not\subseteq \FP$ in the ROM.

\section{Average-case separation between $\PPAD$ and $\FP$}
In this section we prove the following theorem,
giving an exponential separation between $\PPAD$ and $\FP$ in the ROM setting. 

\begin{theorem}
    \label{thm:main-ppad-hardness}
    There exists a $\TFNP$ search problem $F = \left\{ F_N \subseteq \Sigma^N \times \mathcal{O}\right\}$ in $\PPAD$ satisfying that: every decision tree $T$ with depth $\exp\left((\log N)^{o(1)}\right)$ solves $F$ with probability at most
    \begin{equation*}
        \Pr_{\bm{u} \sim \Sigma^N} \left[ (\bm{u}, T(\bm{u})) \in F_N \right] \le \exp\left(-(\log N)^{\Omega(1)}\right).
    \end{equation*}
    In particular, $\PPAD \not\subseteq_{\rm ROM} \FP$.
\end{theorem}

\newcommand{\Singleton}{\textsc{Singleton}}
\newcommand{\hashsize}{\tau}
We start by defining the hard problem. Let $G = ([N], E)$ be an $\ell$-uniform hypergraph, i.e. $E \subseteq \binom{[N]}{\ell}$. We first describe a simplified problem: given a labeling $h\colon [N] \to [\hashsize]$ find an edge $e = \{v_1,\dots,v_\ell\} \in E$ such that no label occurs exactly once in the list $h(v_1),\dots,h(v_\ell)$. That is, for each $i \in [\hashsize]$ either $i \not\in \{h(v_1),\dots,h(v_\ell)\}$ or it occurs in the list at least twice. The actual problem has two differences: we allow labels to be voided ($\bot$), and we plant a special edge $e^* = [\ell]$ and fix all labels in $[\ell]$ to $\bot$. This makes $e^*$ a trivial solution, so the problem asks for \emph{another} solution.

\begin{figure}
    \centering
    \input{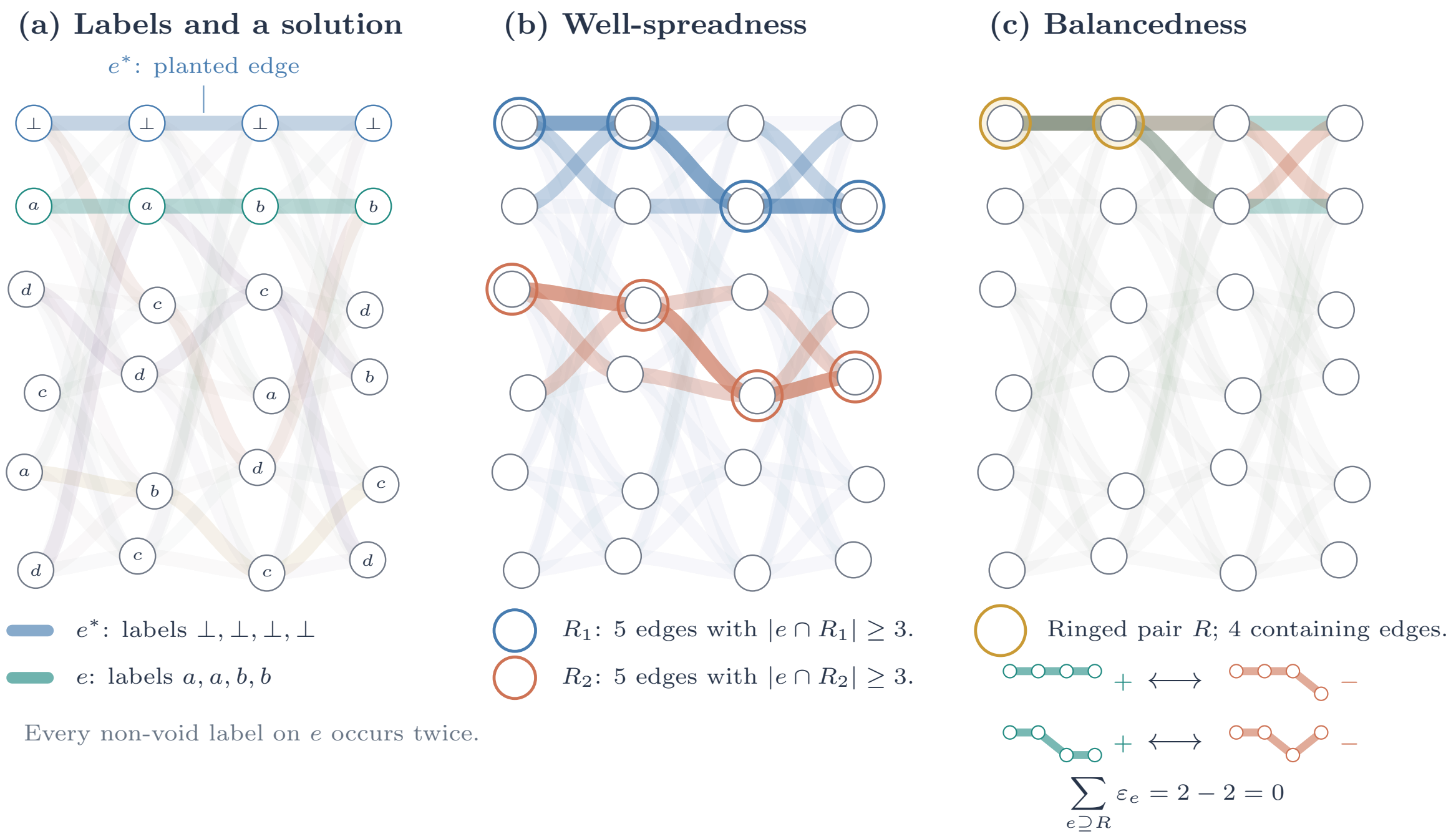}
    \caption{This figure illustrates the setting of the problem $\Singleton$.}
    \label{fig:hypergraphs}
\end{figure}

Formally, let $\calH$ be the set of functions $h\colon [N] \to (\{\bot\} \cup [\hashsize])$ satisfying $h([\ell]) = \{\bot\}$. Then the problem $\Singleton_G \subseteq \calH \times (E \setminus \{e^*\})$ is as follows: given $h\in \calH$, find an edge $e=\{v_1,\dots,v_{\ell}\} \in E \setminus \{e^*\}$ such that all non-$\bot$ elements in the list $h(v_1), \dots, h(v_\ell)$ occur at least twice. We will show that there exists $G$ that simultaneously satisfies the following properties:
\begin{description}[noitemsep]
    \item[Hardness] For $\bm h \sim \calH$, every decision tree $T$ that makes at most $2^{\ell^{1/4}}$  queries to $\bm h$ is very unlikely to solve $\Singleton_G$:
    \[ \Pr[(\bm h,T(\bm h)) \in \Singleton_G] \le \exp(-\ell^{\Omega(1)}).\]
    \item[$\PPAD$-inclusion] $\Singleton_G \in \PPAD$;
\end{description}

In \cref{sec:hardness} we prove that $\Singleton_G$ is hard whenever the set of edges $E$ is well-spread:

We say that a subset family $E \subseteq \binom{[N]}{\ell}$ is \emph{$(r, a, b)$-well-spread} if for every set $S \subseteq [N]$ of size at most $a$ the number of edges $e \in E$ such that $|e \cap S| \ge r$ is at most $b$. See \cref{fig:hypergraphs} for an illustration.

Then in \cref{sec:ppad-containment} we show that $\Singleton_G \in \PPAD$ if $E$ is balanced: $E \subseteq \binom{[N]}{\ell}$ is $r$-\emph{balanced} if there exists a \emph{signing} $\epsilon \in \{\pm 1\}^E$ such that for every $R \in \binom{[N]}{r}$ we have $\sum_{e \supseteq R} \epsilon_e = 0$.

Finally, in \cref{sec:ppad-construction} we show that there is an efficient construction of $G$ satisfying both well-spreadness and balancedness.

\subsection{$\PPAD$ hardness}
\label{sec:hardness}

\begin{theorem}
\label{thm:ppad-hardness}
Fix $\hashsize$ to be an even integer in $[\ell/4, \ell/2.6]$. Suppose that $E$ is a $(3\hashsize/2,q,b)$-well-spread set and $T\colon \mathcal{H} \to E \setminus \{e^*\}$ is a decision tree of depth at most $q$.  Additionally assume that for every $e \in E \setminus \{e^*\}$ we have $|e \cap [\ell]| \le \hashsize$. Then the probability that
$(\bm h, T(\bm h)) \in \Singleton_{([N],E)}$ is at most
$O(b) \cdot \exp(-\Omega(\tau)) $.
\end{theorem}

\begin{proof}
Associated with each path in $T$ is a partial
assignment $\rho \in (\{\star, \bot\} \cup [\hashsize])^N$ to the underlying variables that are queried
along the path. 
Also let $\fix(\rho)$ denote the set of queries made along path $\rho$: 
\(\fix(\rho) \coloneqq \rho^{-1}([\hashsize] \cup \{\bot\}).\)
                 
For every vertex $v \in T$, let $\rho_v$ denote the partial path/restriction
from the root of $T$ to vertex $v$, and 
for a path $\rho$ let ${\rm Leaf}(\rho)$ denote the leaf vertex on this path.

Let $T(\rho)$ be the output of $T$ in ${\rm Leaf}(\rho)$.
Thus the overall probability that $T$ succeeds on $\bm h \sim \calH$ is:
\[\sum_{\rho \text{ a root-to-leaf path in $T$}} \Pr[\bm h~{\rm is~consistent~with~}\rho] \cdot \Pr [(\bm h, T(\rho)) \in \Singleton_G \mid \bm h\text{ is consistent with }\rho]\]

For a path $\rho$ in $T$ and an edge $e \in E$, we say
that $e$ is {\bf exposed} by $\rho$ if there are at least $3\tau/2$ 
nodes in $e$ whose $h$-values are fixed by $\rho$.

The following Lemma states that for any path $\rho$ in $T$
and for any $e \neq e^*$ such that at most $3\tau/2$ $h$-values of $e$ are fixed by $\rho$, the probability that $e$ is a solution is exponentially small.
We prove \cref{lem:main} in \cref{sec:main-lemma-proof}.

\begin{restatable}{lemma}{ppadhardnesslemma}\label{lem:main}
Let $G, e^*, \ell, \tau$ be as in \cref{thm:ppad-hardness}. Let $\rho$ be a partial assignment.
Fix an edge $e \neq e^*$ and suppose that
$|\fix(\rho) \cap e| \leq 3\tau/2$.
Then, given $\ell \ge 2.6\tau$ and $\ell = O(\tau)$, we have
\[\Pr[(\bm h, e) \in \Singleton_G \mid \bm h\text{ extends }\rho] \leq \exp(-\Omega(\tau)).\]
\end{restatable}

Equipped with \cref{lem:main}, we proceed with the lower bound.
There are two cases to consider.

\paragraph{Case I.} The first case is if the path $\rho$ in $T$ is such that $T(\rho)$ is {\it not} exposed.
This is the easy case, and we can apply
\cref{lem:main} directly to deduce that
$\Pr[(\bm h, T(\rho)) \in \Singleton_G \mid \bm h\text{ extends }\rho] \leq \exp(-\Omega(\tau))$.

\paragraph{Case II.} The second case is the harder and more interesting case, where $\rho$ is a path in $T$
and $T(\rho)$ has been exposed by $\rho$.
Note that every path $\rho$ could output an exposed edge $e$, so here we want
to argue that the probability (over $\bm h$) that
$T(\rho)$ is exposed along its path {\it and} that $(\bm h, T(\rho)) \in \Singleton_G$ is
very small. In other words, we need to show that very few assignments $h$ lead to exposed \emph{solutions}.

The rest of the proof will focus on Case II.
First, the following claim bounds the number of
exposed edges on any given path of $T$.

\begin{claim}
\label{cl:application-of-spreadness}
For every path $\rho$ in $T$, at most $b$ edges are
exposed by $\rho$.
\end{claim}
\begin{proof}
Recall that $\fix(\rho)$ is the set of queries made along path $\rho$, so since $|\fix(\rho)| \le q$ applying $(3\tau/2, q, b)$-well-spreadness to this set implies that at most $b$ edges $e \in E$ satisfy $|e \cap \fix(\rho)| \ge 3\tau/2$. 
\end{proof}

For each path $\rho$ in $T$, we define an ordered list
of non-special edges that become exposed:
Let ${\rm List}(\rho)$ consist of all pairs
$(v,e)$ with $e \neq e^*$ such that $e$ {\it first} becomes exposed along $\rho$ at vertex $v$.
By \cref{cl:application-of-spreadness}, $|{\rm List}(\rho)|\le b$.
Also we note that the vertices $v$ may not be distinct as more than one
edge could be exposed by the same query vertex $v$.
Assume that ${\rm List}(\rho)$ is ordered according to time 
so the edges that are exposed earlier occur before the edges that are exposed
later; when more than one edge is exposed at the same vertex,
we order them lexicographically.
Thus, for every path $\rho$, ${\rm List}(\rho)$ is an ordered list of length at most $b$.

Now let $\rho$ be a path, and let $(v,e)$ be one of the pairs in
${\rm List}(\rho)$. By definition, this means that at vertex $v$ along $\rho$, $e$
first becomes exposed. Let $\rho_v$ denote the partial
transcript of $\rho$ from the root of $T$ up to and including vertex $v$.
Thus $\rho_v$ fixes exactly $3\tau/2$ $h$-values in $e$
and its other $\ell-3\tau/2$ coordinates are still not queried.

Now we can apply \cref{lem:main} here to obtain:
\[\Pr[(\bm h, e) \in \Singleton_G \mid \rho_v] \leq \exp(-\Omega(\tau)).\]
Therefore, conditioned on extending $\rho_v$, the paths in $T$ for which
$e$ is good have total probability mass at most $\exp(-\Omega(\tau))$.

Now consider the entire collection
of ordered lists $\{{\rm List}(\rho)\}_{\rho \in T}$, over all root-to-leaf paths $\rho \in T$.
Fix a position $j \in [b]$; this is a position
in all of the ordered lists.
Define
\[V_j \coloneqq \{(v,e) \mid \text{$\rho$ is a path in $T$ and $(v,e)$ is the $j^{th}$ element of the list ${\rm List}(\rho)$}\}.\]
Note that although we have shown that along any \emph{single} path $\rho$, there are at most $b$ exposed edges, since $V_j$ contains the set of $j^{th}$ exposed edges across \emph{all} $\rho$'s, the number of elements in $V_j$ could be very large.
Now for a vertex $v \in T$, define $E_v$ to be the event that $h$ is consistent with $\rho_v$:
\[E_v \coloneqq \{h \in \calH \mid h\text{ extends }\rho_v\}.\]

\begin{claim}
\label{cl:event-disjoint}
For every $j \leq b$ and distinct pairs $(v,e),(v',e') \in V_j$, the events $E_v$ and $E_{v'}$ are mutually exclusive,
and therefore
\[\sum_{(v,e) \in V_j} \Pr [E_v] \leq 1.\]
\end{claim}

\begin{proof}
Let $(v,e) \neq (v',e') \in V_j$ where $(v, e)$ is the $j^{th}$ element
of the list ${\rm List}(\rho)$, and $(v', e')$ is the $j^{th}$ element
on the list ${\rm List}(\rho')$.
We want to show that $\rho_v$ and $\rho_{v'}$ must be inconsistent,
and therefore the events $E_v,E_{v'}$ are mutually exclusive.
If not then assume without loss of generality that $\rho_{v'}$ extends the
partial assignment $\rho_v$, and therefore $\rho_v$ corresponds to
a subpath (possibly equal) of $\rho_{v'}$.
But this implies that the first $j$ elements must be identical
in both lists ${\rm List}(\rho)$ and ${\rm List}(\rho')$, contradicting
our assumption that they disagree on the $j^{th}$ element.
\end{proof}

Now fix $j \in [b]$.
By \cref{cl:event-disjoint}, $\sum_{(v,e) \in V_j} \Pr [E_v] \leq 1$,
and therefore the probability (over assignments $h$) that 
the $j$th exposed edge on the sampled path exists and is a solution is:
\[\sum_{(v,e)\in V_j} \Pr [E_v] \cdot \Pr [(\bm h, e) \in \Singleton_G \mid E_v].\]

We can use \cref{lem:main} to bound each second term by $\exp(-\Omega(\tau))$,
and by \cref{cl:event-disjoint}, the sum of the first terms is at most 1.
Therefore by a union bound over all $j \in [b]$, the probability that some non-special exposed edge on the sampled path is a solution is at most $b \cdot \exp(-\Omega(\tau))$.
\end{proof}

\subsubsection{Proof of the main lemma}
\label{sec:main-lemma-proof}
Finally, we prove \cref{lem:main}, that we restate here for convenience.
\ppadhardnesslemma*

The key probabilistic fact we need here is McDiarmid's inequality:
\begin{theorem}[\cite{McD89}]
\label{thm:mcdiarmid}
    Let $f\colon \Sigma^m \to \mathbb{R}$ be a function with $c$-bounded differences, i.e. such that for all $x,x' \in \Sigma^m$ that differ in a single coordinate $|f(x) - f(x')| \le c$. Then 
    \[ \Pr_{\bm x \sim \Sigma^m}[|f(\bm x) - \Exp[f(\bm x)]| \ge \lambda] \le 2\exp(-\lambda^2/(c^2 m)).\]
\end{theorem}

\begin{proof}[Proof of \cref{lem:main}]
    Let $e = \{u_1,\dots,u_\ell\}$ and assume wlog that $\bm h(u_1),\dots,\bm h(u_k)$ are either fixed by $\rho$ or by the initial assignment of $h(1)=\dots=h(\ell) = \bot$. By assumption, $k \le 2.5\tau$, since at most $3\tau/2$ are fixed by $\rho$ and at most $\tau$ by the initial assignment. Denote the assignment $\rho$ extended to $[\ell]$ by $\rho'$.
    
    Then $\bm h(u_{k+1}), \dots, \bm h(u_\ell)$ are i.i.d. uniform over $[\tau] \cup \{\bot\}$ conditioned on $\rho$. Let $D \subseteq [\tau]$ be the set of values $d$ such that there is at most one $i \in [k]$ with $\rho'(u_i) = d$. The size of $D$ is at least $\tau / 4$: each element in $[\tau] \setminus D$ has to occur at least twice among the non-$\bot$ $h(u_1),\dots,h(u_k)$, so $|[\tau]\setminus D|\le 3\tau/4$, since the only fixed non-$\bot$ values are assigned by $\rho$.
    
    We show that at least one label remains a singleton. Define $f_d(x_{k+1},\dots,x_\ell)$ to be $1$ if $d$ occurs exactly once in $\rho'(u_1),\dots,\rho'(u_k),x_{k+1},\dots,x_\ell$, and $0$ otherwise. Then for $I_d \coloneqq \{i\le k \mid \rho'(u_i)=d\}$ we get
    \[\Pr[f_d(\bm h(u_{k+1}), \dots, \bm h(u_\ell)) = 1] 
    = \begin{cases}
        (1-1/(\tau + 1))^{\ell-k} & \text{ If }|I_d| = 1; \\
        (\ell - k)/(\tau + 1) \cdot (1-1/(\tau + 1))^{\ell-k - 1} & \text{ If } I_d = \emptyset.
    \end{cases}
    \]
    In both cases the probability is $\exp(-\Theta((\ell-k)/\tau)) = \Omega(1)$, since $\ell-k = \Theta(\tau)$.

    Defining $f = f_1 + \dots + f_{\tau}$ observe that $f$ has $2$-bounded differences: changing one value of $h(u_i)$ may create at most two singletons. Thus, since $\Exp[f(\bm h(u_{k+1}), \dots, \bm h(u_\ell))] = \Omega(\tau)$, \cref{thm:mcdiarmid} gives $\Pr[f(\bm h(u_{k+1}), \dots, \bm h(u_\ell)) = 0] \le \exp(-\Omega((\tau/4)^2/(\ell-k))) = \exp(-\Omega(\tau)).$
\end{proof}

\subsection{$\PPAD$ containment}
\label{sec:ppad-containment}
Recall that $E \subseteq \binom{[N]}{\ell}$ is $r$-balanced if there is a signing $\epsilon \in \{\pm 1\}^E$ such that for every $I \in \binom{[N]}{\le r}$ we have $\sum_{e \in E_I} \epsilon_e = 0$ for $E_I \coloneqq \{e \in E \mid e \supseteq I\}$. We say that $E$ is \emph{efficiently} $r$-balanced if the mapping $e \mapsto \epsilon_e$ is efficiently computable, and for every $I$ there is a poly-time computable involution $\sigma_I\colon E_I \to E_I$ such that $\epsilon_{\sigma_I(e)} = -\epsilon_e$, i.e. given $I$ and $e$ there is a poly-time algorithm that computes $\sigma_I(e)$. Wlog we assume $\epsilon_{e^*} = 1$.

\begin{theorem}
\label{thm:ppad-containment}
    Let $G=([N], E)$ with $E \subseteq \binom{[N]}{\ell}$  be explicit with $\ell = \poly(\log N)$. Suppose that $E$ is \emph{efficiently} $\hashsize$-balanced. Then $\Singleton_G \in \PPAD$. 
\end{theorem}

We reduce $\Singleton_G$ to a uniform family of unsatisfiable polynomial equations over the reals,
with integer coefficients. We show that for $\hashsize$-balanced graphs $G$ this family has a simple
(uniform) unary Nullstellensatz ($\uNS$) refutation.  
The general theorem from \cite{GHJ24} then immediately implies $\Singleton_G \in \PPAD$ in the \emph{black-box} model. Our proof does not formally rely on this theorem. 
We will show that \emph{efficient} $\hashsize$-balancedness allows the argument to hold in the uniform setting.

\paragraph{Polynomial encoding of $\Singleton_G$.} 
We define an unsatisfiable system of polynomials,
${\cal P}_G$. The underlying variables of it
are $h_{v,j}$ where $v \in [N]$ and $j \in [w]$, where $w = \lceil \log (\tau + 1)\rceil$.
The (semantically boolean) variables $h_{v,1},\dots,h_{v,w}$ encode the value of $h(v) \in \{\bot\} \cup [\hashsize]$: the binary representations of elements in $[\hashsize]$ encode those elements, all other strings encode $\bot$. Since the encoded function $h \in \calH$, we also define $h_{i,j}=0$ for all $i \in [\ell],\, j \in [w]$.

We need to represent the fact that $(h,e) \not\in \Singleton_G$ as a polynomial in $\{h_{v,j}\}_{v \in [N], j \in [w]}$ for a fixed $e=\{v_1,\dots,v_\ell\}$. We want to count occurrences of each non-$\bot$ value in $h(v_1),\dots,h(v_\ell)$. 
That is, for $\beta \in [\hashsize]$, define
 \[n_{\beta}(e;h) \coloneqq \bigl|\{v \in e \mid h(v)=\beta\}\bigr|,\]
 $n_{\beta}(e;h)$ counts the number of times the value $\beta$ appears in the sequence.

The function $n_{\beta}(e;h)$ can be expressed as a low degree polynomial in the variables $\{h_{v,j}\}$:
writing $[h(v)=\beta]$ to represent the product of literals $h_{v,j}$ expressing $h(v)=\beta$, $n_{\beta}(e;h)$ is computed by
the polynomial $\sum_{v\in e} [h(v)=\beta]$ of degree at most
$w$.

Now we define the following system of polynomial equations:
 \[{\cal P}_G = \{p_e(h)=0 \mid e \in E \setminus \{e^*\}\}, ~~~~ \text{where}~~~ p_e(h) \coloneqq \prod_{\beta \in [\hashsize]} \left(1-\sum_{v \in e} [h(v)=\beta]\right).\]

Note that each $p_e$ is a polynomial of degree $\hashsize w$.
Also, with respect to boolean variables $h_{v,j}$, the polynomial $1-\sum_{v \in e} [h(v)=\beta]$ is zero exactly when $\beta$ occurs once. Therefore we have:
\[
 p_e(h)\ne0
 \quad\Longleftrightarrow\quad
 \text{no value in $[\hashsize]$ occurs exactly once.}
\]

\paragraph{Unary Nullstellensatz and $\PPAD$.} 
G\"o\"os, Hollender, Jain, Maystre, Pires, Robere, and Tao
\cite[Lemma~5]{GHJ24} prove that $\PPAD^{\rm dt}$-inclusion is equivalent to having a \emph{Unary Nullstellensatz} ($\uNS$) proof. More precisely, suppose that $a_1, \dots, a_M \in \mathbb{Z}[x_1,\dots,x_N]$ are multilinear polynomials each depending on $\poly(\log N)$ variables. Then a Nullstellensatz refutation is a collection of polynomials $c_i \in \mathbb{Z}[x_1, \dots, x_N]$ and $d_1,\dots,d_N \in \mathbb{Z}[x_1, \dots, x_N]$ such that $\sum_{i \in [M]} a_i \cdot c_i + \sum_{i \in [N]} (x_i^2 - x_i) d_i = 1$. Then Nullstellensatz degree is the maximum degree of $a_i c_i$ and \emph{unary} Nullstellensatz size is the sum of magnitudes of all coefficients in $c_1,\dots,c_M$. 
\begin{lemma}[{\cite[Lemma~5]{GHJ24}}]
\label{lem:nullstellensatz-PPAD}
    Let $S \coloneqq \{(x,i) \in \{0,1\}^N \times [M] \mid a_i(x) \neq 0\}$, and suppose there exists a unary Nullstellensatz refutation of $\{a_i\}_{i \in [M]}$ of degree $\poly(\log N)$ and size $2^{\poly(\log N)}$ then $S \in \PPAD^{\rm dt}$. \footnote{\cite{GHJ24} only formulate this for CNF search, but it is easy to see that a $\poly(\log N)$-local multilinear polynomial can be converted to a quasipolynomial collection of clauses.}
\end{lemma}

\paragraph{Totality of $\Singleton_G$ and $\PPAD$ upper bound.}
First we show that $\Singleton_G$ is total by proving that
the underlying system of polynomial equations, ${\cal P}_G$ is unsatisfiable. Then, using the general equivalence between $\PPAD$ and unary Nullstellensatz we show that $\Singleton_G$ has an efficient $\PPAD$ formulation. 

\begin{lemma}[Totality of $\Singleton_G$]\label{lem:identity}
In the setting of \cref{thm:ppad-containment} the system ${\cal P}_G$ of polynomial equations is unsatisfiable, equivalently $\Singleton_G$ is total. Furthermore, ${\cal P}_G$ has the following simple
$\uNS$ refutation (modulo the boolean axioms $x^2 - x$ for all variables $x$): 
\[1=- \sum_{e \in E \setminus \{e^*\}}\epsilon_e p_e .\]
\end{lemma}

\begin{proof}
Expand the $\tau$ factors defining $p_e$:
\[p_e = \sum_{B \subseteq [\hashsize]; \kappa \in e^B} (-1)^{|B|} \prod_{\beta \in B} [h(\kappa_\beta) = \beta].\]
That is, each term specifies a subset $B \subseteq [\hashsize]$, and a list of $|B|$ nodes $\kappa$ that all belong to $e$.
Now define the set of nodes $I$, associated with this term: 
\[I \coloneqq \{\kappa_{\beta} \mid \beta \in B\}\subseteq [N], ~~~ |I| \leq |B| \leq \tau .\]
If $|I| < |B|$, the coefficient of the term in $p_e$ is $0$ (by boolean axioms). If $|I|=|B|$ the coefficient is $(-1)^{|B|}$ if $e \supseteq I$, otherwise it is $0$. 
Therefore, since $|I| \le \tau$, the $\tau$-balancedness implies that the coefficient of the term in $\sum_{e \in E} \epsilon_e p_e$ is $(-1)^{|B|} \cdot \sum_{e \supseteq I} \epsilon_e = 0$. 

On the other hand, for $e^* = [\ell]$, by definition we have $n_{\beta}(e^*;h)=0$ since $h(v)=\bot$ for all $v \in e^*$.
Therefore all of its counts $n_{\beta}(e^*;h)$ are zero, and consequently
we have $p_{e^*}(h)=1$ and $\epsilon_{e^*}=1$.
Removing this polynomial $p_{e^*}(h)$ from the sum we have
$1 = -\sum_{e \in E \setminus \{e^*\}} \epsilon_e p_e(h)$.

Now this implies that ${\cal P}_G = \{p_e(h) \mid e \in E \setminus \{e^*\}\}$ is unsatisfiable:
suppose for sake of contradiction that there is an assignment to
the underlying variables $h$ such that for every
$e \in E \setminus \{e^*\}$, $p_e(h)=0$.
Then $\sum_{e \in E \setminus \{e^*\}} \epsilon_e p_e(h)=0$, but this contradicts
the fact that the sum is identically $1$ (as argued above).

Thus, we have not only shown that ${\cal P}_G$ is
unsatisfiable, but have given a unary Nullstellensatz refutation
$\sum_{e \in E \setminus \{e^*\}} \epsilon_e p_e =-1$, where each
$\epsilon_e \in \{\pm 1\}$.
Therefore, the sum of the absolute values of the coefficients in $p_e$ is at most $(1+\ell^2)^\tau$ for each $e \in E$,
and therefore the refutation has degree $\poly(\ell)$ and unary size $|E|(1+\ell^2)^\tau\leq 2^{O(\ell\log N+\tau\log\ell)}=2^{\polylog N}$.
\end{proof}

\paragraph{Efficiency of inclusion.} By \cref{lem:nullstellensatz-PPAD} and \cref{lem:identity}, $\Singleton \in \PPAD^{\rm dt}$. In this section we make the inclusion explicit. Using the efficiency of $\tau$-balancedness we can complete the proof following the one by \cite{GHJ24} exactly. Let us introduce the canonical $\PPAD$-complete problem:
\begin{definition}
    \textsc{End-of-Line} problem is defined with two arrays $\Succ, \Pred \in [N]^N$; solutions are
\begin{itemize}[noitemsep]
    \item $1$ if $\Pred_1 \neq 1$ or $\Succ_1 = 1$ or $\Pred_{\Succ_1}\neq1$; \hfill (distinguished node is not a source)
    \item $i \in [N] \setminus \{1\}$ if $\Succ_{\Pred_i} \neq i$.\hfill (non-canonical source)
    \item $i \in [N]$ if $\Pred_{\Succ_i} \neq i$. \hfill (sink)
\end{itemize}
\end{definition}

\begin{proof}[Proof of \cref{thm:ppad-containment}]
 We instantiate the proof of \cite[Lemma~5]{GHJ24} in our simpler case (the refutation simply adds the axiom polynomials with coefficients $\{\pm 1\}$) and verify that the inclusion is efficient. The nodes of the \textsc{End-of-Line} formulation are pairs $(e, u)$ with $e \in E$ and $u \in (\{\bot\} \cup e)^\tau$. Each pair corresponds to an \emph{occurrence} of a monomial $m_u \coloneqq \prod_{i \in [\tau]\colon u_i \neq \bot} [h(u_i) = i]$ in $p_e$. We do not open up the representations of $[h(u_i) = i]$.  For each $(e,u)$ the coefficient of the corresponding monomial in $p_e$ is $c(e,u) \coloneqq (-1)^{|\{i \mid u_i \neq \bot\}|}$. The edges in the \textsc{End-of-Line} instance are either \emph{monomial-preserving}, they express the cancellation of two occurrences of the same monomial with opposite signs; or hyperedge-preserving, they express the semantic behavior of the polynomial $p_e(h)$. See \cref{fig:ppad-inclusion-illustration} for an example.
\begin{description}
    \item[Monomial-preserving] For $(e,u)$ identify the set $S$ of non-$\bot$ elements of $u$. By $\hashsize$-balancedness there exists a mapping $\sigma_S\colon E_S \to E_S$ that matches up edges satisfying $e \supseteq S$ with the property $\epsilon_e = -\epsilon_{\sigma_S(e)}$. Then we define $A(e,u) \coloneqq (\sigma_S(e),u)$.
    \item[Hyperedge-preserving] Fix a hyperedge $e \in E$. For all nodes $(e,u)$ if $m_u(h) = 1$ we define $B_h(e,u) \coloneqq (e, u')$ such that $m_u$ and $m_{u'}$ have opposite coefficients in $p_e(h)$ and $B_h(e,u') = (e,u)$. The mapping $B_h$ can be computed efficiently, since the set of $u$ for which monomials $m_u \in p_e$ evaluate to $1$ is a simple product set: $\prod_{i \in [\tau]} (\{\bot\} \cup \{v \in e \mid h(v) = i\})$: given a monomial $m_u$ we can compute its rank in the lexicographic order among all monomials with the same coefficient and map it to a monomial with the same rank, but opposite coefficient. If no such monomial exists, define $B_h(e,u) = \bot$. Thus, $B_h$ is not correctly defined iff there are different numbers of monomials with $+1$ and $-1$ coefficient, i.e. $p_e(h) \neq 0$.
\end{description}
\begin{figure}
    \centering
    \scalebox{0.7}{\definecolor{ppInk}{HTML}{3F4A5B}
\definecolor{ppMuted}{HTML}{758195}
\definecolor{ppLine}{HTML}{E9EDF3}
\definecolor{ppFrame}{HTML}{E3EAF0}
\definecolor{ppPanel}{HTML}{FFFFFF}
\definecolor{ppSpecialBg}{HTML}{FCEAE4}
\definecolor{ppOuter}{HTML}{9EB5D8}
\definecolor{ppOuterStrong}{HTML}{6D91C2}
\definecolor{ppInner}{HTML}{91C9B8}
\definecolor{ppInnerStrong}{HTML}{5DAB91}
\definecolor{ppPos}{HTML}{DCE8F8}
\definecolor{ppNeg}{HTML}{E2F2EC}
\definecolor{ppNodeStroke}{HTML}{95A2B4}
\definecolor{ppStrongNode}{HTML}{5B667C}
\definecolor{ppSource}{HTML}{F3C9B9}
\definecolor{ppSourceStroke}{HTML}{B9796B}
\definecolor{ppSink}{HTML}{DED1F5}
\definecolor{ppSinkStroke}{HTML}{8870B2}
\begin{tikzpicture}[x=1cm,y=1cm,
  every node/.style={font=\sffamily,text=ppInk},
  arrowm/.style={postaction={decorate},decoration={markings,mark=at position .52 with {\arrow{Stealth[length=3.0pt,width=2.6pt]}}}},
  ppOuterFaint/.style={draw=ppOuter, line width=1.0pt, opacity=.65, line cap=round, arrowm},
  ppInnerFaint/.style={draw=ppInner, line width=1.15pt, opacity=.69, line cap=round, arrowm},
  ppOuterPath/.style={draw=ppOuterStrong, line width=2.1pt, line cap=round, arrowm},
  ppInnerPath/.style={draw=ppInnerStrong, line width=2.1pt, line cap=round, arrowm},
  ppOrdinary/.style={circle,inner sep=0pt,minimum size=7.2pt,draw=ppNodeStroke,line width=.55pt},
  ppActivePath/.style={circle,inner sep=0pt,minimum size=8.6pt,fill=white,draw=ppStrongNode,line width=1.15pt},
  ppSourceNode/.style={circle,inner sep=0pt,minimum size=12.6pt,fill=ppSource,draw=ppSourceStroke,line width=1.55pt},
  ppSinkNode/.style={circle,inner sep=0pt,minimum size=12.6pt,fill=ppSink,draw=ppSinkStroke,line width=1.55pt}
]
\newcommand{\ppPanel}[3]{%
 \begin{scope}[shift={(#2,0)}]
   \filldraw[rounded corners=9pt,fill=ppPanel,draw=ppFrame,line width=.8pt] (-.05,-6.66) rectangle (10.05,1.62);
   \fill[ppSpecialBg,rounded corners=6pt] (1.00,-6.45) rectangle (2.12,.68);
   \node[font=\sffamily\small\bfseries] at (5.0,1.28) {#3};
   \foreach \c/\col in {1/{e^*},2/{234^-},3/{345^+},4/{135^-},5/{156^+},6/{456^-},7/{246^+},8/{126^-}} {%
     \pgfmathsetmacro{\cx}{1.56+1.075*(\c-1)}
     \node[font=\sffamily\scriptsize,text=ppMuted] at (\cx,.33) {$\col$};
     \draw[ppLine,line width=.55pt] (\cx-.5375,-.06)--(\cx-.5375,-6.45);
   }
   \draw[ppLine,line width=.55pt] (9.625,-.06)--(9.625,-6.45);
   \foreach \r/\mon in {1/{1},2/{x_4},3/{x_5},4/{y_5},5/{y_6},6/{x_4y_5},7/{x_4y_6},8/{x_5y_6}} {%
     \pgfmathsetmacro{\cy}{-.38-.78*(\r-1)}
     \node[font=\sffamily\scriptsize,text=ppMuted] at (.56,\cy) {$\mon$};
     \draw[ppLine,line width=.55pt] (1.0225,\cy-.39)--(9.625,\cy-.39);
     \foreach \c in {1,...,8} {%
       \pgfmathsetmacro{\cx}{1.56+1.075*(\c-1)}
       \coordinate (#1-\c-\r) at (\cx,\cy);
     }
   }
 \end{scope}
}
\ppPanel{L}{0}{(a)\quad $h^{(0)}=(0,0,0,1,1,2)$}
\ppPanel{R}{10.73}{(b)\quad $h^{(1)}=(0,0,0,1,2,2)$}
\draw[ppOuterFaint] (L-5-1) -- (L-6-1);
\draw[ppOuterFaint] (L-5-8) -- (L-6-8);
\draw[ppOuterFaint] (L-6-2) -- (L-7-2);
\draw[ppOuterFaint] (L-6-3) -- (L-5-3);
\draw[ppOuterFaint] (L-6-5) -- (L-5-5);
\draw[ppOuterFaint] (L-7-1) -- (L-8-1);
\draw[ppOuterFaint] (L-7-7) -- (L-6-7);
\draw[ppOuterFaint] (L-8-5) -- (L-7-5);
\draw[ppInnerFaint] (L-5-3) .. controls +(.32,0) and +(.32,0) .. (L-5-8);
\draw[ppInnerFaint] (L-5-5) .. controls +(.32,0) and +(.32,0) .. (L-5-1);
\draw[ppInnerFaint] (L-6-1) .. controls +(.32,0) and +(.32,0) .. (L-6-5);
\draw[ppInnerFaint] (L-6-7) .. controls +(.32,0) and +(.32,0) .. (L-6-2);
\draw[ppInnerFaint] (L-6-8) .. controls +(.32,0) and +(.32,0) .. (L-6-3);
\draw[ppInnerFaint] (L-7-2) .. controls +(.32,0) and +(.32,0) .. (L-7-7);
\draw[ppInnerFaint] (L-7-5) .. controls +(.32,0) and +(.32,0) .. (L-7-1);
\draw[ppInnerFaint] (L-8-1) .. controls +(.32,0) and +(.32,0) .. (L-8-5);
\draw[ppOuterPath] (L-1-1) -- (L-2-1);
\draw[ppInnerPath] (L-2-1) .. controls +(.32,0) and +(.32,0) .. (L-2-2);
\draw[ppOuterPath] (L-2-2) -- (L-3-2);
\draw[ppInnerPath] (L-3-2) .. controls +(.32,0) and +(.32,0) .. (L-3-1);
\draw[ppOuterPath] (L-3-1) -- (L-4-1);
\draw[ppInnerPath] (L-4-1) .. controls +(.32,0) and +(.32,0) .. (L-4-3);
\draw[ppOuterPath] (L-4-3) -- (L-3-3);
\node[ppActivePath] at (L-2-1) {};
\node[ppActivePath] at (L-2-2) {};
\node[ppActivePath] at (L-3-1) {};
\node[ppActivePath] at (L-3-2) {};
\node[ppActivePath] at (L-4-1) {};
\node[ppActivePath] at (L-4-3) {};
\node[ppOrdinary,fill=ppPos] at (L-5-1) {};
\node[ppOrdinary,fill=ppNeg] at (L-5-3) {};
\node[ppOrdinary,fill=ppNeg] at (L-5-5) {};
\node[ppOrdinary,fill=ppPos] at (L-5-8) {};
\node[ppOrdinary,fill=ppNeg] at (L-6-1) {};
\node[ppOrdinary,fill=ppPos] at (L-6-2) {};
\node[ppOrdinary,fill=ppPos] at (L-6-3) {};
\node[ppOrdinary,fill=ppPos] at (L-6-5) {};
\node[ppOrdinary,fill=ppNeg] at (L-6-7) {};
\node[ppOrdinary,fill=ppNeg] at (L-6-8) {};
\node[ppOrdinary,fill=ppPos] at (L-7-1) {};
\node[ppOrdinary,fill=ppNeg] at (L-7-2) {};
\node[ppOrdinary,fill=ppNeg] at (L-7-5) {};
\node[ppOrdinary,fill=ppPos] at (L-7-7) {};
\node[ppOrdinary,fill=ppNeg] at (L-8-1) {};
\node[ppOrdinary,fill=ppPos] at (L-8-5) {};
\node[ppSourceNode] at (L-1-1) {};
\node[ppSinkNode] at (L-3-3) {};
\draw[ppOuterFaint] (R-3-1) -- (R-4-1);
\draw[ppOuterFaint] (R-4-4) -- (R-3-4);
\draw[ppOuterFaint] (R-5-1) -- (R-6-1);
\draw[ppOuterFaint] (R-6-2) -- (R-7-2);
\draw[ppOuterFaint] (R-6-4) -- (R-5-4);
\draw[ppOuterFaint] (R-7-1) -- (R-8-1);
\draw[ppOuterFaint] (R-7-7) -- (R-6-7);
\draw[ppOuterFaint] (R-8-5) -- (R-7-5);
\draw[ppInnerFaint] (R-3-4) .. controls +(.32,0) and +(.32,0) .. (R-3-1);
\draw[ppInnerFaint] (R-4-1) .. controls +(.32,0) and +(.32,0) .. (R-4-4);
\draw[ppInnerFaint] (R-5-4) .. controls +(.32,0) and +(.32,0) .. (R-5-1);
\draw[ppInnerFaint] (R-6-1) .. controls +(.32,0) and +(.32,0) .. (R-6-4);
\draw[ppInnerFaint] (R-6-7) .. controls +(.32,0) and +(.32,0) .. (R-6-2);
\draw[ppInnerFaint] (R-7-2) .. controls +(.32,0) and +(.32,0) .. (R-7-7);
\draw[ppInnerFaint] (R-7-5) .. controls +(.32,0) and +(.32,0) .. (R-7-1);
\draw[ppInnerFaint] (R-8-1) .. controls +(.32,0) and +(.32,0) .. (R-8-5);
\draw[ppOuterPath] (R-1-1) -- (R-2-1);
\draw[ppInnerPath] (R-2-1) .. controls +(.32,0) and +(.32,0) .. (R-2-2);
\draw[ppOuterPath] (R-2-2) -- (R-3-2);
\draw[ppInnerPath] (R-3-2) .. controls +(.32,0) and +(.32,0) .. (R-3-6);
\draw[ppOuterPath] (R-3-6) -- (R-6-6);
\draw[ppInnerPath] (R-6-6) .. controls +(.32,0) and +(.32,0) .. (R-6-5);
\draw[ppOuterPath] (R-6-5) -- (R-5-5);
\node[ppActivePath] at (R-2-1) {};
\node[ppActivePath] at (R-2-2) {};
\node[ppOrdinary,fill=ppPos] at (R-3-1) {};
\node[ppActivePath] at (R-3-2) {};
\node[ppOrdinary,fill=ppNeg] at (R-3-4) {};
\node[ppActivePath] at (R-3-6) {};
\node[ppOrdinary,fill=ppNeg] at (R-4-1) {};
\node[ppOrdinary,fill=ppPos] at (R-4-4) {};
\node[ppOrdinary,fill=ppPos] at (R-5-1) {};
\node[ppOrdinary,fill=ppNeg] at (R-5-4) {};
\node[ppOrdinary,fill=ppNeg] at (R-6-1) {};
\node[ppOrdinary,fill=ppPos] at (R-6-2) {};
\node[ppOrdinary,fill=ppPos] at (R-6-4) {};
\node[ppActivePath] at (R-6-5) {};
\node[ppActivePath] at (R-6-6) {};
\node[ppOrdinary,fill=ppNeg] at (R-6-7) {};
\node[ppOrdinary,fill=ppPos] at (R-7-1) {};
\node[ppOrdinary,fill=ppNeg] at (R-7-2) {};
\node[ppOrdinary,fill=ppNeg] at (R-7-5) {};
\node[ppOrdinary,fill=ppPos] at (R-7-7) {};
\node[ppOrdinary,fill=ppNeg] at (R-8-1) {};
\node[ppOrdinary,fill=ppPos] at (R-8-5) {};
\node[ppSourceNode] at (R-1-1) {};
\node[ppSinkNode] at (R-5-5) {};
\end{tikzpicture}}
    \caption{An \textsc{End-of-Line} instance for $G=([6],E)$ with $E=\{123,234,345,135,156,456,246,126\}$, $e^*=123$, and $\epsilon_e=(-1)^{|e\cap\{4,5,6\}|}$, writing $abc=\{a,b,c\}$.
    Here $x_v=[h(v)=1]$, $y_v=[h(v)=2]$, and $0$ denotes $\bot$; only monomial occurrences evaluating to $1$ are shown.
    Blue edges ($A$) cancel equal monomials across opposite-sign hyperedges; green edges ($B_h$) match opposite coefficients within a hyperedge.
    Highlighted paths connect the source $(e^*,\bot^2)$ (peach) to sinks (purple) on non-special solution edges.}
    \label{fig:ppad-inclusion-illustration}
\end{figure}
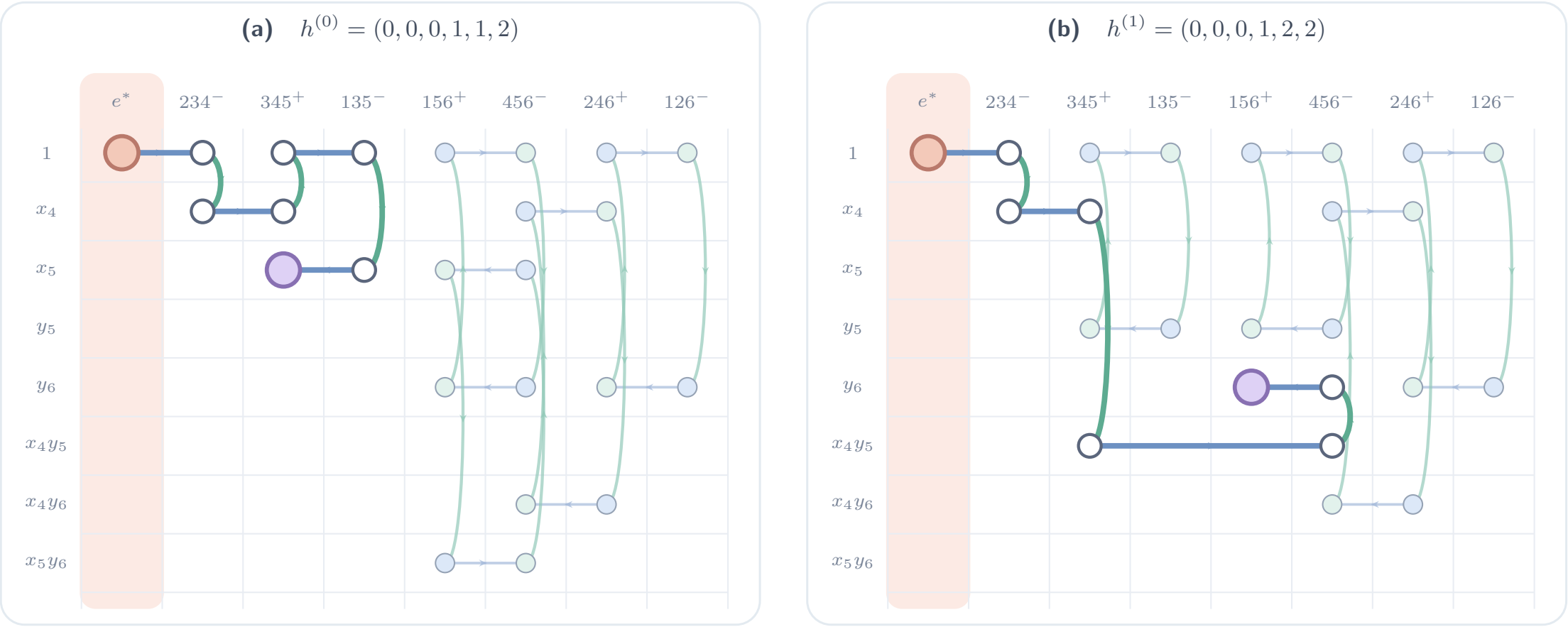

Finally we are ready to define successors and predecessors using the matching $A$ and $B_h$: if the monomial $m_u$ evaluates to $0$ by $h$, create a self-loop in the node $(e,u)$; otherwise if $\epsilon_e \cdot c(e,u) \cdot m_u(h) = 1$, take $\Succ_{e,u} \coloneqq A(e,u)$ and $\Pred_{e,u} \coloneqq B_h(e,u)$, if $B_h(e,u) = \bot$ set $\Pred_{e,u} = (e,u)$. If $\epsilon_e \cdot c(e,u) \cdot m_u(h) = -1$ swap $\Succ$ and $\Pred$.

By the definition of the matchings $A$ and $B_h$, all the nodes $(e,u)$ with $m_u(h) = 1$ are matched, unless $p_e(h) \neq 0$, hence, an unmatched node in the constructed \textsc{End-of-Line} instance reveals a solution. The exception is the node $(e^*, \bot^\tau)$, that we set to be the canonical source.\end{proof}

\subsection{Construction of the graph}
\label{sec:ppad-construction}

In this section, we construct $G = ([N], E)$ that satisfies the requirements of both \cref{thm:ppad-hardness} and \cref{thm:ppad-containment}:

\begin{theorem}
\label{thm:graph-existence}
    Suppose $\ell$ is large enough power of $4$. Then there exists $N=2^{\poly(\ell)}$ and an explicit $G=([N],E)$ such that $E \subseteq \binom{[N]}{\ell}$ is 
    \begin{itemize}[noitemsep]
        \item $(3\hashsize/2, \exp((\log N)^{\Omega(1)}), \exp(o(\ell)))$-well-spread;
        \item efficiently $\hashsize$-balanced;
        \item  every $e \in E \setminus \{e^*\}$ satisfies $|e \cap e^*| \le \hashsize$;
    \end{itemize} where $\hashsize = 3\ell/8$.
\end{theorem}

\paragraph{The code.} The construction of $G$ comes from a folded Reed-Solomon error correcting code. We identify $[N]$ with a set $[\ell] \times \Gamma$, identifying a codeword $c \in \Gamma^\ell$ with a hyperedge $e_c \coloneqq \{(i,c_i) \mid i \in [\ell]\}$. The zero codeword $0^\ell$ corresponds to the planted edge $e^* = e_{0^\ell} = [\ell]$.

Then well-spreadness of $E$ corresponds exactly to list-recovery of the code $C$: we say that $C \subseteq \Gamma^\ell$ is $(\delta,a,b)$-list recoverable if for every $S_1,\dots,S_\ell \subset \Gamma$ with $|S_i| \le a$, the number of codewords $(c_1,\dots,c_\ell) \in C$ with $|\{i \in [\ell]\mid c_i \in S_i\}| \ge (1-\delta)\ell$ is at most $b$. 
The following claim is trivial:
\begin{claim}
\label{cl:list-recover-to-spread}
    If $C \subseteq \Gamma^\ell$ is $(\delta, a, b)$-list recoverable then $E \coloneqq \{e_c \mid c \in C\}$ is $((1-\delta)\ell, a, b)$-well-spread.
\end{claim}

In order to verify the $\hashsize$-balancedness we need to fix a specific construction of a list-recoverable $C$. 
\newcommand{\ReedSolomon}{\textsf{ReedSolomon}}
\newcommand{\FoldedRS}{\textsf{FoldedRS}}

Given a field $\mathbb{F}$ with $|\mathbb{F}| = n + 1$ and a generator $\gamma \in \mathbb{F}^*$ define 
\[\ReedSolomon_{\mathbb{F}, \gamma, d} \coloneqq \{c \in \mathbb{F}^n \mid c_i = f(\gamma^{i-1}) \text{ where } f \in \mathbb{F}[x] \text { with } \deg(f) \le d\},\]
and $\FoldedRS^m_{\mathbb{F},\gamma,d} \coloneqq \{((c_1,\dots, c_{m}), (c_{m+1},\dots,c_{2m}), \dots, (c_{n-m+1},\dots, c_n)) \mid c \in \ReedSolomon_{\mathbb{F}, \gamma, d}\}$.

The following fact is a specialization of \cite[Section~3.6]{Rudra07} used (with slightly different parameters) in \cite{YZ24}.
\begin{corollary}
\label{cor:fewer-params}
Let $\ell$ be a large enough power of $4$ and $\Gamma = \mathbb{F}_{\ell^2}^{\ell - 1}$. Then $\FoldedRS^{\ell - 1}_{\mathbb{F}_{\ell^2}, \gamma, 3\ell (\ell-1)/8} \subseteq \Gamma^{\ell + 1}$ is $(9/16, \ell^{\ell^{1/4}}, \ell^{2\sqrt{\ell}})$-list recoverable.
\end{corollary}

We are now ready to prove the theorem.
\begin{proof}[Proof of \cref{thm:graph-existence}]
    By applying \cref{cor:fewer-params} we get $\Gamma = \mathbb{F}_{\ell^2}^{\ell - 1}$, and $C' \coloneqq \FoldedRS^{\ell-1}_{\mathbb{F}_{\ell^2}, \gamma, 3\ell (\ell - 1)/8} \subseteq \Gamma^{\ell + 1}$ that is $(9/16, \ell^{\ell^{1/4}}, \ell^{2\sqrt{\ell}})$-list recoverable. Let $C \coloneqq \{(c_1,\dots, c_\ell) \mid (c_1,\dots,c_{\ell + 1}) \in C'\}$ be the punctured version of $C'$. Since the codewords in $C'$ can be uniquely recovered by their marginals from $C$, the list-recoverability property is retained with a slightly weaker relative parameter: $1-(1-9/16)(\ell+1)/\ell=9/16-7/(16\ell)$. Hence by \cref{cl:list-recover-to-spread} we have that $E_C \coloneqq \{e_c \mid c \in C\}$ is $(7\ell/16 + O(1), \ell^{\ell^{1/4}}, \ell^{2\sqrt{\ell}})$-well-spread, which trivially implies the required well-spreadness since $3\tau / 2 = 9\ell / 16 \ge 7\ell/16 + O(1)$ for large enough $\ell$.

    Let us check the distance property of $E_C$. For two edges $e \neq e' \in E_C$ let $c$ and $c'$ be the codewords from $C$ corresponding to them. Then $|e \cap e'| \le |\{i \in [\ell]\mid c_i = c'_i\}| \le d/m = \tau$. The last inequality holds, since otherwise two polynomials agree on $>d$ points which implies $c=c'$. 

    Finally, we establish efficient $\tau$-balancedness of $E_C$. Take a point $x = \gamma^{m\ell + 1} \in \mathbb{F}_{\ell^2}$ and fix a function $\kappa \colon \mathbb{F}_{\ell^2} \to \{\pm 1\}$ that satisfies $\sum_{y \in \mathbb{F}_{\ell^2}} \kappa(y) = 0$. Then for an edge $e$ that corresponds to a codeword defined by the polynomial $p_e$ define $\epsilon_{e} \coloneqq \kappa(p_e(x))$. Now fix some $R \in \binom{[\ell] \times \Gamma}{\tau}$ (the case $|R| < \tau$ is handled below). We need to show that $\sum_{e \supseteq R} \epsilon_e = 0$. If $R$ contains two points whose first coordinates coincide, then no edge contains $R$ and the equality is trivial. Then $R$ has at most one point in each set $\{i\} \times \Gamma$, so the condition $e \supseteq R$ fixes the polynomial $p_e$ in $\tau m = d$ points. Fixing $p_e(x) = y$ to any value $y \in \mathbb{F}_{\ell^2}$ fixes the polynomial $p_e$, so there exists exactly one such polynomial $q_y$ for every $y$. Hence $\sum_{e \supseteq R} \epsilon_e = \sum_{y \in \mathbb{F}_{\ell^2}} \kappa(y) = 0$. In order to show that there is an efficient involution that matches the edges of opposite signs, choose $\kappa(y) \coloneqq (-1)^{\text{the first bit in the representation of }y}$. Then in order to match-up the opposite-sign edges, given $e$ and a valid $R$ compute $y = p_e(x)$ and take $y'$ be the result of flipping the first bit in $y$. Then let $e'$ be the result of the interpolation of a polynomial with $p_{e'}(x) = y'$ and the $d$ points defined by $R$. 

    Now let us verify $\tau$-balancedness for $R \in \binom{[\ell] \times \Gamma}{<\tau}$, again we may assume wlog that $R$ contains at most one point in each set $\{i\} \times \Gamma$, let $I \coloneqq \{i \in [\ell] \mid R \cap \{i\} \times \Gamma \neq \emptyset\}$. Fix a canonical $I' \supseteq I$ of size exactly $\tau$, and define $\sigma_R(e) \coloneqq \sigma_{R'}(e)$ where $R' \coloneqq (I' \times \Gamma) \cap e$. 
\end{proof}

\paragraph{Proof of \cref{cor:fewer-params}}
We use the following fact.
\begin{fact}[restatement of {\cite[Lemma~4.3]{YZ24}}, originally from {\cite[Section~3.6]{Rudra07}}]
\label{fact:YZcode}
    Suppose $m$ divides $n = |\mathbb{F}| - 1$. Let $\alpha, \beta$ be positive integers and  $s \le m$, $\delta \in (0,1)$ be a real and suppose the following inequalities hold:
    \[ \frac{(1-\delta) n}{m} \ge \left(1 + \frac{s}{\beta}\right) \cdot \frac{(n \alpha d^s)^{1/(s+1)}}{m - s + 1} ~~~~~ (\beta + s) \left(\frac{n \alpha}{d}\right)^{1/(s+1)} < |\mathbb{F}|.\] 
    Then $\FoldedRS^m_{\mathbb{F}, \gamma, d}$ is $(\delta, \alpha, |\mathbb{F}|^s)$-list recoverable.
\end{fact}

\begin{proof}[Proof of \cref{cor:fewer-params}]
Instantiate \cref{fact:YZcode} with the following parameters: 
\[ |\mathbb{F}| = \ell^2; ~~~ 
   m = \ell - 1; ~~~
   \tau = 3\ell/8; ~~~
   s = \sqrt{\ell}; ~~~
   \beta = \ell; ~~~ d = \tau m.\]
Then for $\delta \in (0,1)$, $\alpha \in \mathbb{Z}_{>0}$, it suffices to verify that:
    \[(1-\delta) \ge 3/8 (1 + o(1)) \alpha^{1/(1 + \sqrt{\ell})} ~~~~
    (1+o(1)) (8\alpha/3)^{1/(1+\sqrt{\ell})} < \ell, \]
for the conclusion of \cref{fact:YZcode} to hold. Setting 
    $\alpha = \ell^{\ell^{1/4}}$ and $\delta = 9/16$, we get the claim.
\end{proof}

\subsection{Assembling the proof of \cref{thm:main-ppad-hardness}}
\begin{proof}[Proof of \cref{thm:main-ppad-hardness}]
Let $\ell$ be a sufficiently large power of $4$ and use the graph constructed in \cref{thm:graph-existence}, with parameters
\(
    N=\ell|\Gamma|=\ell^{2\ell-1},
    \tau=3\ell/8,
    a=\ell^{\ell^{1/4}},
    b=\ell^{2\sqrt{\ell}}.
\)
By \cref{thm:ppad-hardness}, every decision tree of depth at most $a$ solves $\Singleton_G$ with probability at most
\[
    O(b)\exp(-\Omega(\tau))
    =\exp\bigl(O(\sqrt{\ell}\log\ell)-\Omega(\ell)\bigr)
    =\exp(-\Omega(\ell)).
\]
Since $\log N=\Theta(\ell\log\ell)$, this covers every depth $\exp((\log N)^{o(1)})$ and gives success at most $\exp(-(\log N)^{\Omega(1)})$. Efficient $\tau$-balancedness and $\ell=\poly(\log N)$ give $\Singleton_G\in\PPAD$ by \cref{thm:ppad-containment}.

Set $\Sigma=\{\bot\}\cup[\tau]$ and define $(u,e)\in F_N$ iff $(h_u,e)\in\Singleton_G$, where $h_u(i)=\bot$ for $i\in[\ell]$ and $h_u(i)=u_i$ otherwise. Then $F_N\in\PPAD$, and uniform $u\in\Sigma^N$ induces uniform $h_u\in\calH$. Fixing the ignored coordinates $u_1,\dots,u_\ell$ and averaging transfers the same decision-tree lower bound to $F_N$, proving the theorem.
\end{proof}

\subsection{Corollary: $\EoPL$ is not in $\FP$}
We will use the following claim that allows us to combine two $\TFNP$ search problems. The proof is straightforward.

\begin{claim} \label{claim:or-composition}
    Let $F = \left\{ F_N \subseteq \Sigma^N \times \mathcal{O} \right\}$ and $G = \left\{ G_N \subseteq \Sigma^N \times \mathcal{O}' \right\}$ be two arbitrary $\TFNP$ search problems. Define $H = \left\{ H_N \subseteq \Sigma^N \times (\mathcal{O} \cup \mathcal{O}') \right\}$ by
    \begin{equation*}
        (u, v) \in H_N \iff (u, v) \in F_N \text{ or } (u, v) \in G_N.
    \end{equation*}
    Then,
    \begin{itemize}[noitemsep]
        \item $H$ is efficiently reducible to either of $F$ and $G$.
        \item If any depth-$q$ decision tree solves $F$ and $G$ with probability at most $\epsilon_F$ and $\epsilon_G$ over $\bm u \sim \Sigma^N$, respectively, then any depth-$q$ decision tree solves $H$ with probability at most $\epsilon_F + \epsilon_G$ over $\bm u \sim \Sigma^N$.
    \end{itemize}
\end{claim}

The following is a direct corollary from the hardness of $\PPAD$ (\cref{thm:main-ppad-hardness}), the hardness of $\PLS$ (\cite{BG20})\footnote{\cite{BG20} proved the separation in the random oracle model. However, it can be verified that their proof implies a decision-tree lower bound in the black-box setting.}, the collapse $\PPAD \cap \PLS = \EoPL$ (\cite[Theorem~1]{GHJ22}), and \cref{claim:or-composition}. We note that $\epsilon_{\PLS}$ supplied by \cite[Theorems~36 and~43]{BG20} is $o(1)$.

\begin{corollary} \label{cor:eopl-hardness}
    There exists a $\TFNP$ search problem $F = \left\{ F_N \subseteq \Sigma^N \times \mathcal{O}\right\}$ in $\EoPL$ satisfying that: every decision tree $T$ with depth at most $\exp\left((\log N)^{o(1)}\right)$ solves $F$ with probability at most
    \begin{equation*}
        \Pr_{\bm{u} \sim \Sigma^N} \left[ (\bm{u}, T(\bm{u})) \in F_N \right] = o(1).
    \end{equation*}
    In particular, $\EoPL \not\subseteq_{\rm ROM} \FP$.
\end{corollary}

\section{Efficient search for $\UEoPL$ in the black-box ROM}
\label{sec:ueopl}
\newcommand{\Jump}{\textsc{Jump}}
\newcommand{\fail}{\mathsf{fail}}
\newcommand{\qc}{\mathsf{Q}}
\algblockdefx[UEOPLLOOP]{RepeatLoop}{EndRepeatLoop}{\textbf{loop}}{\textbf{end loop}}

In this section we show that $\UEoPL$ is easy in the random oracle model: every instance that is generated from the oracle by shallow decision trees can be solved, with high probability over the oracle, by a shallow decision tree. We use the definition of $\UEoPL$ from~\cite[Section~3.2]{GHJ24}, which we reproduce here.

\begin{definition}[$\UEoPL_N$ {\cite[Section~3.2]{GHJ24}}]
\label{def:ueopl}
Let $N\geq 2$. As input we are given a successor $\Succ_u\in[N]\cup\{\bot\}$ for each $u\in[N]\times[N]$ and a predecessor $\Pred_u\in[N]\cup\{\bot\}$ for each $u\in\{2,\ldots,N\}\times[N]$. A node $(i,j)\in[N-1]\times[N]$ is \emph{active} if $\Succ_{ij}=k\neq\bot$ and $\Pred_{i+1,k}=j$, otherwise it is \emph{inactive}; a node $(i,j)\in\{N\}\times[N]$ is active if $\Succ_{ij}\neq\bot$ and inactive otherwise. A node $u$ is a \emph{proper sink} if $u$ is inactive but some active node has $u$ as a successor. A node $(i,j)$ is a \emph{proper source} if it is active, $i<N$, and no active node has $(i,j)$ as a successor. The goal is to output any of the following
\begin{enumerate}[noitemsep]
    \item $(1,1)$, if $(1,1)$ is inactive, \hfill (inactive distinguished source)
    \item $(N,j)$, if $(N,j)$ is active, \hfill (active sink)
    \item $(i,j)$, if $(i,j)$ is a proper sink, \hfill (proper sink)
    \item $(i,j)$, if $(i,j)\neq(1,1)$ and $(i,j)$ is a proper source, \hfill (proper source)
    \item $(i,j)$ and $(i,j')$, if $j\neq j'$ and both nodes are active. \hfill (two parallel lines)
\end{enumerate}
Formally, $\UEoPL_N$ is the set of all pairs $(x,z)$ such that $x$ is an input as above and $z$ is one of the outputs listed for $x$. We call $x$ an \emph{instance} and $z$ a \emph{solution} of $x$.
\end{definition}

Here the successor $\Succ_v=k\neq\bot$ of a node $v=(i,j)$ with $i<N$ is interpreted as naming the node $(i+1,k)$ on the next row, and we also call this node the successor of $v$. A solution always exists: if $(1,1)$ is active, then following successors from $(1,1)$ traces a path that advances by one row with every step and ends in a proper sink or in an active node of the last row.

A proper sink $u=(i,j)$ is locally verified by checking that $i>1$, $\Pred_u=p\neq\bot$, $\Succ_{i-1,p}=j$, and $u$ is inactive. The two pointer equalities certify an incoming edge from an active node, and inactivity is checked with at most two further pointer evaluations.

\paragraph{Instances in the random oracle model.}
An instance is generated from the oracle $u\in\{0,1\}^{\mathbb{N}}$ by a decision forest $\omega$: every successor $\Succ_v$ and predecessor $\Pred_v$ of the instance $\omega(u)$ is computed by a decision tree that queries bits of $u$ and outputs an element of $[N]\cup\{\bot\}$. We write $q$ for the maximum depth of the decision trees in $\omega$.

\begin{theorem}
\label{thm:ueopl}
Let $N\geq 2$ and $q\geq 1$, and let $\omega$ be a decision forest that generates instances of $\UEoPL_N$ and computes every pointer in depth at most $q$. Then there is a decision tree $T$ of depth $\poly(q)\cdot\log N$ such that
\[
    \Pr_{\bm u\sim\{0,1\}^{\mathbb{N}}}\big[(\omega(\bm u),T(\bm u))\in\UEoPL_N\big]\geq\frac{2}{3}.
\]
Moreover, for every $u\in\{0,1\}^{\mathbb{N}}$, the output $T(u)$ is either a solution of $\omega(u)$ or the failure symbol $\fail$.
\end{theorem}

\paragraph{Remark on upper bounds in the decision tree setting.}

\cref{thm:ueopl} bounds the number of oracle queries, not the computation: the tree $T$ is shown to exist, and nothing is claimed about computing it. This is likely unavoidable, since $\UEoPL$ is hard in the white-box random oracle model under a complexity assumption~\cite{CHKPRR19}.

\paragraph{Restrictions.}
A \emph{restriction} $\rho$ assigns values to finitely many bits of the oracle, and $\cube(\rho)\subseteq\{0,1\}^{\mathbb{N}}$ is the set of oracles that agree with $\rho$. We write $\Pr_\rho$ and $\Exp_\rho$ for probability and expectation over $\bm u\sim\cube(\rho)$. A decision tree on $\cube(\rho)$ queries only bits that are not fixed by $\rho$.

\paragraph{Decision tree families.} Throughout, a \emph{family of decision trees} $f = (f_y)_{y \in Y}$ with $f_y\colon \{0,1\}^{\mathbb{N}} \rightarrow \{0,1\}$ \emph{accepts} the label $y \in Y$ on $u$ if $f_y(u)=1$; the failure symbol $\fail$ is never accepted.
We say that a decision tree $D\colon \{0,1\}^{\mathbb{N}} \rightarrow Y \cup \{\fail\}$ \emph{certifies} the label $y \in Y$ on $u$ if, before outputting $y$, it has queried all the bits fixed by some accepting leaf of $f_y$ that $u$ reaches; in that case, $f$ accepts $y$ on every oracle that agrees with the queried bits.

\medskip

The proof of \cref{thm:ueopl} relies on the following lemma, which does not refer to $\UEoPL$.

\begin{lemma}
\label{lem:dichotomy}
For every $q\geq 1$ there are $a=O(q^2)$ and $\eta=\Theta(1/q)$ such that the following holds.
Let $f=(f_y)_{y\in Y}$ be a finite family of decision trees $f_y\colon\{0,1\}^{\mathbb{N}}\to\{0,1\}$ of depth at most $q$ and 
let $D\colon\{0,1\}^{\mathbb{N}}\to Y\cup\{\fail\}$ be a decision tree of depth $t$ such that
\[
    \Pr\nolimits_{\bm u\sim\{0,1\}^{\mathbb{N}}}\big[f\text{ accepts }D(\bm u)\text{ on }\bm u\big]\geq 1-\eta.
\]
Then at least one of the following holds.
\begin{enumerate}[noitemsep]
    \item There is a decision tree $T_1\colon\{0,1\}^{\mathbb{N}}\to Y\cup\{\fail\}$ of depth at most $a$ such that $\Pr_{\bm u}[T_1(\bm u)\neq\fail]\geq 1/2$, and 
    $T_1$ certifies $T_1(u)$ on $u$ whenever $T_1(u)\neq\fail$.
    \item There is a decision tree $T_2\colon\{0,1\}^{\mathbb{N}}\to\binom{Y}{2}\cup\{\fail\}$ of depth $O(t+a)$ such that $\Pr_{\bm u}[T_2(\bm u)\neq\fail]\geq\eta$, and 
    $T_2$ certifies both labels of $T_2(u)$ on $u$ whenever $T_2(u)\neq\fail$.
\end{enumerate}
In both cases the tree queries only bits that $D$ or some $f_y$ queries.
\end{lemma}

\begin{remark}
\cref{lem:dichotomy} is a decision-tree form of \cite[Corollary~5.4]{BCG26}. The idea that a search problem with (almost) unique solutions cannot be hard relative to a random oracle goes back to Rudich~\cite{Rudich}; his conjecture was proved by Kahn, Saks, and Smyth~\cite{KSS10} via the dual BKR inequality, which \cite{GRY25} and then \cite{BCG26} applied to verifiable delay functions and proofs of work. The proof below does not use the BKR inequality.
\end{remark}

\paragraph{Intuition behind \cref{lem:dichotomy}.}

In \cref{sec:ueopl-proof} we prove our $\UEoPL$ upper bound (\cref{thm:ueopl}) by applying \cref{lem:dichotomy} recursively, where $Y$ will correspond to the set of nodes in a row of the $\UEoPL$ instance, and the decision tree family $(f_y)_{y \in Y}$ will be the depth-$2q$ decision trees checking whether node $y$ is active.
Therefore ``$f$ accepts $y \in Y$ on $\bm u$'' means that  node $y$ in this row is active on input $\bm u$, and ``$D$ certifies $y$ on $\bm u$'' means that $D$ outputs $y$ after reading a certificate that $y$ is active.
In this context,  \cref{lem:dichotomy} gives a win-win dichotomy applied to this row of nodes. 
Informally it states: 
Let $D$ be any decision tree (of potentially very large depth $t \gg q$) that outputs an active node in this row with probability at least $1-\eta$.
Then at least one of the following holds:
(1) There is a shallow decision tree $T_1$ (of depth at most $O(q^2)$) such that $T_1(\bm u)$ outputs an active node in this row with probability at least $1/2$, and furthermore $T_1(\bm u)$ always outputs either an active node or $\fail$; or  
(2) There is a decision tree $T_2$ of depth $O(t+a)$ such that $T_2(\bm u)$ outputs {\it two} distinct active nodes in this row with probability at least $\eta$, and furthermore $T_2(\bm u)$ always outputs either 2 active nodes or $\fail$.

Why should we expect such a win-win situation? 
Think of each accepting leaf of an activity tree $f_y$ as a certificate: a small restriction $\rho$ that guarantees that $y$ is active for all $u$ consistent with $\rho$.
The tree $D$ (however deep it is) is assumed to almost always output an active node, so 
almost every oracle $u$ is active for at least one node, namely the one that $D(u)$ outputs.
 Now there are two cases: either (1) the certificates corresponding to accepting leaves of $(f_y)_{y \in Y}$ are almost  disjoint, or (2) they overlap substantially.
In case (1), $D$ was never needed: 
we can construct a shallow tree $T_1$ that finds the unique active node in this row by applying a robust version of the classic result (due to Nisan) showing that any Boolean function represented by a disjoint $q$-DNF %
can be transformed into a depth $O(q^2)$ decision tree.
In case (2) since the certificates overlap, on many input oracles $u$, $D(u)$ outputs one active node while $u$ is also consistent with a different certificate. Such $u$'s have at least 2 active nodes, which in $\UEoPL$ is a solution! 
In this second case using a sensitivity argument it is shown that the tree $T_2$ obtained by running $D$ twice, once on $u$ and once on a slightly perturbed copy of $u$, tends to output two active nodes.
 In this second case even though the depth of $T_2$ is close to that of $D$, we get a big win since $T_2$ outputs an actual solution. As we explain below, we avoid incurring the expensive cost of running this high depth tree recursively by delaying running it until the end where it is only run once.

\subsection{Proof of \cref{thm:ueopl}}
\label{sec:ueopl-proof}

Fix $\omega$ and $q$ as in \cref{thm:ueopl}.  Throughout the algorithm,
$\rho$ denotes the restriction that records the answers to all queries made so
far.  It is a single global object: every subroutine reads it, and every query
that any subroutine makes extends it.  For an algorithm $A$ that queries the
oracle, we write $\qc(A)$ for the number of queries that $A$ makes, and call
$\Exp_\rho[\qc(A)]$ the expected query complexity of $A$ on $\cube(\rho)$.  A
property of the instance $\omega(u)$ \emph{holds on $\cube(\rho)$} if it holds
for every $u\in\cube(\rho)$.
 
\paragraph{Activity trees.}
For a node $v$, let $\act_v(u)\in\{0,1\}$ indicate whether $v$ is active in
$\omega(u)$.  This is computed by a decision tree of depth at most $2q$, which
evaluates $\Succ_v$ and then, if $v$ is not in the last row and $\Succ_v\neq\bot$, the
predecessor of the successor of $v$.  For a row $b\in[N]$, let
$\act^b\coloneqq (\act_{(b,j)})_{j\in[N]}$ be the family of these trees for the nodes of
row $b$.  Throughout the proof, $a=O(q^2)$ and $\eta=\Theta(1/q)$ are the
parameters that Lemma~\ref{lem:dichotomy} provides for verifier depth $2q$.
 
We apply \cref{lem:dichotomy} to a tree $D$ on $\cube(\rho)$ and to the trees
$\act^b$ with the bits of $\rho$ fixed, which does not increase their depth.
These trees, and hence $T_1$ and $T_2$, query no bit that $\rho$ fixes, so for
every event $E$ determined by their outputs,
$\Pr_{u\sim\cube(\rho)}[E]=\Pr_{u\sim\{0,1\}^{\mathbb{N}}}[E]$ after renaming the free bits.
In particular, a label that $T_1$ or $T_2$ certifies is accepted on the cube of
the restriction obtained after running the tree; for the activity trees this
means that the node is active on that cube.
 
\paragraph{High-level description of the algorithm.}
The backbone of the algorithm is the recursive subroutine $\Jump(v,r)$, given in
\cref{alg:ueopl-jump}, which jumps from a node $v=(i,j)$ in row $i$ to the target row 
$b\coloneqq \min\{i+2^r,N\}$.  It is called only on nodes that are \emph{active on
$\cube(\rho)$}, and it returns one of three things: an active node of row $b$, a
solution, or a decision tree that finds a solution with probability at least
$\eta$ (\cref{clm:ueopl-output}).
The third kind of output is explained below.
 
For $r=0$ the subroutine reads the successor of $v$ and tests whether it is
active, which gives either a proper sink or an active node of the next row.  For
$r\ge1$, the obvious way to reach row $b$ is by two jumps of $2^{r-1}$ rows (the
solid arrows in \cref{fig:ueopl-jump}): let $C$ be the decision tree that runs
$\Jump(v,r-1)$ and then, from the node $v'$ that this returns, $\Jump(v',r-1)$.
But running $C$ at every level would double the number of queries with each
level, and $\log N$ levels would bring us back to $\Theta(N)$ queries.  So, as a
rule, we do not run $C$.  Instead we use it as the tree $D$ of
\cref{lem:dichotomy}, applied with the activity trees of row $b$ as
verifiers: since $C$ almost always outputs an active node of row $b$, the lemma
provides one of two trees.
\begin{itemize}[leftmargin=*,noitemsep]
 \item In the compression case it provides a tree $T_1$ of depth $a$, which
   does not depend on $r$, that outputs an active node of row $b$ with
   probability at least $1/2$ (the dashed arrow in \cref{fig:ueopl-jump}).  We
   then jump by running $T_1$ instead of $C$.  Since $T_1$ outputs $\fail$ on
   some oracles, this may fail; but then we have learned at least one new bit,
   so we repeat the loop with the enlarged restriction.
 \item In the collision case it provides a tree $T_2$ that outputs two active
   nodes of row $b$, which is a solution, with probability at least $\eta$.
   This is as good as a jump (in fact better), except for its cost: the depth of $T_2$ is of
   the order of $1/\eta$ times the query complexity of $C$.  Running it inside
   the recursion would multiply the cost of every level by $1/\eta$, which
   compounds over the $\log N$ levels.  So \Jump\ does not run $T_2$, and instead it
   returns the tree $T_2$ itself as its output, and every enclosing call
   returns it as well: a run of $C$ that outputs a tree ends the loop of the
   next level in 
   line~\ref{line:ueopl-expand}
   which returns the tree, and so on up to
   \cref{alg:ueopl-main}.  Thus the whole recursion unwinds at once, with no
   further queries, and \cref{alg:ueopl-main} runs $T_2$ a single time.  If
   that run fails, \cref{alg:ueopl-main} restarts $\Jump((1,1),r^*)$ from
   the enlarged restriction.  In this way the factor $1/\eta$ is paid once, at
   the top level, rather than once per level.
\end{itemize}
The only situation in which $C$ itself is run is when it is likely, with
probability at least $\eta/2$, to output a solution or a tree
(line~\ref{line:ueopl-expand}).
 Either output ends the entire computation of
$\Jump(v,r)$ as just described, so this happens $O(1/\eta)$ times per level in
expectation (\cref{clm:ueopl-queries}).  In every other situation $C$ outputs an
active node of row $b$ with probability more than $1-\eta/2$; cutting it off
after $t=\lceil 2\Exp_\rho[\qc(C)]/\eta\rceil$ queries (line~\ref{line:ueopl-D})
costs another $\eta/2$ by Markov's inequality and yields the tree $D$ of
bounded depth that \cref{lem:dichotomy} needs.
 
It is important for the accounting that lines
~\ref{line:ueopl-C},~\ref{line:ueopl-t},~\ref{line:ueopl-D} and \ref{line:ueopl-T2}
of \cref{alg:ueopl-jump} make no oracle
queries: they define trees and compute probabilities and expectations on
$\cube(\rho)$.  The recursion is executed only in line~\ref{line:ueopl-expand}, %
and
queries are made only in lines~\ref{line:ueopl-base},~\ref{line:ueopl-expand} and \ref{line:ueopl-T1}.
 
\medskip

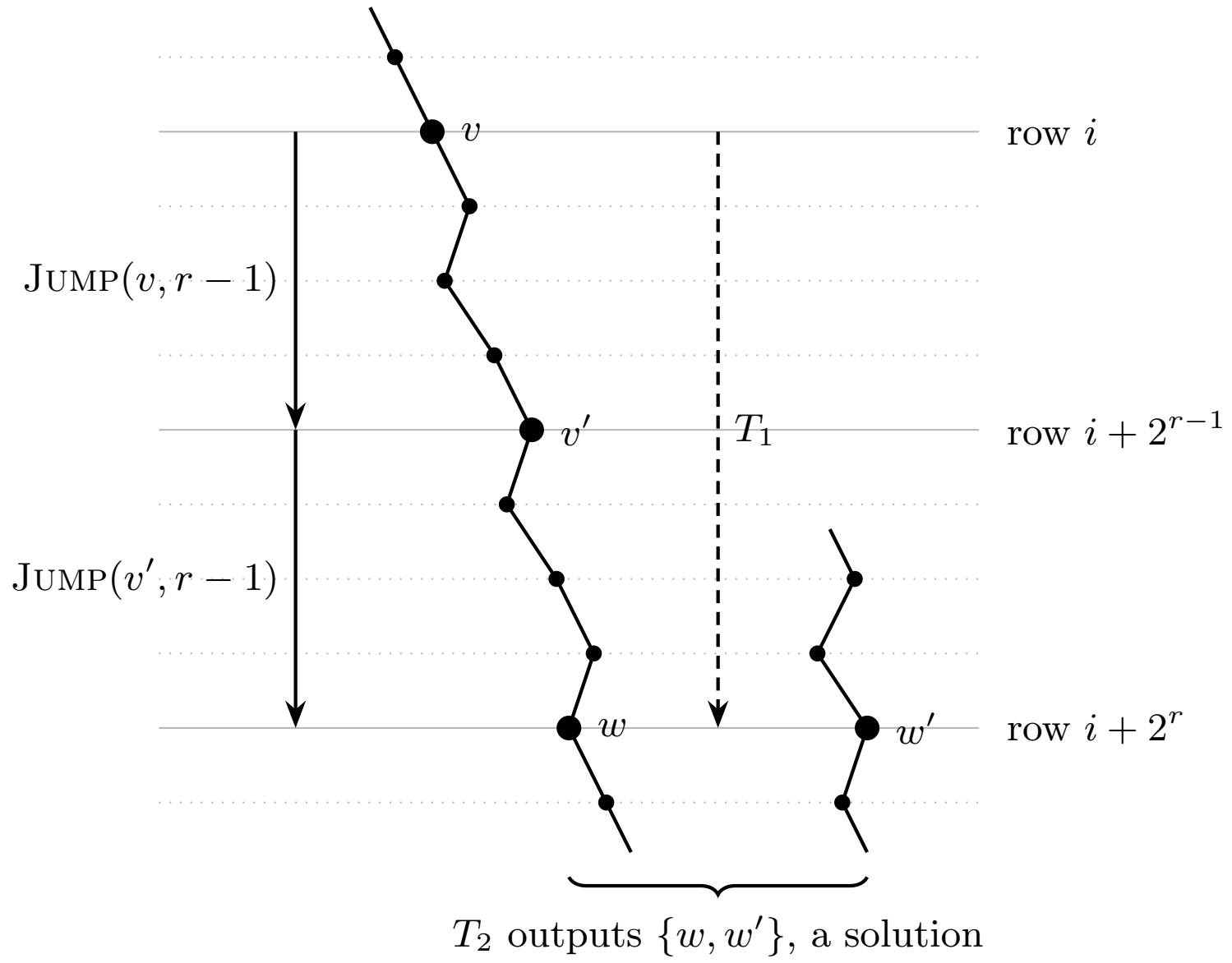
\begin{figure}[t]
\centering
\begin{tikzpicture}[x=1cm,y=1cm,
  act/.style={circle,fill,inner sep=1.3pt},
  big/.style={circle,fill,inner sep=2pt},
  row/.style={gray!55,thin},
  jump/.style={-{Stealth[length=2.2mm]},thick},
  lbl/.style={font=\small}]
  \foreach \y/\name in {0/{row $i$},-2.4/{row $i+2^{r-1}$},-4.8/{row $i+2^{r}$}} {
    \draw[row] (1.4,\y) -- (8.0,\y);
    \node[lbl,anchor=west] at (8.1,\y) {\name};
  }
  \foreach \y in {0.6,-0.6,-1.2,-1.8,-3.0,-3.6,-4.2,-5.4} \draw[row,dotted] (1.4,\y) -- (8.0,\y);
  \draw[thick] (3.1,1.0) -- (3.3,0.6) -- (3.6,0) -- (3.9,-0.6) -- (3.7,-1.2) -- (4.1,-1.8) -- (4.4,-2.4) -- (4.2,-3.0) -- (4.6,-3.6) -- (4.9,-4.2) -- (4.7,-4.8) -- (5.0,-5.4) -- (5.2,-5.8);
  \foreach \p in {(3.3,0.6),(3.9,-0.6),(3.7,-1.2),(4.1,-1.8),(4.2,-3.0),(4.6,-3.6),(4.9,-4.2),(5.0,-5.4)} \node[act] at \p {};
  \node[big,label={[lbl]right:$v$}] (v) at (3.6,0) {};
  \node[big,label={[lbl]right:$v'$}] (vp) at (4.4,-2.4) {};
  \node[big,label={[lbl]right:$w$}] (w) at (4.7,-4.8) {};
  \draw[thick] (6.8,-3.2) -- (7.0,-3.6) -- (6.7,-4.2) -- (7.1,-4.8) -- (6.9,-5.4) -- (7.1,-5.8);
  \foreach \p in {(7.0,-3.6),(6.7,-4.2),(6.9,-5.4)} \node[act] at \p {};
  \node[big,label={[lbl]right:$w'$}] (wp) at (7.1,-4.8) {};
  \draw[jump] (2.5,0) -- node[lbl,left]{$\Jump(v,r-1)$} (2.5,-2.4);
  \draw[jump] (2.5,-2.4) -- node[lbl,left]{$\Jump(v',r-1)$} (2.5,-4.8);
  \draw[jump,densely dashed] (5.9,0) -- node[lbl,right]{$T_1$} (5.9,-4.8);
  \draw[decorate,decoration={brace,mirror,amplitude=4pt},thick] (4.7,-6.0) -- (7.1,-6.0) node[lbl,midway,below=5pt]{$T_2$ outputs $\{w,w'\}$, a solution};
\end{tikzpicture}
\caption{One iteration of the loop in \cref{alg:ueopl-jump}: the two jumps on the left form $C$, $T_1$ (dashed) jumps directly, and $T_2$ outputs two active nodes $w,w'$ of row $i+2^r$. The dots indicate active nodes at each layer.}
\label{fig:ueopl-jump}
\end{figure}

\begin{algorithm}[H]
\renewcommand{\algorithmicrequire}{\textbf{Input:}}
\renewcommand{\algorithmicensure}{\textbf{Output:}}
\algtext*{EndIf}
\algtext*{EndRepeatLoop}
\caption{Subroutine $\Jump$ in \cref{alg:ueopl-main}.}
\label{alg:ueopl-jump}
\begin{algorithmic}[1]
\Require a node $v=(i,j)$ that is active on $\cube(\rho)$, and $r\geq 0$
\Ensure an active node of row $\min\{i+2^r,N\}$, a solution, or a decision tree
\State \textbf{if} $i=N$ \textbf{then return} the node $v$.
\State Set $b\coloneqq\min\{i+2^r,N\}$.
\If{$r=0$}
    \State Query $\Succ_v$ and $\act_{v'}$, where $v'\coloneqq(i+1,\Succ_v)$ is the successor of $v$. \label{line:ueopl-base}
    \State \Return the solution $v'$ if $v'$ is inactive; the node $v'$ otherwise.
\EndIf
\RepeatLoop
    \State \parbox[t]{\dimexpr\linewidth-\algorithmicindent\relax}{Let $C$ be the decision tree on $\cube(\rho)$ that runs $\Jump(v,r-1)$ and, if it outputs a node $v'$, runs $\Jump(v',r-1)$; $C$ outputs the output of the last run.\strut} \label{line:ueopl-C}
    \State \textbf{if} $\Pr_\rho[C\text{ outputs a solution or a tree on }\bm{u}]\geq\eta/2$ \textbf{then} run $C$ and \textbf{return} its output. \label{line:ueopl-test}\label{line:ueopl-expand}
    \State Set $t\coloneqq\lceil 2\Exp_\rho[\qc(C)]/\eta\rceil$. \label{line:ueopl-t}
    \State Let $D$ be $C$ cut off after $t$ queries, with output $\fail$ if $C$ is cut off or does not output a node. \label{line:ueopl-D}
    \State Apply \cref{lem:dichotomy} on $\cube(\rho)$ to $\act^b$ and $D$. \textbf{if} it provides no tree $T_1$ \textbf{then return} the tree $T_2$. \label{line:ueopl-T2}
    \State Run $T_1$. \textbf{if} $T_1$ outputs a node $v'$ \textbf{then return} the node $v'$. \label{line:ueopl-T1}
\EndRepeatLoop
\end{algorithmic}
\end{algorithm}

\begin{algorithm}[H]
\renewcommand{\algorithmicrequire}{\textbf{Input:}}
\renewcommand{\algorithmicensure}{\textbf{Output:}}
\algtext*{EndRepeatLoop}
\caption{The main algorithm.}
\label{alg:ueopl-main}
\begin{algorithmic}[1]
\Require $u\in\{0,1\}^{\mathbb{N}}$
\Ensure a solution of $\omega(u)$
\State Query $\act_{(1,1)}$. \textbf{if} $(1,1)$ is inactive \textbf{then return} $(1,1)$.
\State Set $r^*\coloneqq\lceil\log_2 N\rceil$.
\RepeatLoop
    \State Run $\Jump((1,1),r^*)$.
    \begin{itemize}[nosep,label=\textbullet,leftmargin=\dimexpr\algorithmicindent+2em\relax]
        \item If it outputs a node or a solution $z$: \textbf{return} $z$.
        \item If it outputs a decision tree $G$: run $G$, and \textbf{return} its output if it is a solution.
    \end{itemize}
\EndRepeatLoop
\end{algorithmic}
\end{algorithm}

Let $r\geq 0$, let $\rho$ be a restriction, and let $v=(i,j)$ be a node that is active on $\cube(\rho)$. The following four claims are about $\Jump(v,r)$ started from $\rho$. \cref{clm:ueopl-halts,clm:ueopl-output} are proved together by induction on $r$; \cref{clm:ueopl-queries,clm:ueopl-depth} then follow.

\begin{claim}
\label{clm:ueopl-halts}
$\Jump(v,r)$ halts on every $u\in\cube(\rho)$.
\end{claim}

\begin{claim}
\label{clm:ueopl-output}
$\Jump(v,r)$ outputs one of the following, where $\rho'$ is the restriction right before it halts:
\begin{itemize}[noitemsep]
    \item a node of its target row that is active on $\cube(\rho')$;
    \item a solution on $\cube(\rho')$;
    \item a decision tree $G$ on $\cube(\rho')$ such that $G(u)$ is a solution of $\omega(u)$ or $\fail$ for every $u\in\cube(\rho')$, and $\Pr_{\rho'}[G(\bm u)\neq\fail]\geq\eta$.
\end{itemize}
\end{claim}

\begin{proof}[Proof of \cref{clm:ueopl-halts}]
The cases $i=N$ and $r=0$ are immediate, so let $i<N$ and $r\geq 1$. By \cref{clm:ueopl-halts,clm:ueopl-output} for level $r-1$,
\begin{equation}
\label{eq:ueopl-C}
\parbox{0.86\linewidth}{the tree $C$ of line~\ref{line:ueopl-C} halts, and it outputs a solution or a tree as in \cref{clm:ueopl-output}, or a node of row $b$ that is active on the cube of the restriction at that time.}
\end{equation}
Consider an iteration of the loop, at a restriction $\rho$, that reaches line~\ref{line:ueopl-t}. Then $C$ outputs a solution or a tree with probability less than $\eta/2$ over $\cube(\rho)$, and by Markov's inequality it makes more than $t$ queries with probability less than $\eta/2$. So, by~\eqref{eq:ueopl-C}, the tree $D$ of line~\ref{line:ueopl-D} has depth at most $t$ and outputs an active node of row $b$ with probability more than $1-\eta$, and \cref{lem:dichotomy} applies on $\cube(\rho)$ to $\act^b$ and $D$. If $\Jump(v,r)$ does not halt in this iteration, then $T_1$ was run and output $\fail$. Since $\Pr_\rho[T_1(\bm u)\neq\fail]\geq 1/2$, the tree $T_1$ cannot output $\fail$ without making a query, so the iteration fixes at least one more bit of the oracle. The instance $\omega(u)$ depends on only finitely many bits of $u$, and no other bit ever needs to be queried, so there are only finitely many iterations.
\end{proof}

\begin{proof}[Proof of \cref{clm:ueopl-output}]
If $i=N$, then the output is the active node $v$ of row $b=N$. If $i<N$ and $r=0$, the query to $\Succ_v$ fixes its successor $v'$ in row $b$ on $\cube(\rho')$. The node $v$ remains active on $\cube(\rho')$, since $\rho'$ extends the starting restriction. If the activity test finds $v'$ inactive, its queried transcript certifies inactivity on $\cube(\rho')$, and the active node $v$ witnesses an incoming edge to $v'$. Thus $v'$ itself is a proper sink, and hence a solution, throughout $\cube(\rho')$. Otherwise the activity test certifies that $v'$ is active on that cube. If $i<N$ and $r\geq 1$, then $\Jump(v,r)$ halts in line~\ref{line:ueopl-expand}, \ref{line:ueopl-T2} or~\ref{line:ueopl-T1}.
\begin{itemize}[noitemsep]
    \item In line~\ref{line:ueopl-expand} it outputs the output of $C$, which is as claimed by~\eqref{eq:ueopl-C}.
    \item In line~\ref{line:ueopl-T2} it outputs $T_2$, which by \cref{lem:dichotomy} outputs a pair of distinct active nodes of row $b$, that is, a solution, with probability at least $\eta$, and $\fail$ otherwise.
    \item In line~\ref{line:ueopl-T1} it outputs a node of row $b$ that $T_1$ certifies by \cref{lem:dichotomy}, hence active on $\cube(\rho')$.\qedhere
\end{itemize}
\end{proof}

\begin{claim}
\label{clm:ueopl-queries}
$\Exp_\rho[\qc(\Jump(v,r))]=O(a(r+1)/\eta)$.
\end{claim}

\begin{proof}
For $k\leq r$, let $\bm N_k$ be the number of times that $\Jump(\cdot,k)$ is run during $\Jump(v,r)$, and let $\bm M_k\leq\bm N_k$ be the number of those times in which $C$ is run in line~\ref{line:ueopl-expand}. Then $\bm N_r=1$, and $\bm N_{k-1}\leq 2\bm M_k$ for $k\in[r]$, because $\Jump(\cdot,k-1)$ is run only inside such a $C$, and at most twice there. By the test in line~\ref{line:ueopl-expand}, each time $C$ is run it outputs a solution or a tree with probability at least $\eta/2$, and then $\Jump(v,r)$ halts without running anything further. Hence $\Exp_\rho[\bm M_k]=O(1/\eta)$ for every $k\in[r]$, and so $\Exp_\rho[\sum_{k\leq r}\bm N_k]=O(1+r/\eta)$.

Each time $\Jump(\cdot,k)$ is run, the queries that it makes outside of line~\ref{line:ueopl-expand} are the $O(q)=O(a)$ queries of line~\ref{line:ueopl-base} and the queries of $T_1$ in line~\ref{line:ueopl-T1}. Each time $T_1$ is run, it makes at most $a$ queries and, with probability at least $1/2$, outputs a node, upon which $\Jump(\cdot,k)$ halts. So these are $O(a)$ queries in expectation, and $\Exp_\rho[\qc(\Jump(v,r))]=O(a)\cdot O(1+r/\eta)=O(a(r+1)/\eta)$.
\end{proof}

\begin{claim}
\label{clm:ueopl-depth}
Every tree that $\Jump(v,r)$ outputs has depth $O(a(r+1)/\eta^2)$.
\end{claim}

\begin{proof}
A tree that $\Jump(v,r)$ outputs is the tree $T_2$ in line~\ref{line:ueopl-T2} of $\Jump(\cdot,k)$ for some $k\in[r]$. Its depth is $O(t+a)$, where $t=O(\Exp_\rho[\qc(C)]/\eta)=O(ak/\eta^2)$ by \cref{clm:ueopl-queries}.
\end{proof}

\begin{proof}[Proof of \cref{thm:ueopl}]
Consider an iteration of the loop in \cref{alg:ueopl-main} that starts from a restriction $\rho$. By Claims~\ref{clm:ueopl-halts}--\ref{clm:ueopl-depth}, $\Jump((1,1),r^*)$ makes $O(ar^*/\eta)$ queries in expectation, and it outputs a solution or a tree $G$ of depth $O(ar^*/\eta^2)$ that outputs a solution with probability at least $\eta$. Hence the iteration makes $O(ar^*/\eta^2)$ queries in expectation and returns with probability at least $\eta$ over $\cube(\rho)$. So whenever \cref{alg:ueopl-main} halts on $u$, its output is a solution of $\omega(u)$, and its expected query complexity is at most
\[
    B=O(q)+\frac{1}{\eta}\cdot O\Big(\frac{ar^*}{\eta^2}\Big)=\poly(q)\cdot\log N.
\]
Let $T$ be \cref{alg:ueopl-main} stopped after $3B$ queries, with output $\fail$ if it has not halted by then. This is a decision tree of depth at most $3B$, because an iteration that does not return makes at least one query, as it returns with positive probability over $\cube(\rho)$. By Markov's inequality, $T$ outputs $\fail$ with probability at most $1/3$.
\end{proof}

\subsection{Proof of \cref{lem:dichotomy}}
\label{sec:dichotomy-proof}

Let $f$ and $D$ be as in \cref{lem:dichotomy}; the parameters $a$ and $\eta$ are chosen in the proof of \cref{lem:dichotomy} below. The trees $f_y$ and $D$ query only finitely many bits, which we rename to $1,\dots,K$. Below $\bm u$ is uniform over $\{0,1\}^K$, all trees query only these bits, and for a set $A\subseteq\{0,1\}^K$ we write $\Pr[A]$ for $\Pr[\bm u\in A]$. Let
\[
    \alpha\coloneqq\Pr\big[f\text{ does not accept }D(\bm u)\text{ on }\bm u\big],
    \qquad
    \epsilon\coloneqq\max_E\,\Pr\big[f\text{ accepts }E(\bm u)\text{ on }\bm u,\text{ and }E(\bm u)\neq D(\bm u)\big],
\]
where the maximum is over all decision trees $E\colon\{0,1\}^K\to Y\cup\{\fail\}$ of depth at most $t$.

Let $\Lambda$ be the set of all accepting leaves of the trees $f_y$ that are reached on some input. For $\tau\in\Lambda$ we write $y_\tau$ for the label of the tree that $\tau$ belongs to, and $B_\tau\subseteq\{0,1\}^K$ for the subcube of inputs on which $f_{y_\tau}$ reaches $\tau$; it fixes at most $q$ bits.

\begin{claim}
\label{clm:ueopl-leaf-partition}
There is a set $\Lambda'\subseteq\Lambda$ such that $m(u)\coloneqq|\{\tau\in\Lambda' : u\in B_\tau\}|$ satisfies
\[
    \Exp|m(\bm u)-1|\leq\alpha+O(q\epsilon).
\]
\end{claim}

\begin{claim}
\label{clm:ueopl-search-partition}
There is an absolute constant $c>0$ such that the following holds. Let $(B_\tau)_{\tau\in\Lambda'}$ be a finite family of subcubes of $\{0,1\}^K$, each fixing at most $q$ bits, and let $m(u)\coloneqq|\{\tau\in\Lambda' : u\in B_\tau\}|$. If $\Exp|m(\bm u)-1|\leq c$, then there is a decision tree of depth $O(q^2)$ whose output on $u$ is either $\fail$ or some $\tau$ with $u\in B_\tau$, and which outputs $\fail$ with probability at most $1/2$.
\end{claim}

We prove \cref{clm:ueopl-leaf-partition} in \cref{sec:ueopl-partition-proof} and \cref{clm:ueopl-search-partition} in \cref{sec:ueopl-search-proof}.

\begin{proof}[Proof of \cref{lem:dichotomy}]
Let $c$ be the constant of \cref{clm:ueopl-search-partition}. We choose $\eta=\Theta(1/q)$ small enough that the bound $\alpha+O(q\epsilon)$ of \cref{clm:ueopl-leaf-partition} is at most $c$ whenever $\alpha\leq\eta$ and $\epsilon\leq 2\eta$, and we let $a=O(q^2)$ be the depth bound of \cref{clm:ueopl-search-partition}, which we may assume to be at least $2q$. By the hypothesis on $D$, we have $\alpha\leq\eta$.

Suppose first that $\epsilon\leq 2\eta$. By \cref{clm:ueopl-leaf-partition}, there is a set $\Lambda'$ of accepting leaves with $\Exp|m(\bm u)-1|\leq c$. Let $T_1$ run the decision tree of \cref{clm:ueopl-search-partition} for this family, and output $y_\tau$ if that tree outputs $\tau$. Then $T_1$ has depth at most $a$ and outputs $\fail$ with probability at most $1/2$. If it outputs $y_\tau$, then it has queried all bits fixed by $B_\tau$ and $u\in B_\tau$, so $T_1$ certifies $y_\tau$ on $u$.

Suppose now that $\epsilon>2\eta$, and let $E$ be a tree that attains the maximum in the definition of $\epsilon$. Let $T_2$ run $D$ and $E$; if $D(u)$ and $E(u)$ are distinct labels in $Y$, it then runs $f_{D(u)}$ and $f_{E(u)}$ and outputs the pair $\{D(u),E(u)\}$ if both trees accept; in every other case it outputs $\fail$. Then $T_2$ certifies both labels whenever it does not output $\fail$, it has depth at most $2t+2q=O(t+a)$, and
\[
    \Pr[T_2(\bm u)\neq\fail]\geq\epsilon-\alpha>\eta. \qedhere
\]
\end{proof}

\subsection{Proof of \cref{clm:ueopl-leaf-partition}}
\label{sec:ueopl-partition-proof}

For $\tau\in\Lambda$, let
\[
    A_\tau\coloneqq\{u\in B_\tau : D(u)=y_\tau\},
    \qquad
    \theta_\tau\coloneqq\Pr[A_\tau]/\Pr[B_\tau],
    \qquad
    \Gamma\coloneqq\sum_{\tau\in\Lambda}\Pr[A_\tau](1-\theta_\tau).
\]
The sets $A_\tau$ are pairwise disjoint, since $u\in A_\tau$ means that $\tau$ is the leaf that the tree $f_{D(u)}$ reaches on $u$. Their union is the event that $f$ accepts $D(\bm u)$, and hence $\sum_\tau\Pr[A_\tau]=1-\alpha$.

\begin{definition}
\label{def:ueopl-noisy-copy}
Let $\beta\in[0,1]$ and $u\in\{0,1\}^K$. A \emph{$\beta$-noisy copy} of $u$ is the random string $\bm x$ obtained from $u$ by independently replacing each bit, with probability $\beta$, by a fresh uniform bit.
\end{definition}

\begin{claim}
\label{clm:ueopl-noise}
Let $\bm x$ be a $\beta$-noisy copy of $\bm u$. Then for every $A\subseteq\{0,1\}^K$ with $p\coloneqq\Pr[A]$,
\[
    \Pr[\bm u\in A,\ \bm x\notin A]\geq\beta p(1-p).
\]
\end{claim}

\begin{proof}
Let $g$ be the indicator of $A$, with Fourier coefficients $\hat g(J)$ for $J\subseteq[K]$. Then $\hat g(\varnothing)=p$ and $\sum_J\hat g(J)^2=p$, so
\[
    \Pr[\bm u\in A,\ \bm x\in A]
    =\sum_{J}(1-\beta)^{|J|}\hat g(J)^2
    \leq p^2+(1-\beta)(p-p^2).
\]
Subtracting this from $p$ gives the claim.
\end{proof}

\begin{claim}
\label{clm:ueopl-gamma}
$\Gamma=O(q\epsilon)$.
\end{claim}

\begin{proof}
Let $\bm x$ be a $\beta$-noisy copy of $\bm u$ for $\beta\coloneqq 1/(2q)$, and run $D$ on $\bm x$. The pairs $(\bm u,\bm x)$ and $(\bm x,\bm u)$ have the same distribution. Fix $\tau\in\Lambda$, and let $H_\tau$ be the event that the noise replaces none of the at most $q$ bits fixed by $B_\tau$, so that $\Pr[H_\tau]\geq 1-q\beta=1/2$. Conditioned on $H_\tau$ and on $\bm u\in B_\tau$, we have $\bm x\in B_\tau$, and on the bits not fixed by $B_\tau$ the pair $(\bm u,\bm x)$ is distributed as before. Applying \cref{clm:ueopl-noise} inside the subcube $B_\tau$ to the set $A_\tau$, with the roles of $\bm u$ and $\bm x$ exchanged, we get
\[
    \Pr[\bm x\in A_\tau,\ \bm u\in B_\tau\setminus A_\tau]
    \geq\Pr[B_\tau]\cdot\Pr[H_\tau]\cdot\beta\theta_\tau(1-\theta_\tau)
    \geq\frac{\beta}{2}\Pr[A_\tau](1-\theta_\tau).
\]
The events on the left-hand side are disjoint for different $\tau$, because the sets $A_\tau$ are. On each of them, the label $D(\bm x)=y_\tau$ is accepted on $\bm u$ and differs from $D(\bm u)$. Therefore
\[
    \Pr\big[f\text{ accepts }D(\bm x)\text{ on }\bm u,\text{ and }D(\bm x)\neq D(\bm u)\big]\geq\frac{\beta\Gamma}{2}.
\]
Conditioned on which bits are replaced and on their new values, $D(\bm x)$ is computed from $\bm u$ by a decision tree of depth at most $t$, so this probability is at most $\epsilon$. Hence $\Gamma\leq 2\epsilon/\beta=O(q\epsilon)$.
\end{proof}

\begin{proof}[Proof of \cref{clm:ueopl-leaf-partition}]
Let $\Lambda'\coloneqq\{\tau\in\Lambda : \theta_\tau\geq 1/2\}$. If $m(u)=0$, then either $u$ lies in no set $A_\tau$, or it lies in $A_\tau$ for some $\tau\notin\Lambda'$. Since $1\leq 2(1-\theta_\tau)$ for $\tau\notin\Lambda'$,
\[
    \Pr[m(\bm u)=0]\leq\alpha+\sum_{\tau\notin\Lambda'}\Pr[A_\tau]
    \leq\alpha+2\sum_{\tau\notin\Lambda'}\Pr[A_\tau](1-\theta_\tau).
\]
Since every $u$ lies in at most one of the sets $A_\tau$, we have $\max\{m(u)-1,0\}\leq|\{\tau\in\Lambda' : u\in B_\tau\setminus A_\tau\}|$. Hence, using $\Pr[B_\tau]=\Pr[A_\tau]/\theta_\tau\leq 2\Pr[A_\tau]$ for $\tau\in\Lambda'$,
\[
    \Exp\max\{m(\bm u)-1,0\}
    \leq\sum_{\tau\in\Lambda'}\Pr[B_\tau](1-\theta_\tau)
    \leq 2\sum_{\tau\in\Lambda'}\Pr[A_\tau](1-\theta_\tau).
\]
Adding the two bounds gives $\Exp|m(\bm u)-1|\leq\alpha+2\Gamma$, which is $\alpha+O(q\epsilon)$ by \cref{clm:ueopl-gamma}.
\end{proof}

\subsection{Proof of \cref{clm:ueopl-search-partition}}
\label{sec:ueopl-search-proof}

We prove \cref{clm:ueopl-search-partition} with $c=1/64$. Let
\[
    W(u)\coloneqq\mathbf{1}[m(u)=0]+m(u)\cdot\mathbf{1}[m(u)\geq 2].
\]
Since $W\leq 2|m-1|$ pointwise, we have $\Exp [W(\bm u)]\leq 1/32$. The decision tree is given in \cref{alg:ueopl-search}.

\begin{algorithm}[H]
\renewcommand{\algorithmicrequire}{\textbf{Input:}}
\renewcommand{\algorithmicensure}{\textbf{Output:}}
\algtext*{EndFor}
\caption{The decision tree of \cref{clm:ueopl-search-partition}.}
\label{alg:ueopl-search}
\begin{algorithmic}[1]
\Require $u\in\{0,1\}^K$
\Ensure $\tau\in\Lambda'$ with $u\in B_\tau$, or $\fail$
\State Initialize $\sigma$ as the empty restriction.
\For{$k\in[8(q+1)]$}
    \State \textbf{if} $\Exp_\sigma [W(\bm u)]>1/8$ \textbf{then return} $\fail$. \label{line:ueopl-abort}
    \State Set $\tau^*\coloneqq\arg\min_\tau\Pr_\sigma[m(\bm u)\geq 2\mid\bm u\in B_\tau]$, where $\tau$ ranges over $\Lambda'$ with $\Pr_\sigma[B_\tau]>0$. \label{line:ueopl-choice}
    \State Query all bits that are fixed by $B_{\tau^*}$ but not by $\sigma$, and record the answers in $\sigma$.
    \State \textbf{if} $u\in B_{\tau^*}$ \textbf{then return} $\tau^*$.
\EndFor
\State \Return $\fail$. \label{line:ueopl-timeout}
\end{algorithmic}
\end{algorithm}

\cref{alg:ueopl-search} makes at most $8(q+1)\cdot q=O(q^2)$ queries, and if it outputs $\tau$ then $u\in B_\tau$. There are two cases where it outputs $\fail$:
\begin{enumerate}[noitemsep]
    \item \cref{alg:ueopl-search} halts in line~\ref{line:ueopl-abort}.
    \item \cref{alg:ueopl-search} halts in line~\ref{line:ueopl-timeout}.
\end{enumerate}
We denote these two events by $\mathcal{E}_1$ and $\mathcal{E}_2$ respectively.

\begin{claim}
\label{clm:ueopl-E1}
$\Pr[\mathcal{E}_1]\leq 1/4$.
\end{claim}

\begin{proof}
The restrictions $\sigma$ at which \cref{alg:ueopl-search} halts in line~\ref{line:ueopl-abort} have disjoint cubes and satisfy $\Exp_\sigma W>1/8$. As $W\geq 0$,
\[
    \frac{1}{32}\geq\Exp W(\bm u)\geq\sum_{\sigma}\Pr[\cube(\sigma)]\cdot\Exp\nolimits_\sigma W\geq\frac{1}{8}\Pr[\mathcal{E}_1],
\]
where the sum is over these restrictions.
\end{proof}

\begin{claim}
\label{clm:ueopl-overlap}
Suppose that an iteration of \cref{alg:ueopl-search} reaches line~\ref{line:ueopl-choice} at a restriction $\sigma$. Then $\tau^*$ exists, and $B^*\coloneqq B_{\tau^*}$ satisfies $\Pr_\sigma[m\geq 2\mid B^*]\leq 1/7$.
\end{claim}

\begin{proof}
Let $e\coloneqq\Exp_\sigma W\leq 1/8$. Since $W\geq\mathbf{1}[m\neq 1]$, we have $\Pr_\sigma[m=1]\geq 1-e$. Hence
\[
    \sum_\tau\Pr\nolimits_\sigma[B_\tau]=\Exp\nolimits_\sigma[m]\geq 1-e
    \qquad\text{and}\qquad
    \sum_\tau\Pr\nolimits_\sigma[B_\tau]\cdot\Pr\nolimits_\sigma[m\geq 2\mid B_\tau]=\Exp\nolimits_\sigma\big[m\cdot\mathbf{1}[m\geq 2]\big]\leq e,
\]
where the second sum is over $\tau$ with $\Pr_\sigma[B_\tau]>0$. So $\tau^*$ exists, and $\Pr_\sigma[m\geq 2\mid B^*]\leq e/(1-e)\leq 1/7$.
\end{proof}

For $u$ with $m(u)=1$, let $B(u)$ be the unique member of the family that contains $u$. We say that an iteration of \cref{alg:ueopl-search} \emph{makes progress on $u$} if $m(u)=1$ and it either outputs $\tau^*$ or queries a bit that is fixed by $B(u)$. As $B(u)$ fixes at most $q$ bits, at most $q+1$ iterations make progress on any given $u$.

\begin{claim}
\label{clm:ueopl-progress}
An iteration of \cref{alg:ueopl-search} that reaches line~\ref{line:ueopl-choice} at a restriction $\sigma$ makes progress with probability at least $1/2$ over $\bm u\sim\cube(\sigma)$.
\end{claim}

\begin{proof}
Let $\mathcal{N}$ be the set of $u\in\cube(\sigma)$ such that $m(u)=1$, the member $B(u)$ is not $B^*$, and no bit is fixed to different values by $B(u)$ and $B^*$. For $u\in\cube(\sigma)$, let $x$ be obtained from $u$ by overwriting the bits fixed by $B^*$ with the values that $B^*$ prescribes. If $\bm u\sim\cube(\sigma)$, then $\bm x$ is uniform over $\cube(\sigma)\cap B^*$. If $u\in\mathcal{N}$, then $x$ lies in both $B^*$ and $B(u)$, and so $m(x)\geq 2$. By \cref{clm:ueopl-overlap},
\[
    \Pr\nolimits_\sigma[\mathcal{N}]\leq\Pr\nolimits_\sigma[m(\bm x)\geq 2]=\Pr\nolimits_\sigma[m\geq 2\mid B^*]\leq\frac{1}{7}.
\]
Now let $u\in\cube(\sigma)\setminus\mathcal{N}$ with $m(u)=1$. If $B(u)=B^*$, then the iteration outputs $\tau^*$. Otherwise some bit is fixed to different values by $B(u)$ and $B^*$. This bit is not fixed by $\sigma$, because both $B(u)$ and $B^*$ intersect $\cube(\sigma)$, so the iteration queries it. Since $\Exp_\sigma W\leq 1/8$ and $W\geq\mathbf{1}[m\neq 1]$, the iteration makes progress with probability at least $\Pr_\sigma[m=1]-\Pr_\sigma[\mathcal{N}]\geq 7/8-1/7>1/2$.
\end{proof}

\begin{claim}
\label{clm:ueopl-E2}
$\Pr[\mathcal{E}_2]\leq 1/4$.
\end{claim}

\begin{proof}
For $k\leq 8(q+1)$, let $\mathcal{R}_k$ be the event that the $k$-th iteration reaches line~\ref{line:ueopl-choice}, and let $\mathcal{P}_k\subseteq\mathcal{R}_k$ be the event that it makes progress. Then $\Pr[\mathcal{P}_k]\geq\Pr[\mathcal{R}_k]/2$ by \cref{clm:ueopl-progress}, and $\sum_k\Pr[\mathcal{P}_k]\leq q+1$. If $\mathcal{E}_2$ occurs, then all the events $\mathcal{R}_k$ occur. Hence
\[8(q+1)\cdot\Pr[\mathcal{E}_2]\leq\sum_k\Pr[\mathcal{R}_k]\leq 2\sum_k\Pr[\mathcal{P}_k]\leq 2(q+1). \qedhere
\]
\end{proof}

Given \cref{clm:ueopl-E1,clm:ueopl-E2}, \cref{alg:ueopl-search} outputs $\fail$ with probability at most $1/2$, which proves \cref{clm:ueopl-search-partition}.

\section{$\PLS$ separations from proof complexity}
\label{sec:PLS_separations}

In this section we prove separations between $\TFNP$ classes in the random oracle model. We will work with the following notion of a ROM reduction. 

\paragraph{Public reduction.} Let $F=\{F_n\}, R=\{R_m\}$ be $\TFNP$ problems, where 
$F_n \subseteq \Sigma^{n} \times {\cal O}$ and $R_m \subseteq \Sigma^{m} \times {\cal O}'$. %
A \emph{public (ROM) reduction} from $F_n$ to $R_m$ is given by a family of decision trees $(\omega, f,g)$ satisfying the following properties.
    \begin{enumerate}[noitemsep]
        \item The first family of decision trees, $\omega = \{\omega_1,\ldots,\omega_n\}$ query variables $u_1, \ldots ,u_t$, and together they map a random string $\bm u \in \Sigma^{t}$ to an instance of $F_n$. That is,  for each random string $u$, $\omega(\bm u) = \omega_1(\bm u) \ldots \omega_n(\bm u) \in \Sigma^n$ defines an input to  $F_n$, and thus $\omega$ defines a ``locally-samplable'' distribution over instances of $F_n$.  
    \item The second family of decision trees, $f = \{f_1,\ldots,f_m\}$ query $\bm u$ and together they define a reduction from an instance of $F_n$ to an instance of $R_m$.
   \item The third family of decision trees $g= \{g_o \mid o \in \mathcal O'\}$ queries $\bm u$, and if $o$ is a legal solution to the instance $f(\bm u)$ of $R_m$, the decision tree $g_o(\bm u)$ outputs a candidate solution to $F_n$ on instance $\omega(\bm u)$.
   \end{enumerate}
    The \emph{depth} $d(\omega, f,g)$ of a reduction $(f,g)$ is the maximum depth of any of the decision trees, and the \emph{complexity} is $c(\omega,f,g) \coloneqq d(\omega, f,g) + \log m$.
    The \emph{error} of the reduction is 
    \[ \err(\omega,f,g) \coloneq \Pr_{{\bm u} \sim \Sigma^t} \Big[ \exists o \in {\cal O}': (f({\bm u}),o) \in R_m \land (\omega({\bm u}),g_o({\bm u})) \not \in F_n \Big].\]

    We say that $F$ \emph{public reduces} to $R$ if for every sufficiently large $n$ there is a $\polylog(n)$-complexity reduction for \emph{every} locally-samplable distribution. That is, for every $\polylog(n)$-depth $\omega$ there exists $f,g$ such that $c(\omega,f,g)=\polylog(n)$ and $\err(\omega,f,g) \leq 1/3$.

    Note that up to a blow-up of $d(\omega,f,g)$  this is equivalent to allowing $f$ and $g$ to depend on $\omega(\bm u)$ as well, since each query to a bit of $\omega({\bm u})$ can be simulated by querying the values of ${\bm u}$ that $\omega$ queries for the value of that bit. As well, note that if $\err(\omega,f,g)=0$, then we recover the standard definition of a black-box reduction between $\TFNP$ problems.

\subsection{Separations in the ROM from average-case proof complexity bounds}

        Our $\TFNP$ separations will proceed by a reduction to proof complexity lower bounds. Proof complexity and $\TFNP$ are intimately connected in the black-box model. Characterization theorems show that black-box $\TFNP$ subclasses correspond exactly to the search problems whose totality can be proven efficiently in a corresponding proof system \cite{GKRS19, BussFI23}. For example, recalling \Cref{def:res} of a resolution refutation, the following result shows that resolution is exactly the search problems whose totality can be proven by $\PLS$-style arguments. 
        
        \paragraph{Resolution = $\PLS$ (\cite{BKT14,kamath2019some}).}
For any $F \in \TFNP^{\rm dt}$, $F \in \PLS^{\rm dt}$ iff there is a width-$\poly(\log n)$ pair $(\mathcal{F},S)$ that solves $F$. Quantitatively, a zero-error reduction of complexity $t$ to the canonical $\PLS$ problem yields a resolution refutation of width $O(t+v)$, where $v$ bounds the source-verifier depth; see \cite[Theorem~8.18(1)]{kamath2019some}. We use the same characterization for alphabet-valued decision trees, replacing Boolean literals by cell equalities $[u_i=a]$.\\

        In this section, we extend this connection to ROM reductions. Let $(\omega,f,g)$ be a public reduction from a $\TFNP$ problem $F \subseteq \Sigma^{n} \times \mathcal O$ to a $\TFNP$ problem $R \subseteq \Sigma^{m} \times \mathcal O'$. All our lower bounds will be against the uniform distribution $\omega(u) = u$, and henceforth we restrict attention to this case and omit $\omega$. The main difference between standard black-box reductions and ROM reductions is that the latter allows for error. We will capture this error by a low-width DNF $B$ whose probability of being satisfied over the uniform distribution is $\err(\omega,f,g)$. A width-$w$ \emph{DNF} over an alphabet $\Sigma$ is a disjunction of predicates $B= \bigvee_{i \in [t]} B_i$ such that each $B_i$ is a term depending on at most $w$ variables -- there is a single assignment to the variables of $B_i$ satisfying it. 
        This will give us a new search problem which we denote by $F \lor B$: on a given input, find either a solution to $F$ or a satisfied term of $B$.

We now define $B$. For each target answer $o\in\mathcal O'$, let $H_o$ be the decision tree that first verifies $(f(u),o)\in R$, then runs the decoder $g_o$, and finally checks that $(u,g_o(u))\notin F$. Let $\mathrm{viol}(f,g)$ be the set of accepting root-to-leaf transcripts of these trees, viewed as partial assignments to $u$. Define the violation DNF for the reduction as
\[ B_{f,g} \coloneq \bigvee_{\rho \in \mathrm{viol}(f,g)}\ \bigwedge_{i\in\fix(\rho)}[u_i=\rho_i]. \]
Thus $B_{f,g}(u)=1$ exactly when some valid target answer decodes to an invalid source answer. If the reduction has depth $d$ and the target and source verifiers have depths $v_R$ and $v_F$, respectively, each term has width at most $dv_R+d+v_F$. $B_{f,g}$ captures the allowable error of the public reduction, and will be given to the proof system as ``advice'' to help prove $F$.
The key property of $B_{f,g}$ imposed by the reduction is that 
\[ \Pr_{\bm u \sim \Sigma^n} \big[B_{f,g}(\bm u) =1  \big] = \err(\omega,f,g). \]
Therefore, if the reduction has low error, then $B_{f,g}$ is almost never satisfied and intuitively cannot be too helpful in proving $F$. Abusing notation, let $F \lor B \subseteq \Sigma^n \times (\mathcal O \cup \{\rho \mid \rho \in B\})$ be the search problem of returning a solution to $F$ or a satisfied term of $B$.

\begin{lemma}
\label{lem:ROM_to_proof_complexity}
    Let $F\subseteq\Sigma^n\times\mathcal O$ have solution verifiers of depth at most $w$. Any public reduction $(\mathrm{id},f,g)$ from $F$ to the canonical $\PLS$ problem of complexity at most $w$ yields a DNF $B$ of width $O(w)$ such that
    \[
        \Pr_{\bm u\sim\Sigma^n}[B(\bm u)=1]=\err(\mathrm{id},f,g),
    \]
    and $F\lor B$ has a resolution refutation of width $O(w)$. More precisely, if the reduction has depth $d$ and the source verifiers have depth $v_F$, the DNF $B$ has width at most $3d+v_F$.
\end{lemma}
In particular, if $F\in\TFNP^{\rm dt}$ and $F\lor B$ has no $\polylog n$-width resolution refutation for every width-$\polylog n$ DNF $B$ satisfying $\Pr[B(\bm u)=1]\leq1/3$, then there is no public reduction from $F$ to $\PLS$. This is a lower bound on the randomized resolution width of $F$ from \cite{BKT14} specialized to the uniform distribution.
\begin{proof}
    For each target answer $o \in \mathcal{O}'$, the tree $H_o$ defined above accepts exactly when $o$ is valid for $f(u)$ but $g_o(u)$ is invalid for $F$. Its target-verification stage uses at most $2d$ source queries, its decoder uses at most $d$, and its source verifier uses at most $v_F$. Taking the disjunction of all accepting transcripts gives $B=B_{f,g}$ of width at most $3d+v_F\leq4w$, with satisfaction probability exactly $\err(\mathrm{id},f,g)$.

    Modify the decoder to run the three stages of $H_o$. If the target check fails, return an arbitrary value. Otherwise, return $g_o(u)$ when the source verifier accepts, or the accepting transcript of $H_o$, a satisfied term of $B$, when it rejects. This is a zero-error reduction from $F\lor B$ to the same $\PLS$ instance, of complexity $O(w)$. Solutions to $F\lor B$ are verified in depth $O(w)$, so the quantitative $\PLS$--resolution characterization \cite[Theorem~8.18(1)]{kamath2019some} supplies the width-$O(w)$ refutation.
\end{proof}

    \subsection{Robust strategies and average-case resolution}

In this section, we develop tools for proving average case resolution lower bounds -- that is, $F \lor B$ for every $B$ satisfying the conditions of \Cref{lem:ROM_to_proof_complexity}. We begin by describing the dual object (known as a \emph{Delayer strategy}) for resolution proofs, and later extend it to the average case. For a partial assignment $\rho \in (\Sigma \cup \{\star\})^m$, denote by $\fix(\rho) \coloneq \{ i \in [m] \mid \rho_i \neq \star\}$, and let $\free(\rho) \coloneq [m] \setminus \fix(\rho)$. As well, let $\cube(\rho) \coloneq \{x \in \Sigma^m \mid \forall i \not \in \free(\rho), x_i = \rho_i \}$ be the sub-cube of total assignments which agree with $\rho$.

\paragraph{Strategy.}

    Let $F \subseteq \Sigma^m \times \mathcal O$. A width-$w$ \emph{strategy} for $F$ is a non-empty downward-closed set $\mathcal F \subseteq (\Sigma \cup \{\star \})^m $ such that for every $\rho \in \mathcal F$ with $|\fix(\rho)|\leq w$ the following hold:
    \begin{itemize}[noitemsep]
        \item For every $o \in \mathcal O$, there is an $x \in \cube(\rho)$ such that $(x,o) \not \in F$
        \item \emph{Extension Property:} If $|\fix(\rho)| < w$ and $i \in \free(\rho)$ there is $a \in \Sigma$ such that $\rho \cup \{x_i = a\} \in \mathcal{F}$.
    \end{itemize}

This is the dual object for resolution proofs: 
\begin{lemma}[\cite{AtseriasD08}] There is a width-$w$ resolution refutation of $F$ iff there is no width-$w$ strategy.
\end{lemma}
Hence, to separate $F$ from $\PLS$ it suffices to show that there is a $\polylog n$-width strategy for $F \lor B$ for all suitable $B$. We show that this is implied by constructing a suitably \emph{robust} strategy for $F$. 

\paragraph{Robust strategy.} A width-$w$ strategy for $F$ is $k$-robust if it also satisfies:
\begin{itemize}
    \item \emph{Robust Extension Property:} For every $\rho \in \mathcal{F}$ with $|\fix(\rho)|<w$ and every $i \in [m] \setminus \fix(\rho)$ there exist at least $|\Sigma|-k$ values $a \in \Sigma$ such that $\rho \cup \{x_i = a\} \in \mathcal{F}$.
\end{itemize}

    \begin{lemma}
    \label{lem:robust-delayer}
        Suppose that $F \subseteq \Sigma^m \times \mathcal O$ has a width-$w$ $k$-robust strategy. Then for every width-$w$ DNF $B$ such that $\Pr_{\bm u \sim \Sigma^m}[B(\bm u) = 1] < (1-k/|\Sigma|)^w$ there is an (ordinary) strategy of width $w$ for $F \lor B$.
    \end{lemma}
    
    The remainder of this section is devoted to proving this lemma. Let us identify $\Sigma = [n]$ for simplicity.
   We aim to extract an ordinary strategy for $F \lor B$ from a $k$-robust strategy for $F$. The naive approach would be to take the $k$-robust strategy $\mathcal{F}$ for $F$ and keep only $\rho \in \mathcal{F}$ such that 
    \begin{equation}
    \label{eq:naive-restriction}
        \cube(\rho) \not\subseteq B^{-1}(1) .
    \end{equation}
    Then no $\rho \in \mathcal{F}$ that meets \eqref{eq:naive-restriction} can falsify any clause of $F$ or $B$. The issue with such a naive restriction is with the \extensionProp .
    Let $x_1,\dots,x_m$ be the variables of $F$. 
    Suppose $\rho$ satisfies \eqref{eq:naive-restriction} and $n-k$ extensions of the form $\rho \cup \{x_i = j\}$ are in $\mathcal{F}$. If we show that at least one of the extensions also satisfies \eqref{eq:naive-restriction}, then this proves the \extensionProp. If $k=0$, this would be trivial:
    \begin{equation} 
    \label{eq:total-probability}
    \cube(\rho) \setminus B^{-1}(1) = \bigsqcup_{j \in [n]} (\cube(\rho \cup \{x_i = j\}) \setminus B^{-1}(1)),
    \end{equation}
    so if the left-hand side is non-empty, then at least one of the terms on the right-hand is non-empty. Hence, $\rho$ satisfies the \extensionProp . This becomes false even for $k=1$, as the difference on the left-hand side may be completely contained in one of the parts of the right-hand side. 

    We resolve this by using the linear algebraic embedding of Razborov~\cite{RazborovWFPHP} developed for the pseudo-width method. The high-level idea is to forget about which holes are occupied by pigeons under $\rho$ and instead only track the number of holes that are occupied. To do so, we associate each pigeon $i \in [m]$ with a vector space $V_i = \mathbb{R}^{n-k}$ and choose vectors $v_{i,1}, \dots, v_{i,n}$ such that every $n-k$ of them span the space---this will have the effect of only counting the number of occupied holes. Random vectors or Vandermonde vectors have this property. Then we set $V \coloneqq \bigotimes_{i \in [m]} V_i$ and define $L\colon [n]^m \to V$ as mapping $x \in \Sigma^m$ to $\otimes_{i \in [m]} v_{i,x_i}$. For a set $A \subseteq [n]^m$ extend $L(A) \coloneqq \spn\{L(x) \mid x \in A\}$ and shorthand $L(\rho) \coloneqq L(\cube(\rho))$. We then claim that 
    \[ \mathcal{F'} \coloneqq \{\rho \in \mathcal{F} \mid L(\rho) \not\subseteq L(B^{-1}(1))\}\]
    is a width-$w$ strategy for $F \lor B$. Indeed, the key property of this embedding is that now if we want to extend $\rho$ by mapping $x_i$ then all we need to know is that there are $n-k$-many values in $\Sigma$ where $x_i$ can be mapped to. This is formalized in the next lemma as  the following ``saturated counting'' version of \eqref{eq:total-probability}, stating that \emph{any} $n-k$ extensions of $\rho$ are sufficient to span $L(\rho)$.
    \begin{lemma}
    \label{lem:sat-total-prob-law}
    For every $M \in \binom{\Sigma}{n-k}$ and $i \not\in \fix(\rho)$ we have $L(\rho) = \sum_{j \in M} L(\rho \cup \{x_i = j\})$.
    \end{lemma}
    \begin{proof}
    By definition we have:
    \begin{align*} L(\rho) &= \bigotimes_{t \in \fix(\rho)} \spn(v_{t,\rho(t)}) \otimes V_i \otimes \bigotimes_{\ell \in [m] \setminus(\fix(\rho)\cup \{i\})} V_{\ell}\\& = \bigotimes_{t \in \fix(\rho)} \spn(v_{t,\rho(t)}) \otimes \spn\{v_{i,j} \mid j \in M\} \otimes \bigotimes_{\ell \in [m] \setminus(\fix(\rho)\cup \{i\})} V_{\ell}.  \end{align*}
    Then by the property of tensor products we can rewrite
    \[ L(\rho) = \sum_{j \in M} \bigotimes_{t \in \fix(\rho)} \spn(v_{t,\rho(t)}) \otimes \spn(v_{i,j}) \otimes \bigotimes_{\ell \in [m] \setminus(\fix(\rho)\cup \{i\})} V_{\ell} = \sum_{j \in M} L(\rho \cup \{ x_i = j\}).\qedhere\]
    \end{proof}

    It follows that $\mathcal{F}'$ satisfies the \extensionProp : suppose that $\rho \in \mathcal{F}'$, so  $L(\rho) \not\subseteq L(B^{-1}(1))$, $i \in [m]\setminus\fix(\rho)$. Since $\rho \in \mathcal{F}$ by the $k$-robustness of $\mathcal{F}$ we have that the size of $M = \{j \in [n] \mid \rho \cup \{x_i = j\} \in \mathcal{F}\}$ is at least $n-k$. If for all $j \in M$ we have $L(\rho \cup \{x_i = j\}) \subseteq L(B^{-1}(1))$, then by \cref{lem:sat-total-prob-law} $ \sum_{j \in M} L(\rho \cup \{x_i = j\}) = L(\rho) $, and since $L(B^{-1}(1))$ is a subspace, we have $L(\rho) \subseteq L(B^{-1}(1))$, a contradiction.

    The remaining issue is the non-triviality of $\mathcal{F}'$; indeed, the \extensionProp is satisfied by the empty strategy. Thus, we need to show that $\star^m \in \mathcal{F}'$, equivalently $V = L(\star^m) \not\subseteq L(B^{-1}(1))$. We show that for simple small sets $A$, $\dim(L(A))$ is also small:

    \begin{lemma}
    \label{cl:correctness-of-razborov-emb}
        For every $w$-DNF $B$ we have
        \[ \frac{\dim(L(B^{-1}(1)))}{\dim(V)} \le (1-k/n)^{-w} \cdot \Pr_{\bm u \sim [n]^m}[B(\bm u) = 1]. \]
    \end{lemma}

    Given \cref{cl:correctness-of-razborov-emb}, we can conclude the proof of \cref{lem:robust-delayer}: $\Pr[B(\bm u) = 1] < (1-k/n)^w$, so $\dim(L(B^{-1}(1))) < \dim(V)$, and $\mathcal{F}' \ni \star^m$. 

    \begin{proof}[Proof of \Cref{cl:correctness-of-razborov-emb}]
        Let $B_1,\ldots, B_h$ be the terms of $B$ and let $B_{\leq i} = \bigvee_{j \leq i } B_j$. By induction on $i$, we will bound how much the dimension can increase from $B_{\leq i-1}$ to $B_{\leq i}$. Denote $Z_i \coloneqq B_i^{-1}(1)$ and $Z_{\le i} \coloneqq \bigcup_{j\le i} Z_j$, $Z \coloneqq B^{-1}(1)$.  We will prove that the following holds:
\begin{align}
    \dim(L(Z_{\leq i}))-\dim(L(Z_{< i})) \leq \left( \frac{n-k}{n}\right)^{m-w} \Big|Z_i \setminus Z_{<i}\Big|. \label{eq:main}
\end{align}

We postpone the proof of this inequality and first complete the proof of the lemma. 
Summing over all $i$, 
\[
\dim(L(Z)) \leq \Big(\frac{n-k}{n}\Big)^{m-w} \Big(\sum_{i} |Z_i\setminus Z_{<i}|\Big) 
= \Big(\frac{n-k}{n}\Big)^{m-w} |Z|.\]
Hence, 
\[\frac{\dim(L(Z))}{\dim(V)} \leq \frac{((n-k)/n)^{m-w}}{(n-k)^m}|Z| =  \Big(\frac{n-k}{n} \Big)^{-w} |Z| n^{-m} = (1-k/n)^{-w} \cdot \Pr_{\bm u \sim [n]^m}[B(\bm u) = 1]. \]

To prove \autoref{eq:main} we will use the following claim which bounds the dimension of a set as a function of its size. 

\begin{claim} \label{clm:sizeToDimension} For any partial assignment  $\rho\colon[m] \rightarrow [n] \cup \{\star\}$ and any $A \subseteq \cube(\rho)$, \[\dim(L(A)) \geq \Big(\frac{n-k}{n} \Big)^{m-|\fix(\rho)|}|A|.\]
    In particular, $\dim(L(Z_i)) \geq ((n-k)/n)^{m-w}$. 
    \end{claim}

    \begin{proof}
    First, we prove that for any $A \subseteq [n]^m$, $\dim(L(A)) \geq |A| ((n-k)/n)^m$. For each $i \in [m]$ choose $\bm B_i \sim \binom{[n]}{n-k}$ and note that by our embedding, the vectors in $L(\bm B_i)$ are all linearly independent. Let $\bm B=\prod_{i \in [m]} \bm B_i$. Then, 
    \[ \dim(L(A \cap \bm B)) = |A \cap \bm B|,\]
by linear independence. That is, $\bm B_i$ is restricting the $i$-th variable to have a value only in $\bm B_i$. Hence
\[ \dim( L(A)) \geq \Exp[\dim(L(A \cap \bm B))] = \Exp [|A \cap \bm B|] = |A| \Big(\frac{n-k}{n}\Big)^m,\]
where the final equality is the probability that each element in $\bm B$ is in $A$. By the same argument, for any partial restriction $\rho$ and $A \subseteq \cube(\rho)$, \[\dim(L(A)) \geq \Big(\frac{n-k}{n}\Big)^{m-|\fix(\rho)|}|A|. \qedhere\]
\end{proof} 

We now prove \autoref{eq:main}. First note that without loss of generality we can assume every term in $B$ has width exactly $w$: a term $B_i$ of width $t$ can be transformed into an equivalent DNF of width $t+1$ by taking a variable $x_j$ not mentioned in $B_i$ and defining $\bigvee_{a \in [n] }B_i \wedge [\![x_j = a]\!]$. With this simplification,
\begin{align*}
    \left( \frac{n-k}{n}\right)^{m-w} |Z_i \setminus Z_{<i} | &= \left( \frac{n-k}{n}\right)^{m-w} (|Z_i| - |Z_i \cap Z_{<i} |) \\ 
    &=\left( \frac{n-k}{n}\right)^{m-w} (n^{m-w} - |Z_i \cap Z_{<i} |)\\
    &\geq (n-k)^{m-w} -  \dim(L(Z_i \cap Z_{<i})) \tag{Applying \autoref{clm:sizeToDimension} }\\
    &=\dim(L(Z_i)) - \dim(L(Z_i \cap Z_{<i})) \\
    &\geq \dim(L(Z_{\le i})) - \dim(L(Z_{<i})),
\end{align*}
where the final inequality follows by applying the next \cref{lem:tensor-dim} with $B = Z_i \cap Z_{<i}$, $A =  Z_i \setminus Z_{<i}$, and $C=  Z_{<i} \setminus Z_i$. 
\end{proof}

\begin{lemma}
\label{lem:tensor-dim}
    For any disjoint sets $A,B,C$,
    \[\dim(L(A \cup B)) -\dim(L(B)) \geq \dim(L(A \cup B \cup C)) - \dim(L(B \cup C)).\]
\end{lemma}
\begin{proof}
Let $A,B,C $ be disjoint sets. Then,
    \begin{align*}
        \dim(L(A \cup B)) - \dim(L(B)) &= \dim(L(A)) - \dim(L(A) \cap L(B)) \\ &\geq \dim(L(A))- \dim(L(A) \cap L(B \cup C))\\
        &=\dim(L(A \cup B \cup C)) - \dim(L(B \cup C)),
    \end{align*}
    where the final line follows by adding $B \cup C$ to both sides. 
\end{proof}

\subsection{$\PWPP$ is not in $\PLS$}\label{subsec:pwpp-notin-pls}

In this section we show that $\PWPP$ is not contained within $\PLS$ under public reductions.

    \begin{theorem}\label{thm:PWPP_not_in_PLS}
        Let $(\omega,f,g)$, with $\omega(\bm u)=\bm u$, be any public reduction of complexity $w \le n^{1/3}$ from $\Pigeon$ on $[n]^{2n}$ to the canonical $\PLS$-complete problem of \cref{glossary}. Then, $\err(\omega,f,g) =1-n^{-\Omega(1)}$.
    \end{theorem}

    Consider the problem $\Pigeon \subseteq [n]^m \times \binom{[m]}{2}$, for $m > n$, of finding a collision $x_i = x_j$ in a given $x \in [n]^m$. For $m =2n$ this is $\PWPP$-complete.
    Let $(\omega,f,g)$, with $\omega(\bm u)=\bm u$, be a public reduction from $\Pigeon$ to the canonical $\PLS$ problem, of depth $d$ and complexity $w=d+\log M$, where $M\geq4$ is the number of target pointers. A collision is verified with two source queries. Thus \cref{lem:ROM_to_proof_complexity} gives a DNF $B$ of width at most $3d+2\leq3w$ and a resolution refutation of $\Pigeon\lor B$ of width $O(w)$, such that
    $\Pr_{{\bm u} \sim [n]^m}[B({\bm u})=1] = \err(\omega,f,g)$.
    Choose an absolute constant $C$ such that both widths are at most $W=\lceil Cw\rceil$. The following lemma forces
    \[
        \err(\omega,f,g)\ge (1-W/n)^W\geq1-W^2/n=1-O(w^2/n),
    \]
    which proves \cref{thm:PWPP_not_in_PLS} for $w\leq n^{1/3}$ and sufficiently large $n$.

    \begin{lemma}
    \label{lem:pwpp-lb-random-res}
        If $B$ is a width-$w$ DNF such that
        $\Pr_{{\bm u} \sim [n]^m} [B({\bm u})=1] < (1-w/n)^w$ then there is a width-$w$ strategy for $\Pigeon \lor B$. 
    \end{lemma}
    \begin{proof}%
        By \cref{lem:robust-delayer} it suffices to show that $\Pigeon$ has a $w$-robust  strategy of width $w$. Define $\mathcal{F}$ to be the set of all partial injections of width at most $w$. $\mathcal{F}$ is downward-closed by definition and no injection falsifies $\Pigeon$, so it remains to show the query property. Take $\rho \in \mathcal{F}$ and arbitrary $i \in \free(\rho)$. For every $j \in [n] \setminus \range(\rho)$ (there are $n-w$ of those) the partial assignment $\rho \cup \{p_i = j\}$ is a partial injection, so if $|\fix(\rho)| < w$, $\rho \cup \{p_i = j\} \in \mathcal{F}$ as required.
    \end{proof}

As a consequence, we can also obtain average-case resolution \emph{size} lower bounds. In terms of \cref{def:res} the size is $|\mathcal{F}|$. Notice that we can relax \cref{def:res} so either $\rho$ fixes a solution or $i(\rho)$ is defined, regardless of the width of $\rho$. Then the terminal elements of $\mathcal{F}$ do not need to have the same width. The following corollary's proof is a variant of \cite[Proposition~3.7]{PT19}.
\begin{corollary}
    For any DNF $B$ satisfying $\Pr_{\bm u \sim [n]^m}[B(\bm u)=1] < (1-w/n)^w - 1/n$, any resolution refutation of $\Pigeon \lor B$ has size at least $n^{w}$.
\end{corollary}
\begin{proof}
    Suppose $\mathcal{F}$ is the (arbitrary width) refutation of $\Pigeon \lor B$ of size $<n^w$. Remove all elements of $\mathcal{F}$ of width $>w$ from $\mathcal{F}$ and form a DNF $B'$. Then $\Pr[B'(\bm u) = 1] < n^{w} \cdot n^{-(w+1)} = 1/n$, so there is a width-$(w+1)$ resolution refutation of $\Pigeon \lor B \lor B'$, which contradicts \cref{lem:pwpp-lb-random-res}.
\end{proof}

\subsection{$\PPA$ is not in $\PLS$}\label{subsec:ppa-notin-pls}

The preceding $\PPAD\not\in \FP$ separation already implies $\PPA\not\in \FP$.
Here we strengthen the conclusion: we construct a $\PPA$ search problem that has no low-depth average-case reduction to $\PLS$.

The problem is again to find an edge in the hypergraph $G = ([N], E)$ supplied by \cref{thm:graph-existence}: given a labeling $h\colon [N] \to \{0,1\}^r$ find $e \in E$ such that $\oplus_{v \in e} h(v) = 0^r$. We again plant a special edge $e^* = [\ell]$ and set $h(1)=\dots=h(\ell)=0^r$, denoting $\calH$ the set of such functions, and forbid $e^*$ as a solution:
\[\ZeroEdge_G \coloneqq \{(h,e)\in \calH \times (E \setminus \{e^*\});\; \oplus_{v \in e} h(v) = 0^r\}. \]
The reason for this change is in order to make $\ZeroEdge_G$
hard even for $\PLS$.
\begin{theorem}
\label{thm:ppa-vs-pls}
    There exists $G = ([N], E)$ such that $\ZeroEdge_G \in \PPA$ and there is no public reduction from $\ZeroEdge_G$ to $\PLS$ of reduction complexity $\exp((\log N)^{o(1)})$. 
\end{theorem}
The theorem immediately follows from \cref{lem:delayer-strategy-ppa,lem:ppa-containment} below:
\begin{lemma}
\label{lem:delayer-strategy-ppa}
    Suppose $G = ([N], E)$ and $E \subseteq \binom{[N]}{\ell}$ is $(\ell,w + \ell,b)$-well-spread. Then there exists a $b$-robust width-$w$ strategy for $\ZeroEdge_G$.
\end{lemma}
\begin{lemma}
\label{lem:ppa-containment}
    If $G = ([N], E)$ with $E \subseteq \binom{[N]}{\ell}$ is explicit and efficiently $r$-balanced, with $\ell=\poly(\log N)$, then $\ZeroEdge_G \in \PPA$.
\end{lemma}
\begin{proof}[Proof of \cref{thm:ppa-vs-pls}]
    The graph supplied by \cref{thm:graph-existence} satisfies the preconditions of \cref{lem:delayer-strategy-ppa,lem:ppa-containment} simultaneously. Then by \cref{lem:robust-delayer} and the robust strategy given by \cref{lem:delayer-strategy-ppa} we get that for every DNF $B$ of width $w=\exp((\log N)^{o(1)})$ and $\Pr[B(\bm u) = 1] \le 1/3$ there is no width-$w$ refutation for $\ZeroEdge_G \lor B$. Hence by \cref{lem:ROM_to_proof_complexity}, no public reduction from $\ZeroEdge_G$ to $\PLS$ exists. 
\end{proof}

\subsubsection*{$\PLS$ hardness: proof of \cref{lem:delayer-strategy-ppa}}

    Let $\mathcal{L}$ be the set of all partial assignments $\rho\colon [N]\setminus[\ell] \to \{0,1\}^r \cup \{\star\}$.  We say that $\rho \in \mathcal{L}$ \emph{reveals an edge} $e \in E$ if $\rho(v) \neq \star$ for all $v \in e \setminus [\ell]$ and  $\oplus_{v \in e \setminus [\ell]} \rho(v) = 0$. Then we simply define \[\mathcal{F} \coloneqq \{\rho \in \mathcal{L} \mid |\fix(\rho)| \le w\text{ and $\rho$ does not reveal any edge $e \in E \setminus \{e^*\}$}\}.\]
    Take any $\rho \in \mathcal{F}$ of width less than $w$ and any $v \in ([N] \setminus [\ell]) \setminus \fix(\rho)$. By well-spreadness we get that at most $b$ edges $F \subseteq E$ intersect $\fix(\rho) \cup [\ell] \cup \{v\}$ in at least $\ell$ nodes. By the fact that $\rho$ does not reveal any edge, the sum of labels in all edges that do not contain $v$ is non-zero. For each edge $e \in F$ such that $v\in e$ let $f_e \coloneqq \oplus_{z \in e \setminus [\ell] \setminus \{v\}} \rho(z)$ be the value of $h(v)$ that makes $e$ a solution to $\ZeroEdge_G$. Then for $\alpha \in \{0,1\}^r \setminus \{f_e \mid e \in F; v \in e\}$ the assignment $\rho \cup \{ h(v) = \alpha\} \in \mathcal{F}$, so since $|F| \le b$ the extension property of a $b$-robust strategy is satisfied.
\subsubsection*{$\PPA$ containment: proof of \cref{lem:ppa-containment}}
\label{subsec:ppa-system}

We follow exactly the same strategy as in the $\PPAD$ case: encode $\ZeroEdge$ as a system of polynomials which is unsatisfiable over the Boolean cube and show that it has a short $\mathbb{F}_2$-Nullstellensatz proof, which implies that it lies within $\PPA$ by \cite{GKRS19}. 
For every $a\in [N] \setminus [\ell]$, and $j\in[r]$, introduce
one Boolean variable $h_{a,j}$. 
Write
\[
 h(a)=(h_{a,1},\ldots,h_{a,r})\in\{0,1\}^r.
\]
Set $h_{a,j} \coloneqq 0$ for all $a \in [\ell]$ and $j \in [r]$; these are constants, not variables.
Thus $h(1)=\dots=h(\ell)=0^r$. 
Let $h=\{h_{a,j} \mid a \in [N]\setminus [\ell],\; j \in [r]\}$.
Then the number of Boolean variables in $h$ is
$n \coloneqq (N - \ell)r$.

For an edge $e \in E$, define
\[
 F_e(h)=\sum_{v \in e} h(v)\in\F_2^r,
 \qquad
 F_e(h)_j=\sum_{v \in e} h_{v,j}.
\]
where the first sum is coordinate-wise XOR and the second sum
is over $\F_2$. Define

\begin{equation}\label{eq:ppa-system}
 \mathcal P_G=\{p_e(h) \mid e \in E \setminus \{e^*\}\},\qquad
 p_e(h)=\prod_{j=1}^r\left(1+\sum_{v \in e} h_{v,j}\right).
\end{equation}

Every polynomial is over $\F_2$. 
The set of polynomials in ${\cal P}_G$ will be our
system of (unsatisfiable) polynomial equations. 
On a Boolean assignment,
\[
 p_e(h)=1\quad\Longleftrightarrow\quad F_e(h)=0^r.
\]

The next lemma shows that if $E$ is $r$-balanced, then the
system of polynomial equations $(\mathcal P_G)$ is unsatisfiable,
and therefore $\ZeroEdge_G$
is total. Secondly, we exhibit a simple $\NS_2$ refutation
for ${\mathcal P_G}$, which by \cite{GKRS19} implies a \emph{black-box} containment $\ZeroEdge_G \in \PPA^{\rm dt}$. 

\begin{lemma}[$\NS_2$ refutation of ${\cal P}_G$] 
\label{lem:ppa-totality}
The system $\mathcal P_G$ is unsatisfiable and has the $\NS_2$
refutation
 \[1=\sum_{e \in E \setminus \{e^*\}} p_e(h) \mod 2.\]
Furthermore, the $\NS_2$ certificate, $\sum_{e \in E \setminus \{e^*\}} p_e$ has degree $r$ and at most $|E|(\ell + 1)^r$ monomial occurrences before cancellation.
\end{lemma}
\begin{proof}
Let $p_{e^*}(h)$ be defined as above (although it isn't one of the
equations in ${\cal P}_G$); that is, $p_{e^*}(h) = \prod_{j \in [r]} (1 + F_{e^*}(h)_j)$.
We first show that by the balancedness of $E$ we have:
$\sum_{e \in E} p_e(h) = 0$ over $\mathbb{F}_2$.
To see this first note that when we expand $p_e(h)$ we get the sum of terms of the form
\[h_{a_1,j_1} h_{a_2,j_2} \ldots h_{a_s,j_s},\]
where $s \leq r$, all $j_k$'s are distinct, and $a_1,\dots,a_s \in e$. 
Now, summing over all $e$ and expanding each $p_e(h)$ we want to show that every term appears an even number of times in
$\sum_{e\in E} p_e(h)$.
Each term appears exactly once in each $p_e(h)$ where $e$ contains the set of nodes mentioned in it.  By the $r$-balancedness the number of edges that contain a fixed subset of size at most $r$ is even. 

On the other hand, since $h(1)=\dots=h(\ell)=0$, the polynomial $p_{e^*}(h) = p_{[\ell]}(h)$ is equal to $1$:
\[p_{e^*}(h) = \prod_{j \in [r]} (1 + F_{e^*}(h)_j) = \prod_{j \in [r]} (1 + \sum_{i\in [\ell]} h(i)_j) = \prod_{j \in [r]} (1 + 0) =1.\]
Therefore since $p_{e^*}(h)=1$ and $\sum_{e \in E} p_e(h) =0$,
it follows that $\sum_{e \in E \setminus \{e^*\}} p_e(h)=1$ over $\mathbb{F}_2$.
\end{proof}
We have shown not only that ${\cal P}_G$ is unsatisfiable, but
that there is a simple $\NS_2$ refutation, $\sum_e p_e q_e=1$, (where the $q_e$'s are all $1$) obtained by simply taking the sum of all polynomials in ${\cal P}_G$.
Lastly, each nonconstant $p_e$ has degree $r$, and there are at most $(\ell + 1)^r$ terms in each $p_e$. Therefore, the proof size (number of monomials before cancellation) is
at most $|E|(\ell + 1)^r = 2^{\poly(\ell)}$.

\paragraph{Translating from $\NS_2$ to $\PPA$.}
To finish the proof of $\PPA$ inclusion, we instantiate the conversion of a $\F_2$-Nullstellensatz refutation to a $\PPA$ instance by
\cite[Lemma~13]{GKRS19}. Their generic proof does not necessarily imply an efficient instance. We use the fact that $G$ is \emph{efficiently} $r$-balanced. Recall that it means that for every set $S \subseteq [N]$ of size $\le r$ the edges $e \supseteq S$ can be efficiently paired-up. 

First recall the canonical $\PPA$-complete problem. 
\begin{definition}
    \textsc{Lonely} problem is defined on a single array $\Mate \in ([2N + 1])^{2N + 1}$ 
    \begin{itemize}[noitemsep]
        \item $i \in [2N+1]$ if $\Mate_{\Mate_i} \neq i$, \hfill (mismatch)
        \item $i \in [2N+1]$ if $\Mate_i = i$. \hfill (lonely node)
    \end{itemize}
\end{definition}

Every monomial \emph{occurrence} in the refutation $\sum_{e \in E \setminus \{e^*\}} p_e(h) + 1 = 0$ can be encoded by a pair $(e, u)$ with $e \in E$ and $u \in (e \cup \{\bot\})^r$, let $V$ be the set of these encodings, excluding $(e^*, \bot^r)$. Then $|V|=|E|(\ell+1)^r-1$ is odd by balancedness for the empty set. A pair $(e,u)$ corresponds to $m_u \coloneqq \prod_{i \in [r] \colon u_i \neq \bot} h_{u_i, i}$ (if $u_i \in e^*$, then $h_{u_i, i}$ is a constant $0$). We define the instance $\mate\colon \mathcal{H} \to V^V$ to reflect the cancellation of the different occurrences of the same monomial. Take some $(e,u)$ such that $m_u(h) = 0$, i.e. the corresponding monomial evaluates to $0$ in the refutation. Such monomials are not useful for finding a solution to $\ZeroEdge_G$, so we match them up using \emph{syntactic} properties of $\NS_2$ refutation. Let $S \subseteq e$ be the set of non-$\bot$ coordinates in $u$. By efficient $r$-balancedness there exists an involution $\sigma_S \colon E_S \to E_S$ that matches-up the edges $e$ containing the set $S$. Then we define $\mate_{e,u}(h) = (\sigma_S(e), u)$. Now we need to match the monomial occurrences that evaluate to $1$ in the refutation. Take $e \in E \setminus \{e^*\}$. If $p_e(h) = 1$, we just define $\mate_{e,\bot^r}(h) = (e,\bot^r)$ and pair-up all the other nodes corresponding to monomials evaluating to $1$ on $h$. If any such node is a solution to the $\PPA$-instance, $(h,e) \in \ZeroEdge_G$ by definition. If $p_e(h) = 0$, the number of monomials in $p_e$ that evaluate to $1$ is even, so we can pair them arbitrarily. To make the pairings above efficient observe that the set of tuples $u \in (e \cup \{\bot\})^r$ such that $m_u(h) = 1$ is a product set $\prod_{i \in [r]} (\{\bot\} \cup \{v \in e \mid h_{v,i} = 1\})$, so we can efficiently pair the elements up in the lexicographic order (excluding $\bot^r$ in the case $p_e(h) = 1$). Then all solutions correspond to solutions of $\ZeroEdge_G$ as required.

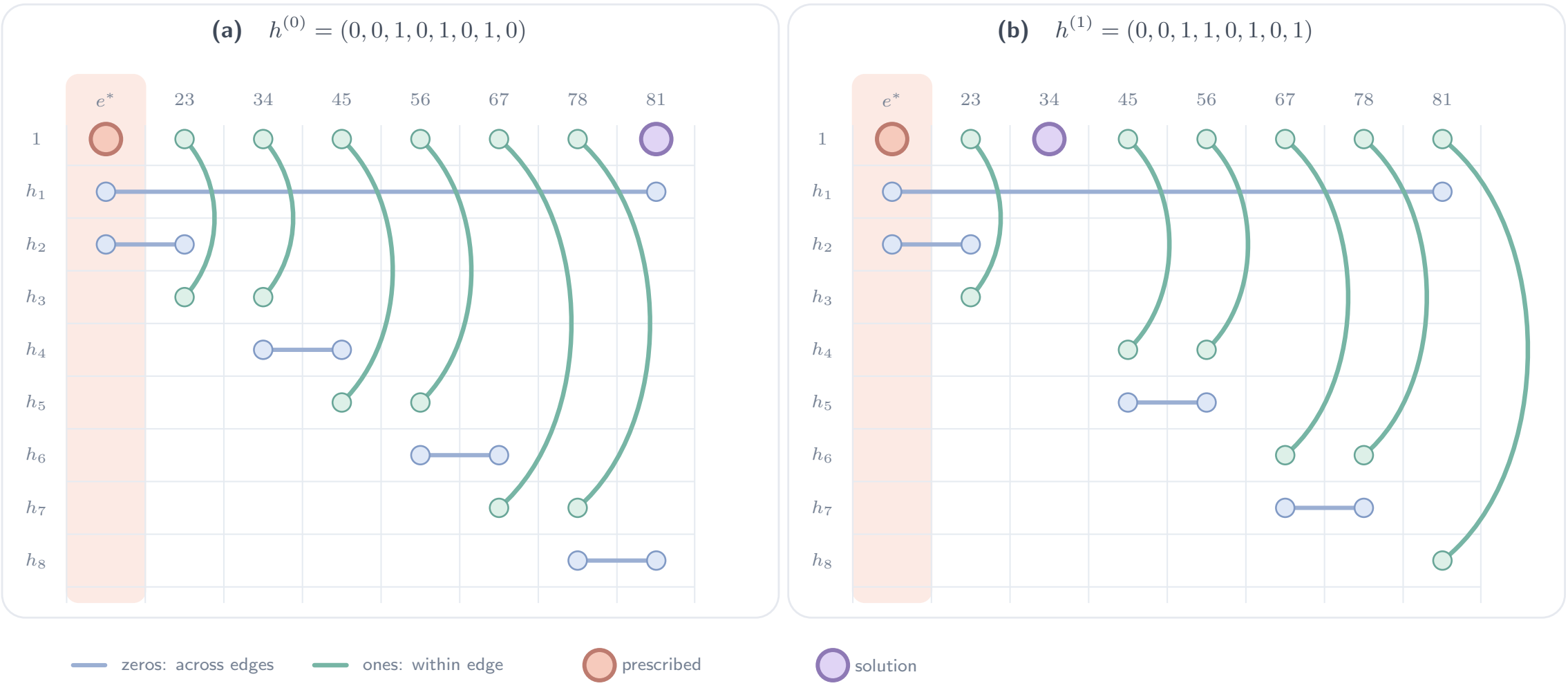
\begin{figure}
    \centering
    \scalebox{0.7}{\definecolor{zlInk}{HTML}{3E4857}
\definecolor{zlMuted}{HTML}{748092}
\definecolor{zlGrid}{HTML}{E6EBF1}
\definecolor{zlFrame}{HTML}{E6E9EF}
\definecolor{zlSpecial}{HTML}{FCEAE4}
\definecolor{zlBlue}{HTML}{9BAFD3}
\definecolor{zlBlueFill}{HTML}{DFE8F7}
\definecolor{zlBlueEdge}{HTML}{819AC5}
\definecolor{zlGreen}{HTML}{77B5A5}
\definecolor{zlGreenFill}{HTML}{DDF0E8}
\definecolor{zlGreenEdge}{HTML}{69A698}
\definecolor{zlPeach}{HTML}{F6CABC}
\definecolor{zlPeachEdge}{HTML}{BD7B6F}
\definecolor{zlLavender}{HTML}{E0D4F5}
\definecolor{zlLavenderEdge}{HTML}{8E79B5}

\begin{tikzpicture}[
  x=1cm,y=1cm,
  every node/.style={font=\sffamily, text=zlInk},
  zlzeropair/.style={draw=zlBlue, line width=1.65pt, line cap=round},
  zlonepair/.style={draw=zlGreen, line width=1.75pt, line cap=round},
  zlzero/.style={circle, fill=zlBlueFill, draw=zlBlueEdge, line width=.70pt,
      inner sep=0pt, minimum size=7.2pt},
  zlone/.style={circle, fill=zlGreenFill, draw=zlGreenEdge, line width=.70pt,
      inner sep=0pt, minimum size=7.2pt},
  zlstart/.style={circle, fill=zlPeach, draw=zlPeachEdge, line width=1.60pt,
      inner sep=0pt, minimum size=11.8pt},
  zlsolution/.style={circle, fill=zlLavender, draw=zlLavenderEdge, line width=1.60pt,
      inner sep=0pt, minimum size=11.8pt}
]

\newcommand{\zlCyclePanel}[3]{%
  \begin{scope}[shift={(#2,0)}]
    \filldraw[rounded corners=9pt, fill=white, draw=zlFrame, line width=.75pt]
      (.13,-6.80) rectangle (10.75,1.58);
    \fill[zlSpecial, rounded corners=5pt]
      (1.00,-6.60) rectangle (2.10,.63);
    \node[font=\sffamily\small\bfseries] at (5.15,1.24) {#3};
    \foreach \c/\edgeLabel in {1/{e^*},2/{23},3/{34},4/{45},5/{56},6/{67},7/{78},8/{81}} {%
      \pgfmathsetmacro{\cx}{1.55+1.075*(\c-1)}
      \node[font=\sffamily\scriptsize, text=zlMuted] at (\cx,.28)
         {$\edgeLabel$};
      \draw[zlGrid, line width=.55pt]
         (\cx-.5375,-.07)--(\cx-.5375,-6.60);
    }
    \draw[zlGrid, line width=.55pt] (9.6000,-.07)--(9.6000,-6.60);
    \foreach \r/\mon in {1/{1},2/{h_1},3/{h_2},4/{h_3},5/{h_4},6/{h_5},7/{h_6},8/{h_7},9/{h_8}} {%
      \pgfmathsetmacro{\cy}{-.26-.72*(\r-1)}
      \node[font=\sffamily\scriptsize, text=zlMuted]
        at (.60,\cy) {$\mon$};
      \draw[zlGrid, line width=.55pt]
         (1.0125,\cy-.36)--(9.6000,\cy-.36);
      \foreach \c in {1,...,8} {%
        \pgfmathsetmacro{\cx}{1.55+1.075*(\c-1)}
        \coordinate (#1-\c-\r) at (\cx,\cy);
      }
    }
  \end{scope}
}

\zlCyclePanel{A}{0}{(a)\quad $h^{(0)}=(0,0,1,0,1,0,1,0)$}
\zlCyclePanel{B}{10.75}{(b)\quad $h^{(1)}=(0,0,1,1,0,1,0,1)$}

\draw[zlzeropair] (A-1-2)--(A-8-2); %
\draw[zlzeropair] (A-1-3)--(A-2-3); %
\draw[zlzeropair] (A-3-5)--(A-4-5); %
\draw[zlzeropair] (A-5-7)--(A-6-7); %
\draw[zlzeropair] (A-7-9)--(A-8-9); %
\draw[zlonepair] (A-2-1) to[out=-48,in=48,looseness=.97] (A-2-4);
\draw[zlonepair] (A-3-1) to[out=-48,in=48,looseness=.97] (A-3-4);
\draw[zlonepair] (A-4-1) to[out=-44,in=44,looseness=.92] (A-4-6);
\draw[zlonepair] (A-5-1) to[out=-44,in=44,looseness=.92] (A-5-6);
\draw[zlonepair] (A-6-1) to[out=-41,in=41,looseness=.89] (A-6-8);
\draw[zlonepair] (A-7-1) to[out=-41,in=41,looseness=.89] (A-7-8);
\foreach \c/\r in {1/2,8/2,1/3,2/3,3/5,4/5,5/7,6/7,7/9,8/9}
  \node[zlzero] at (A-\c-\r) {};
\foreach \c/\r in {2/1,2/4,3/1,3/4,4/1,4/6,5/1,5/6,6/1,6/8,7/1,7/8}
  \node[zlone] at (A-\c-\r) {};
\node[zlstart] at (A-1-1) {};
\node[zlsolution] at (A-8-1) {};

\draw[zlzeropair] (B-1-2)--(B-8-2); %
\draw[zlzeropair] (B-1-3)--(B-2-3); %
\draw[zlzeropair] (B-4-6)--(B-5-6); %
\draw[zlzeropair] (B-6-8)--(B-7-8); %
\draw[zlonepair] (B-2-1) to[out=-48,in=48,looseness=.97] (B-2-4);
\draw[zlonepair] (B-4-1) to[out=-43,in=43,looseness=.92] (B-4-5);
\draw[zlonepair] (B-5-1) to[out=-43,in=43,looseness=.92] (B-5-5);
\draw[zlonepair] (B-6-1) to[out=-40,in=40,looseness=.89] (B-6-7);
\draw[zlonepair] (B-7-1) to[out=-40,in=40,looseness=.89] (B-7-7);
\draw[zlonepair] (B-8-1) to[out=-38,in=38,looseness=.88] (B-8-9);
\foreach \c/\r in {1/2,8/2,1/3,2/3,4/6,5/6,6/8,7/8}
  \node[zlzero] at (B-\c-\r) {};
\foreach \c/\r in {2/1,2/4,4/1,4/5,5/1,5/5,6/1,6/7,7/1,7/7,8/1,8/9}
  \node[zlone] at (B-\c-\r) {};
\node[zlstart] at (B-1-1) {};
\node[zlsolution] at (B-3-1) {};

\begin{scope}[shift={(1.1,-7.45)}]
  \draw[zlzeropair] (0,0)--(.44,0);
  \node[font=\sffamily\scriptsize, anchor=west, text=zlMuted]
     at (.54,0) {zeros: across edges};
  \draw[zlonepair] (3.30,0)--(3.74,0);
  \node[font=\sffamily\scriptsize, anchor=west, text=zlMuted]
     at (3.84,0) {ones: within edge};
  \node[zlstart] at (7.20,0) {};
  \node[font=\sffamily\scriptsize, anchor=west, text=zlMuted]
     at (7.38,0) {prescribed};
  \node[zlsolution] at (10.39,0) {};
  \node[font=\sffamily\scriptsize, anchor=west, text=zlMuted]
     at (10.57,0) {solution};
\end{scope}

\end{tikzpicture}}
    \caption{\small This picture illustrates the translation with $r=1$ and $G$ being the ordinary $8$-node cycle with $e^* = \{1,2\}$. The problem $\ZeroEdge_G$ in this case is given a boolean labeling $h \in \{0,1\}^8$ with $h(1)=h(2)=0$ to find an edge with equal nodes labels. One can view this instance of $\ZeroEdge_G$ as a Tseitin formula for the dual of $G$, so the $\NS_2$-refutation is simply $\sum_{uv \in E} (1 + h_u + h_v) = 0$.
    Each cell of the table corresponds to some pair edge-monomial (monomials have degree $\le 1$ here), but only the cells $(e,u)$ with $u \in \{\bot\} \cup e$ are nodes of the $\PPA$ instance.}
    \label{fig:zeroedge}
\end{figure}

\section*{Acknowledgements}
\addcontentsline{toc}{section}{Acknowledgements}
Ziyi Guan is funded by the Swiss National Science Foundation (SNSF) through a Postdoc.Mobility fellowship. Oliver Korten is supported by the National Science Foundation under Grant No. DMS-2424441. Artur Riazanov is supported by the Simons
Foundation as a Junior Fellow in the Simons Society of
Fellows (20599). Noah Fleming is supported by the Swedish Research Council under grant number 2025-06762.
Mika G\"o\"os and Weiqiang Yuan are supported by the Swiss State Secretariat for Education, Research, and Innovation (SERI) under contract number MB22.00026.

The authors thank Nir Bitansky, Tal Malkin, Yaroslav Alekseev, Gilbert Maystre, Eylon Yogev, and Yiding Zhang for discussions related to this project.

\section*{AI disclosure}
\addcontentsline{toc}{section}{AI disclosure}
The separation from $\PWPP \not\subseteq \PLS$ was obtained by the human authors, using ChatGPT~5.5 for assistance with writing the proof. 
The human authors obtained the collapse of $\UEoPL$ by asking ChatGPT~6 to refine their initial recursive algorithm using results from \cite{BCG26}.
The separations $\PPA \not\subseteq \PLS$ and $\PPAD \not\subseteq \FP$ were obtained by ChatGPT~6 supplied with the previous two results.
In the generated proof, the use of Nullstellensatz equivalence was hidden. This paper is written by humans; all proofs were significantly simplified.

\DeclareUrlCommand{\Doi}{\urlstyle{sf}}
\renewcommand{\path}[1]{\small\Doi{#1}}
\bibliographystyle{alphaurl}
\bibliography{refs}

\appendix

\section{$\TFNP$ glossary}\label{glossary}
In this section, we give formal definitions of all $\TFNP$ classes we use. With the exception of the entire class $\TFNP$ and the smallest subclass (total) $\FP$ that contains efficiently solvable total relations, all subclasses are defined by canonical complete problems. Then if $F$ is a complete relation for some class $\mathcal{C}$ we say that $F' \in \mathcal{C}$ iff there exists a zero-error public reduction (see \cref{sec:PLS_separations}) from $F'$ to $F$.

\paragraph{$\PPP$ \textsc{(Pigeon)}.}
We are given $\Pigeon\in[N]^{N+1}$; a solution is  
\begin{itemize}[noitemsep]
    \item $i \neq j \in [N+1]$ such that $\Pigeon_i = \Pigeon_j$.\hfill (collision)
\end{itemize}

\paragraph{$\PWPP$ \textsc{(WeakPigeon)}.} We are given $\Pigeon \in [N]^{2N}$; a solution is
\begin{itemize}[noitemsep]
    \item $i \neq j \in [2N]$ such that $\Pigeon_i = \Pigeon_j$.\hfill (collision)
\end{itemize}
\paragraph{$\PPA$ \textsc{(Lonely)}.} We are given $\Mate\in [2N+1]^{2N+1}$; solutions are
\begin{itemize}[noitemsep]
    \item $i \in [2N+1]$ such that $\Mate_i = i$;\hfill (lonely node)
    \item $i \in [2N+1]$ such that $\Mate_{\Mate_i} \neq i$.\hfill (mismatch)
\end{itemize}
\paragraph{$\PPAD$ (\textsc{End of Line}).} We are given $\Succ, \Pred \in [N]^N$; solutions are
\begin{itemize}[noitemsep]
    \item $1$ if $\Pred_1 \neq 1$ or $\Succ_1 = 1$ or $\Pred_{\Succ_1}\neq1$; \hfill (distinguished node is not a source)
    \item $i \in [N] \setminus \{1\}$ if $\Succ_{\Pred_i} \neq i$.\hfill (non-canonical source)
    \item $i \in [N]$ if $\Pred_{\Succ_i} \neq i$. \hfill (active sink)
\end{itemize}

\paragraph{A common grid formulation.}
For $\PLS$, $\EoPL$, and $\UEoPL$, use the $N\times N$ grid, $N\geq2$, of \cref{def:ueopl}. Every node has a successor $\Succ_{ij}\in[N]\cup\{\bot\}$; for $i<N$, a non-$\bot$ value $k$ names $(i+1,k)$. For $\PLS$ a node is \emph{active} if $\Succ_{ij} \neq \bot$ (for $\EoPL$ and $\UEoPL$ we restrict the activity requirement). A \emph{proper sink} is an inactive node reached from an active node.

\paragraph{$\PLS$ \textsc{(Sink of Dag)}.}
Solutions are
\begin{itemize}[noitemsep]
    \item $(1,1)$ if it is inactive;\hfill (no canonical source)
    \item $(N,j)$ if it is active;\hfill (sink)
    \item $(i,j)$ with $i<N$ if it is active, but $(i+1, \Succ_{ij})$ is inactive.\hfill (predecessor of a proper sink)
\end{itemize}

In the third case we output the active predecessor, which witnesses an incoming edge to the inactive sink. This makes the solution locally checkable even though $\PLS$ has no predecessor pointers, as in~\cite[Section~3.2]{GHJ24}.

\paragraph{$\EoPL$ \textsc{(End of Potential Line)}.}
Add pointers $\Pred_{ij}\in[N]\cup\{\bot\}$ for rows $2,\ldots,N$, and {\bf redefine} activity throughout: for $i<N$, node $(i,j)$ is active iff $\Succ_{ij}=k\neq\bot$ and $\Pred_{i+1,k}=j$; in row~$N$, activity still means $\Succ_{N j}\neq\bot$. Keep the first two $\PLS$ solution types with this activity definition, replace the third by the proper sink itself, and add proper sources:
\begin{itemize}[noitemsep]
    \item $(i,j)$ if it is a proper sink; \hfill (proper sink)
    \item $(i,j)\neq(1,1)$ with $i<N$ if it is active and no active node has it as a successor. \hfill (proper source)
\end{itemize}
In particular, every active node in row~1 except $(1,1)$ is a proper source. A proper sink itself is locally checkable here because its predecessor pointer identifies the only possible active predecessor; see \cref{def:ueopl}.

\paragraph{$\UEoPL$ \textsc{(Unique End of Potential Line)}.}
Use the same instance, activity, and four solution types as $\EoPL$, and add just one further type:
\begin{itemize}[noitemsep]
    \item $(i,j)$ and $(i,j')$ with $j\neq j'$ both active.\hfill (parallel active nodes)
\end{itemize}

\section{$\TFNP$ separations in the ROM}
\label{appendix:alphabet-question}

In this section, we will give a formal definition of $\TFNP$ problems and reductions under the random oracle model. Let $\Gamma_{a(n),b(n)}$ be the set of all Boolean functions $\gamma\colon \{0,1\}^{a(n)}\rightarrow \{0,1\}^{b(n)}$.

\begin{definition}[Search problem in an oracle model]
A search problem in the oracle-model of 
$\Gamma_{a(n),b(n)}$
is a family of relations $\mathcal{R} = \{R_n\}_{n \ge 1}$ where $R_n \subseteq \{0,1\}^n \times \Gamma_{a(n),b(n)} \times \{0,1\}^{s(n)}$.
$\mathcal{R}$ is total if for all $n$, $x \in \{0,1\}^n$ and $\gamma_n \in \Gamma_{a(n),b(n)}$ there exists $y$ such that $(x,\gamma_n,y) \in R_n$.
\end{definition}

We say that an algorithm $A$ average-case solves some oracle search problem $\mathcal{R}^{(\cdot)}$ in the random oracle model if for all sufficiently large $n$, when $\gamma_n$ is sampled uniformly over all functions from $a(n)$ bits to $b(n)$ bits,
\begin{equation*}
    \Pr_{\bm x \sim \{0,1\}^n,\bm \gamma_n \sim \Gamma_{a(n),b(n)}}[A^{\bm \gamma_n}(\bm x)~\text{outputs}~y~\text{such that}~(\bm x,\bm \gamma_n,y)\in R_n] \ge \frac23.
\end{equation*}
We say that an oracle $\TFNP$ search problem $\mathcal{R}^{(\cdot)}$ is \emph{efficiently reducible} to $\mathcal{S}^{(\cdot)}$ \emph{in the ROM}, if there are oracle algorithms $f^{(\cdot)}, g^{(\cdot)}$ in $\poly(n)$ time, such that
\begin{equation*}
    \Pr_{\bm x \sim \{0,1\}^n, \bm \gamma_n \sim \Gamma_{a(n),b(n)}}[\forall y,~(f^{\bm \gamma_n}(1^n, \bm x), \gamma_n, y) \in \mathcal{S}^{(\cdot)} \implies (\bm x, \bm \gamma_n, g^{\bm \gamma_n}(1^n, \bm x, y)) \in \mathcal{R}^{(\cdot)} ] \ge \frac23.
\end{equation*}

One may assume that $a(n), b(n), s(n)$ are all bounded by polynomials. For all the \emph{circuit $\TFNP$ problems} we mentioned in this paper, i.e., $\TFNP$ problems, where the input is the description of a circuit whose truth table encodes an object of exponential length, it is natural to define the corresponding oracle $\TFNP$ problem by letting the input be descriptions of \emph{oracle circuits}. For example, the oracle search problem $\PWPP^{(\cdot)}$ has the input $x$ being the description of some oracle circuit $C^{(\cdot)}\colon \{0,1\}^m \to \{0,1\}^{m-1}$, and $(x, \gamma_n, (y_1, y_2)) \in \PWPP^{(\cdot)}$ if and only if $C^{\gamma_n}(y_1) = C^{\gamma_n}(y_2)$ for $y_1 \neq y_2$.

Note that any of the circuit $\TFNP$ problems we considered in this paper is equivalent to \emph{another} problem where the input oracle circuit is promised to be the canonical circuit $\widehat{C}^{\gamma}(x) \coloneqq \gamma(x)$. It is not hard to see that reductions between oracle versions of the $\TFNP$ problems in the ROM can be viewed as \emph{uniform} versions of average-case reductions between the corresponding black-box problems, where each node of the decision trees can be locally printed by a uniform algorithm in $\polylog(N)$ time.

\end{document}